\documentclass{./jfm}

\usepackage{graphicx}
\usepackage{natbib}
\usepackage[breaklinks,colorlinks = true,linkcolor = blue,urlcolor=blue,citecolor=blue]{hyperref}
\usepackage{amsmath}
\usepackage[capitalise]{cleveref}
\usepackage{ar}
\usepackage{tikz}
\usepackage{pgfplots}
\usepackage{tikz-3dplot}
\usepackage{bm}
\usepackage{multirow}
\usepackage{subfig}
\usepackage{overpic}
\usepackage{ulem}
\usepackage{tikz}
\usepackage{mathrsfs}

\def \gradij {A_{ij}}
\newcommand{\aver}[1]{  \left\langle {#1} \right \rangle }

\title{Energy, enstrophy and helicity transfers in polymeric turbulence}

\author[A. Chiarini, R.K. Singh, M.E. Rosti]
{Alessandro Chiarini\aff{1,2}\corresp{\email{alessandro.chiarini@polimi.it}}, 
Rahul K. Singh\aff{1}\corresp{\email{rksphys@gmail.com}}
and 
Marco E. Rosti\aff{1}\corresp{\email{marco.rosti@oist.jp}}
}
\affiliation{
\aff{1} Complex Fluids and Flows Unit, Okinawa Institute of Science and Technology Graduate University, 1919-1 Tancha, Onna-son, Okinawa 904-0495, Japan\\
\aff{2} Dipartimento di Scienze e Tecnologies Aerospaziali, Politecnico di Milano, via La Masa 34, 20156 Milano, Italy
}

\begin{document}
\maketitle

\begin{abstract}
We investigate the scale-by-scale transfers of energy, enstrophy and helicity in homogeneous and isotropic polymeric turbulence using direct numerical simulations. The study relies on the exact scale-by-scale budget equations, derived from the governing model equations, that fully capture the back-reaction of polymers on the fluid dynamics. Polymers act as dynamic sinks and sources and open alternative routes for interscale transfer whose significance is modulated by their elasticity, quantified through the Deborah number ($De$). Polymers primarily deplete the nonlinear energy cascade at small scales, by attenuating intense forward and inverse transfer events. At sufficiently high $De$, a polymer-driven flux emerges and dominates at small scales, transferring on average energy from larger to smaller scales, while allowing for localised backscatter. For enstrophy, polymers inhibit the stretching of vorticity, with fluid-polymer interactions becoming the primary enstrophy source at high $De$. Accordingly, an analysis of the small-scale flow topology reveals that polymers promote two-dimensional straining states and enhance the occurrence of shear and planar extensional flows, while suppressing extreme rotation events. Helicity, injected at large scales, exhibits a transfer mechanism analogous to energy, being dominated by nonlinear dynamics at large scales and by polymer-induced fluxes at small scales. Polymers enhance the breakdown of small-scale mirror symmetry, as indicated by a monotonic increase in relative helicity with $De$ across all scales.
\end{abstract}


\section{Introduction}

A small concentration of polymers in a turbulent flow results in a substantial decrease of drag, as well documented in several experimental, numerical and theoretical works \citep{toms-1948,lumley-1969,berman-1978,benzi-ching-2018}. In a turbulent wall-bounded flow, the polymeric chains influence the energy- momentum transfers in the real space and in the space of scales, giving rise to an increased buffer region that eventually leads to drag reduction \citep[see for example][]{warwaruk-ghaemi-2024}. It is known that large drag reduction may occur without a substantial reduction of the turbulent kinetic energy, which may indeed be either smaller or larger compared to the purely Newtonian counterpart \citep{tsinober-1990}. Polymeric turbulence, therefore, is not necessarily associated with the suppression of turbulence, but rather with qualitative/quantitative changes of its structure. Despite the large interest, however, the way polymeric additives modulate turbulence is not completely understood yet, and their influence on the statistical nature of the velocity fluctuations is still under debate. In this work, we consider the idealised set up of homogeneous and isotropic turbulence (HIT), and investigate the influence of polymeric additives on the cascades of energy, helicity and enstrophy in the space of scales.

\subsection{Polymers in homogeneous and isotropic turbulence}

Two dimensionless numbers are needed to characterise turbulent flows of dilute polymeric solutions in a triperiodic box. They are the Reynolds number $Re$, which estimates the importance of the inertial term compared to the viscous one in the Navier--Stokes equations, and the Deborah number $De$ that quantifies the unique polymer relaxation time scale with respect to the large time scale of the flow. A large $De$  therefore means the polymers are more elastic, and extensible. Turbulent drag reduction has been mainly observed for large Re and $De$, with an evident departure from the Kolmogorov predictions already in the idealised setting of HIT \citep{tabor-gennes-1986,bhattacharjee-thirumalai-1991,fouxon-lebedev-2003}. At a relatively low Reynolds number, $Re_\lambda = u' \lambda/\nu \approx 80$ ($u'$ is the typical velocity fluctuation, $\lambda$ is the Taylor length-scale and $\nu$ is the kinematic fluid viscosity), \cite{perlekar-mitra-pandit-2010} observed that in the presence of the polymers the energy content decreases at the intermediate scales and significantly increases at the smallest scales, in the deep dissipative range. They also observed that the fluid dissipation monotonically decreases as $De$ increases, although they fixed the injected energy using the forcing introduced by \cite{lamorgese-caughey-pope-2005}, hinting that polymers give origin to an alternative route to the classical energy cascade.

At large $Re$ and $De$, the polymer additives modify the nature of the energy cascade and result in a significant alteration of the energy distribution amongst scales. For large enough separation between the energy injection scale $\mathcal{L}$ and the Kolmogorov scale $\eta$, there exists an intermediate scale $r_p$ that separates two different regimes. For $r_p \le r \le \mathcal{L}$ the energy cascade resembles that of a Newtonian flow, with the classical second-order structure function scaling $S_2(r) \sim r^{2/3}$ predicted by the Kolmogorov theory. For $\eta \le r \le r_p$, the elastic range of scales, energy is transferred among scales by the polymeric microstructure as well, and the second-order structure function $S_2$ increases faster than $r^{2/3}$, deviating from the Kolmogorov predictions \citep{deangelis-etal-2005}. Recently, \citet{zhang-etal-2021} and \citet{rosti-perlekar-mitra-2023} provided experimental and numerical evidence that, for large enough $De$, there is a range of scales $ \eta \lesssim r \lesssim r_p$ where $S_2(r) \sim r^\xi$, with $\xi \approx 1.3 \approx 4/3$.
By analysing the scale-by-scale energy budget in Fourier space, \cite{rosti-perlekar-mitra-2023} showed that, for sufficiently large $De$, the energy flux remains dominated by advective nonlinearities at the large scales as in Newtonian turbulence, while polymeric stresses become the dominant contribution at smaller scales. Interestingly, they also observed that $r_p$ has a non-monotonic dependence on $De$, and follows the cross-over scale between the nonlinear flux and the polymer-driven flux. As a consequence, the width of the elastic range of scales does not show a monotonic dependence on the $De$, and is maximum for $De = \mathcal{O}(1)$.
More recently, \cite{singh-rosti-2024} investigated how polymeric turbulence depends on $De$ and $Re$, linking the low-$Re$ regime described by \cite{singh-etal-2024} with the large-$Re$ regime described by \cite{rosti-perlekar-mitra-2023}. They found that in polymeric turbulence the dissipation field is intermittent and shows a qualitative similar distribution as in purely Newtonian turbulence.

However, the influence of the polymers on the average picture of the energy cascade does not provide an exhaustive understanding of polymeric turbulence. On one side, indeed, already in Newtonian turbulence it is well known that the (average) forward energy cascade from larger to smaller scales is accompanied by local backscatter events where energy is transferred from smaller to larger scales. On the other side, despite energy is a central quantity in most of the theories, a complete understanding of polymeric turbulence requires to also focus on other small-scale related quantities such as enstrophy and helicity, which are strongly related to important dynamical processes such as vortex stretching and dissipation. In this work we do a step in this direction. We consider polymeric homogeneous and isotropic turbulence (PHIT) at large $Re$ and over a wide range of $De$, and, introducing a suitable framework, we address the following questions: How do polymers modulate the local direct/inverse energy transfers? How intermittent is the polymer-driven energy transfer route? Is it characterised by local backscatter events? How do polymers modulate the transfers of enstrophy and helicity among scales and the related dynamical processes?

\subsection{Enstrophy and Helicity}

When studying turbulent flows, the kinetic energy has always been of a great interest: it is an invariant of the inviscid Navier--Stokes equations and it is central to most of the theories developed for high Re turbulence \citep{frisch-1995}. However, it is not the only key quantity, and a general description of turbulence also requires the characterisation of quantities related with the smallest scales of the flow. In fact, it is well known that the energy cascade is statistically related to dissipation \citep{kolmogorov-1941}. Under homogeneity and isotropy, in the inertial range of scales, the nonlinear flux and the dissipation $\varepsilon_f$ are related by the celebrated $4/5$-th law, which directly descends from the K\'{a}rm\'{a}n-Howarth equation \citep{frisch-1995,pope-2000}. 
The smallest scales of the flow are characterised by large velocity gradients that lead to large values of the vorticity $\bm{\omega} = \bm{\nabla} \times \bm{u}$, that is central in the definition of other scalar quantities as enstrophy $\omega^2 = \bm{\omega} \cdot \bm{\omega}$ and helicity $ h = \bm{\omega} \cdot \bm{u}$. Although polymers are known to significantly alter the small scales \citep{perlekar-mitra-pandit-2010} by suppressing events of large vorticity and strain \citep{liberzon-etal-2005,liberzon-etal-2006}, an exhaustive characterisation is missing. A complete understanding of how polymers modulate turbulence thus requires a deeper investigation of their influence on the enstrophy and helicity, i.e. on the small-scale velocity gradients.

In HIT, enstrophy is well known to be closely related to dissipation: they have the same mean value up to a constant given by the fluid viscosity $\nu$ \citep{tsinober-2001}. However, they are two different descriptors of the structure of the small scales of turbulence \citep{sreenivasan-antonia-1997}, and extreme events of dissipation and enstrophy are rather different. Extreme dissipation represents intense local strain, while large enstrophy represents strong vortical motions. Despite their intrinsic difference, however, \cite{yeung-donzis-sreenivasan-2012} showed that at sufficiently large $Re$ extreme events of dissipation and enstrophy scale similar, and tend to occur together.
Enstrophy shares some common features with energy as well, such as a transfer from large to small scales, with the dissipation being confined at the small scales. However, the conventional notion of a cascade does not apply for enstrophy: it is not inviscidly conserved because vortex stretching ($\bm{\omega} \cdot \bm{\nabla} \bm{u}$) acts as a source of vorticity at all scales \citep{davidson-etal-2008}. Most of enstrophy is generated at the small scales and does not have to be transferred among space and scales to be dissipated. Nevertheless, vortex stretching is active also in the inertial range of scales (where dissipation is negligible), and this enstrophy has to be transferred from large to small scales to be eventually destroyed.
The importance of vortex stretching in turbulence is well known since the pioneering works of \cite{taylor-1938,betchov-1956,ashurst-etal-1987}. Though, there is a lack of consensus of the role of vortex stretching in the energy cascade. Despite the earlier works \citep[see for example][]{davidson-2004,davidson-etal-2008,doan-etal-2018} providing evidence about the relation between energy cascade and vortex stretching, some recent works have shown that the energy cascade is mainly driven by the self-amplification of the strain-rate field, and that vortex stretching plays a key role only during fluctuations of the cascade about its average value~\citep{carbone-bragg-2020,jhonson-2021}.  

Helicity, a pseudoscalar defined as $h = \bm{\omega} \cdot \bm{u}$, is another inviscid invariant of the three-dimensional ($3D$) Navier--Stokes equations. That is, like energy, helicity is also conserved by the non linearity of the Navier--Stokes equations~\citep{moffat-1969}. 
It admits topological interpretations in relation to the linkages of the flow vortex lines \citep{moffat-1969,moffat-tsinober-1992}: it is related to the knottedness of the vorticity lines, and is a descriptor of the breaking of parity invariance (mirror symmetry). Unlike energy, however, it is not a sign-definite quantity. The invariance of helicity has been associated with the conservation of the linkages of the vortex lines that move with the flow \citep{moffat-1969}, and it is based on two main notions, 
i.e. (i) vortex lines behave like material lines under evolution of the inviscid Navier--Stokes equations, and (ii) the flux of vorticity through any open surface bounded by a curve moving with the fluid is conserved. 
Based on the conservation of total helicity, \cite{brissaud-etal-1973} were the first to envisage the possible simultaneous existence of energy and helicity cascades in $3D$ turbulence, similar to what happens in two-dimensional ($2D$) turbulence where the two invariants are the energy and the enstrophy~\citep{boffetta-ecke-2012,falkovich-etal-2017}. Based on phenomenological arguments, they proposed that two different scenarios are possible, when dealing with helical flows. The first admits a simultaneous cascade of energy and helicity that leads to a $-5/3$ power-law spectrum for both quantities. The second scenario of a pure helicity cascade (i.e. no energy cascade) predicts a range of power-laws for both energy and helicity. \cite{kraichnan-1973} argued that, unlike in $2D$ turbulence where the existence of invariant enstrophy effectively blocks the forward energy cascade, the possibility of a joint energy and helicity cascade is more plausible \citep[see also][]{andre-lesieur-1977} in $3D$ turbulence. This was later confirmed by \cite{polifke-shtilman-1989}, and by the numerical simulations of \cite{borue-orszag-1997}. 
They observed a cascade of helicity from large to small scales, and did not detect any inverse cascade. 
In the following years, several authors studied helicity cascade in helical homogeneous isotropic turbulence.
\cite{chen-etal-2003c} found, using direct numerical simulations, that the energy and helicity fluxes feature a plateau in the inertial range of scales confirming the existence of a joint helicity and energy cascade from large to small scales. They also investigated intermittency in these fluxes and observed that the scaling exponents for the helicity fluxes are smaller compared to those for energy. The helicity flux is thus intrinsically more intermittent than the energy flux, consistent with the observation that helicity essentially behaves like a passive scalar \citep{sreenivasan-antonia-1997,romano-antonia-2001}.
Over the last years, several authors have studied helicity cascades, mainly using the Fourier space statistics \citep[see for example][]{biferale-musacchio-toschi-2013,alexakis-2017,pouquet-etal-2019}. Most of these works exploit an exact decomposition of the velocity field in a helical Fourier basis to properly account for triad interaction between wavenumbers \citep{constantin-majda-1988,waleffe-1992}.

\subsection{The K\'{a}rm\'{a}n-Howarth-Hill or the Generalised Kolmogorov Equation}

It is therefore clear that to fully characterise the influence of the polymeric additives on the Kolmogorov picture of turbulence, one has to consider also enstrophy and helicity besides the turbulent kinetic energy.
In this respect, \cite{baj-etal-2022} introduced a generalised framework which is suitable for investigating the scale-space transfers of energy, enstrophy and helicity in Newtonian turbulent flows. Their framework extends the generalised Kolmogorov equation, or GKE, \citep{marati-casciola-piva-2004,danaila-antonia-burattini-2004,cimarelli-deangelis-casciola-2013,gatti-etal-2020,gattere-etal-2023}, also known as K\'{a}rm\'{a}n-Howarth-Monin-Hill or K\'{a}rm\'{a}n-Howarth-Hill equation \citep{alvesportela-papadakis-vassilicos-2017,yasuda-vassilicos-2018,alvesportela-papadakis-vassilicos-2020,yao-etal-2024}, introduced by \cite{hill-2001,hill-2002}. The GKE is a generalisation of the K\'{a}rm\'{a}n-Howarth equation \citep{frisch-1995,pope-2000}, and is directly derived from the Navier--Stokes equations without any assumptions; it does not require either isotropy nor homogeneity. The GKE is an exact budget equation for the second-order structure function, commonly referred to as scale energy \citep{davidson-pearson-2005}, and characterises the mechanisms of production, transfer and dissipation of energy in the combined space of scales and positions. The GKE and its generalisations have been applied to several flow configurations over the years, mainly to study how inhomogeneity changes the Richardson and Kolmogorov picture of turbulence \citep{casciola-etal-2003, cimarelli-etal-2016, alvesportela-papadakis-vassilicos-2017, mollicone-etal-2018, cimarelli-etal-2021, chiarini-etal-2021, chiarini-etal-2022, apostolidis-etal-2023}. 
\cite{deangelis-etal-2005} used the GKE to investigate the influence of polymeric additives on the energy cascade in homogeneous isotropic turbulence, but at rather small Deborah numbers ($De \le 0.5$), for which the influence of the polymeric stresses is subdominant at all scales.

\subsection{The present study}

In this work we investigate the influence of the polymeric additives on the scale transfers of energy, helicity and enstrophy in homogeneous and isotropic turbulence at a relatively large Re and over a wide range of $De$. First, we extend the formulation of \cite{baj-etal-2022} and provide the exact budget equations for the scale energy, helicity and enstrophy for polymer-laden turbulent flows. Compared to the purely Newtonian case, the resulting budget equations incorporate additional terms that represent sources and sinks stemming from fluid–polymer interactions and capture the polymers’ influence on transfers both across scales and along the flow’s inhomogeneous directions. The new set of equations has been derived without any approximation, and are valid for a generic inhomogeneous and anisotropic turbulent flow. Then, we use these equations to elucidate the influence of the polymers on the production, transfer and dissipation of energy, helicity and enstrophy in a homogeneous isotropic turbulent flow of dilute polymeric suspensions. New insights are provided, with a particular look at the influence of the polymers on the intermittent nature of the transfers, and on the local flow topology. The study relies on the database introduced by \cite{singh-rosti-2024} and obtained by means of direct numerical simulations. The Reynolds number is set at $Re_\lambda \approx 460$, while the Deborah number is varied in the $1/9 \le De \le 9$ range. 

The remainder of the work is structured as follows. In \S\ref{sec:formulation} we introduce the budget equations. Sections \S\ref{sec:strfun}, \S\ref{sec:resbud} and \S\ref{sec:loc-str} deal with our findings, with \S\ref{sec:energy}, \S\ref{sec:enstrophy} and \S\ref{sec:helicity} being respectively devoted to energy, enstrophy and helicity scale budgets. \S\ref{sec:loc-str} examines the influence of polymers on the local flow topology, particularly in relation to their effect on the scale-space budgets. A concluding discussion and perspectives are eventually provided in \S\ref{sec:conclusions}.

\section{Mathematical formulation and numerical method}
\label{sec:formulation}

\subsection{The governing equations}

The governing equations for an incompressible turbulent flow with polymeric additives are
\begin{align}
\frac{\partial u_i}{\partial t} + u_j \frac{\partial u_i}{\partial x_j} = & \ - \frac{1}{\rho} \frac{\partial p}{\partial x_i}
+ \nu \frac{\partial^2 u_i}{\partial x_j \partial x_j} + \frac{1}{\rho} \frac{\partial T_{ij}}{\partial x_j} + f_i, \\
\frac{\partial R_{ij}}{\partial t} + u_k \frac{\partial R_{ij}}{\partial x_k} = & \ \frac{\partial u_i}{\partial x_k} R_{kj} + R_{ik} \frac{\partial u_k}{\partial x_j} - \frac{\left( \mathscr{P} R_{ij} - \delta_{ij} \right)}{\tau_p}, \\
\frac{\partial u_j}{\partial x_j} = & \ 0,
\end{align}
where $u_i$ is the velocity field, $p$ is the reduced pressure, $\rho$ is the fluid density, $\nu$ is the fluid kinematic viscosity, and $f_i$ is the external forcing used to sustain the flow. The presence of the polymers is accounted for in the momentum equation by means of the extra-stress tensor $T_{ij}$, which is related to the conformation tensor $R_{ij}$ as
\begin{equation}
T_{ij} = \mu_p \frac{ \left( \mathscr{P} R_{ij} - \delta_{ij} \right) }{ \tau_p },
\end{equation}
where $\tau_p$ is the polymeric relaxation time, $\mu_p$ is the polymeric viscosity and $\delta_{ij}$ is the Kronecker delta. $\mathscr{P}$ is the Peterlin function and is equal to $\mathscr{P}=1$ for the purely elastic Oldroyd-B model and to $\mathscr{P} = (\ell_{max}^2 -3)/(\ell_{max}^2-R_{ii})$ for the FENE-P model; $\ell_{max}$ is the maximum polymer extensibility, $R_{ii}$ is the instantaneous measure of the end-to-end length of the polymers.

By taking the curl of the momentum equation we obtain the equation for vorticity $\omega_i = \varepsilon_{ijk} \partial u_k/\partial x_j$ as,
\begin{equation}
\frac{\partial \omega_i}{\partial t} + u_j \frac{\partial \omega_i}{\partial x_j} = \omega_j \frac{\partial u_i}{\partial x_j} + \nu \frac{\partial^2 \omega_i}{\partial x_j \partial x_j} + \frac{1}{\rho} \frac{\partial T_{ij}^\omega}{\partial x_j} + f_i^\omega,
\end{equation}
where $T_{i\ell}^\omega = \varepsilon_{ijk} \partial T_{k \ell}/\partial x_j$ , $f_i^\omega = \varepsilon_{ijk} \partial f_k/\partial x_j$, and $\varepsilon_{ijk}$ is the Levi-Civita symbol.

The flow is described by its mean and fluctuating fields, defined after Reynolds decomposition. The mean field is obtained by means of the $\aver{\cdot}$ operator that denotes averaging among realisations, along homogeneous directions, and in time if the flow is statistically stationary. Hereafter, capital letters ($U_i$, $\Omega_i$ and $P$) refer to mean quantities, while small letters ($u_i$, $\omega_i$ and $p$) to the fluctuations around them.

\subsection{The budget equations}

Following the work of \cite{baj-etal-2022}, we consider three specific structure functions, i.e. the velocity structure function $\delta q^2 = \delta u_i \delta u_i$, the vorticity structure function $\delta \omega^2 = \delta \omega_i \delta \omega_i$ and the helicity structure function $\delta h = \delta u_i \delta \omega_i$ where repeated indices imply summation. The three structure functions feature velocity and vorticity increments ($\delta u_i$ and $\delta \omega_i$) between two points $\bm{x}$ and $\bm{x}'$, that can be identified by means of their midpoint $\bm{X} = (\bm{x} + \bm{x}')/2$ and separation vector $\bm{r} = \bm{x}' - \bm{x}$, namely
\begin{equation*}
\delta u_i\left(\bm{X},\bm{r},t\right) = u_i\left(\bm{X}+\frac{\bm{r}}{2},t \right) - u_i \left(\bm{X}-\frac{\bm{r}}{2},t \right),
\end{equation*}
and
\begin{equation*}
\delta \omega_i\left(\bm{X},\bm{r},t\right) = \omega_i\left(\bm{X}+\frac{\bm{r}}{2},t\right) - \omega_i \left(\bm{X}-\frac{\bm{r}}{2},t \right).
\end{equation*}
In the most general case $\aver{\delta q^2}$, $\aver{\delta h}$ and $\aver{\delta \omega^2}$ depend upon seven independent variables $\bm{X}$, $\bm{r}$ and $t$.

The quantity $\aver{\delta q^2}(\bm{X},\bm{r},t)$ is commonly used as a proxy for the kinetic energy associated with eddies of scale $r = |\bm{r}|$ at position $\bm{X}$. This interpretation is however only approximate. Actually, $\aver{\delta q^2}$ reflects the cumulative energy contained in eddies with scales up to $r$, rather than the energy at that scale. This becomes clear when expressing $\aver{\delta q^2}$ in terms of the velocity variance and the correlation function $C_{uu}$. In the idealised case of HIT, we have $\aver{\delta q^2} = 2 \aver{u_i u_i} ( 1 - C_{uu}(r))$, where $C_{uu}= \aver{ u_i(\bm{x}+\bm{r},t) u_i(\bm{x},t) }/\aver{u_i u_i}$: $\aver{\delta q^2} = 0$ for $r=0$ and $\aver{\delta q^2} \rightarrow 2 \aver{u_i u_i}$ as $r \rightarrow \infty$. However, as pointed out by \cite{davidson-pearson-2005}, this is not entirely accurate: $\aver{\delta q^2}$ at a given scale $r$ includes contributions not only from eddies of scale $\le r$, but also from large-scale eddies via their associated enstrophy. A similar interpretation applies to the enstrophy and helicity structure functions. For example, for enstrophy in HIT we have $\aver{\delta \omega^2} = 2 \aver{ \omega_i \omega_i}( 1- C_{\omega \omega}(r) )$, where $C_{\omega \omega } = \aver{ \omega_i (\bm{x}+\bm{r},t) \omega_i (\bm{x},t) }/\aver{ \omega_i \omega_i } \rightarrow 0$ and $\aver{\delta \omega^2} \rightarrow 2 \aver{ \omega_i \omega_i}$ when $r \rightarrow \infty$. The enstrophy structure function $\aver{\delta \omega^2}$ can be (loosely) interpreted as the cumulative enstrophy contributed by eddies of scales up to $r$.

The budget equations for $\aver{\delta q^2}$, $\aver{\delta \omega^2}$ and $\aver{\delta h}$ describe production, transport and dissipation of energy, enstrophy and helicity in the space of scales $\bm{r}$ and positions $\bm{X}$. These equations link the variation in time of the three structure functions at a given scale and position to the instantaneous imbalance among production, transport and dissipation. 
Specifically, they correspond to transport budget equations governing the cumulative energy, enstrophy, and helicity associated with eddies of size up to scale $r$.
The three budget equations are obtained after manipulation of the Navier--Stokes equations for the velocity and the vorticity, without any assumption of homogeneity and isotropy; for the full derivation (for a purely Newtonian fluid) of the equation for $\aver{\delta q^2}$ we refer the interested reader to the appendix of \cite{gatti-etal-2020}.
The derivation of the budget equations for $\aver{\delta q^2}$, $\aver{\delta \omega^2}$ and $\aver{\delta h}$ starts with the evolution equations for velocity and vorticity, and requires a sequence of manipulations that use the incompressibility constraint. A brief recap of the main steps is also provided in \cite{baj-etal-2022}.

In compact form, the budget equations for $\aver{\delta q^2}$, $\aver{\delta \omega^2}$ and $\aver{\delta h}$ can be written as:
\begin{equation}
\frac{\partial \aver{\delta g}}{\partial t} + 
\frac{\partial \langle \psi^{\delta g}_j \rangle}{\partial X_j} + 
\frac{\partial \langle \phi^{\delta g}_j \rangle}{\partial r_j} -
\aver{ \mathcal{P}^{\delta g} } -
\aver{ \mathfrak{p}^{\delta g} } -
\langle \mathcal{D}_f^{\delta g} \rangle -
\aver{ \mathcal{F}^{\delta g} } = -
\aver{ \mathcal{E}^{\delta g} } -
\aver{ \Pi^{\delta g} }
\label{eq:gkhe}
\end{equation}
Here, $\langle \delta g \rangle$ denotes one of the three structure functions. The terms in the corresponding balance equation are defined as follows:
\begin{itemize}
\item $\partial \langle \delta g \rangle / \partial t$ represents the temporal evolution of the structure function.
\item $\partial \langle \psi_j^{\delta g} \rangle / \partial X_j$ accounts for transport in physical space, arising from fluid–fluid and fluid–polymer interactions.
\item $\partial \langle \phi_j^{\delta g} \rangle / \partial r_j$ describes the scale-space transfer associated with these interactions.
\item $\langle \mathcal{P}^{\delta g} \rangle$ denotes the production term due to the interaction between the mean shear and the velocity fluctuations.
\item $\langle \mathfrak{p}^{\delta g} \rangle$ arises from the coupling between the strain-rate and the vorticity fluctuations. This term, associated with vortex stretching, appears only in the equation for $\langle \delta \omega^2 \rangle$.
\item $\langle \mathcal{D}_f^{\delta g} \rangle$ represents viscous diffusion in both physical and scale spaces.
\item $\langle \mathcal{F}^{\delta g} \rangle$ is the contribution from the external forcing.
\item $\langle \mathcal{E}^{\delta g} \rangle$ denotes the two-point average of the fluid dissipation rate.
\item $\langle \Pi^{\delta g} \rangle$ denotes the two-point average of the polymeric source/sink term.
\end{itemize}
We now provide the explicit form of terms appearing in the budget equations for the three structure functions. The components of the flux vector $\boldsymbol{\psi}$ in the physical space are given by:
\begin{align}
   \psi_j^{\delta q^2} = \ & U_j^* \delta q^2 + u_j^* \delta q^2 + 2 \delta p \delta u_j/\rho - \delta T_{ij} \delta u_i/\rho, \\
   \psi_j^{\delta \omega^2} = \ & U_j^* \delta \omega^2 + u_j^* \delta \omega^2 - \delta T_{ij}^{\omega} \delta \omega_i/\rho, \\
   \psi_j^{\delta h} = \ & U_j^* \delta h + u_j^* \delta h - \omega_j^* \delta q^2/2 + \delta p \delta u_j/\rho - (\delta T_{ij}^\omega \delta u_i + \delta T_{ij} \delta \omega_i)/(2 \rho)
\end{align}
where $j = 1, 2, 3$ correspond to the three spatial directions in the case of full inhomogeneity. The superscript $``*"$ denotes an arithmetic average between the two points $\bm{X} \pm \bm{r}/2$. The components of the flux vector $\bm{\phi}$ in the scale-space are given by:
\begin{align}
  \phi_j^{\delta q^2} = \ & \delta U_j \delta q^2 + \delta u_j \delta q^2 - 4 T_{ij}^* \delta u_i/\rho, \\
  \phi_j^{\delta \omega^2} = \ & \delta U_j \delta \omega^2 + \delta u_j \delta \omega^2 - 4 T_{ij}^{\omega,*} \delta \omega_i / \rho, \\
  \phi_j^{\delta h} = \ & \delta U_j \delta h + \delta u_j \delta h - \delta \omega_j \delta q^2/2 - 2 ( T_{ij}^{\omega,*} \delta u_i + T_{ij}^* \delta \omega_i )/\rho
\end{align}
where $j=1,2,3$. The production terms $\mathcal{P}$ are given by:
\begin{align}
  \mathcal{P}^{\delta q^2} = & \ - 2 u_j^* \delta u_i \delta \left( \frac{\partial U_i}{\partial x_j} \right) - 2 \delta u_i \delta u_j \left(\frac{\partial U_i}{\partial x_j} \right)^*, \\
  \mathcal{P}^{\delta \omega^2} = & \ - 2 u_j^* \delta \omega_i \delta \left( \frac{\partial \Omega_i}{\partial x_j} \right) - 2 \delta u_j \delta \omega_i \left( \frac{\partial \Omega_i}{\partial x_j} \right)^* + 2 \omega^* \delta \omega_i \delta \left( \frac{\partial U_i}{\partial x_j} \right)  + 2 \delta \omega_i \delta \omega_j \left( \frac{\partial U_i}{\partial x_j} \right)^* + \nonumber  \\
  & \ + 2 \delta \omega_i \Omega_j^* \delta \left( \frac{\partial u_i}{\partial x_j} \right) + 2 \delta \omega_i \delta \Omega_j \left( \frac{\partial u_i}{\partial x_j} \right)^*, \\
  \mathcal{P}^{\delta h} = & \ - u_j^* \delta u_i \delta \left( \frac{\partial \Omega_i}{\partial x_j} \right)
                     - \delta u_i \delta u_j \left( \frac{\partial \Omega_i}{\partial x_j} \right)^* 
                     + \delta u_i \Omega_j^* \delta \left( \frac{\partial u_i}{\partial x_j} \right) 
                     + \delta u_i \delta \Omega_j \left( \frac{\partial u_i}{\partial x_j} \right)^* + \nonumber \\
                & \ - u_j^* \delta \omega_i \delta \left( \frac{\partial U_i}{\partial x_j} \right)
                    - \delta u_j \delta \omega_i \left( \frac{\partial U_i}{\partial x_j} \right)^* 
                    + \omega_j^* \delta u_i \delta \left( \frac{\partial U_i}{\partial x_j} \right) 
                    + \delta \omega_j \delta u_i \left( \frac{\partial U_i}{\partial x_j} \right)^*.
\end{align}
The additional generation term $\mathfrak{p}^{\delta \omega^2}$, appears only in the budget equation for $\langle \delta \omega^2 \rangle$ and reads:
\begin{align}
  \mathfrak{p}^{\delta \omega^2} = \ 2 \delta \omega_i \omega_j^* \delta \left( \frac{\partial u_i}{\partial x_j} \right)
                                   + 2 \delta \omega_i \delta \omega_j \left( \frac{\partial u_i}{\partial x_j} \right)^*.
\end{align}
The viscous terms $\mathcal{D}_f$, instead, are given by:
\begin{align}
  \mathcal{D}_f^{\delta q^2} = & \  \frac{\nu}{2} \frac{\partial^2 \delta q^2}{\partial X_j^2}  + 2 \nu \frac{\partial^2 \delta q^2}{\partial r_j^2} ,\\
  \mathcal{D}_f^{\delta \omega^2} = & \  \frac{\nu}{2} \frac{\partial^2 \delta \omega^2}{\partial X_j^2} +  2 \nu \frac{\partial^2 \delta \omega^2}{\partial r_j^2} ,\\
  \mathcal{D}_f^{\delta h}   = & \  \frac{\nu}{2} \frac{\partial^2 \delta h}{\partial X_j^2}  +  2 \nu \frac{\partial^2 \delta h}{\partial r_j^2},
\end{align}
while the forcing terms $\mathcal{F}$ take the form:
\begin{align}
  \mathcal{F}^{\delta q^2} = & \ 2 \delta f_i \delta u_i \\
  \mathcal{F}^{\delta \omega^2} = & \ 2 \delta f_i^{\omega} \delta \omega_i,\\
  \mathcal{F}^{\delta h} = & \ \delta f_i^{\omega} \delta u_i + \delta f_i \delta \omega_i.
\end{align}
The two terms at the right-hand side of equation~\eqref{eq:gkhe} correspond to the two-point averages of the fluid dissipation and polymeric source/sink rates. The dissipation terms $\mathcal{E}$ are:
\begin{align}
  \mathcal{E}^{\delta q^2} = & \ 4 \varepsilon_f^{\delta q^2,*} = 4 \nu \left( \frac{ \partial u_i }{\partial x_j} \frac{ \partial u_i}{\partial x_j} \right)^* ,\\
  \mathcal{E}^{\delta \omega^2} = & \ 4 \varepsilon_f^{\delta \omega^2,*} = 4 \nu \left( \frac{\partial \omega_i}{\partial x_j} \frac{\partial \omega_i}{\partial x_j} \right)^* ,\\
  \mathcal{E}^{\delta h} = & \ 4 \varepsilon_f^{\delta h,*} = 4 \nu \left( \frac{\partial u_i }{\partial x_j} \frac{\partial \omega_i}{\partial x_j} \right)^*
\end{align}
and the polymeric source/sink terms $\Pi$ are:
\begin{align}
  \Pi^{\delta q^2} = & \ 4 \pi^{\delta q^2,*}/\rho = \frac{4}{\rho} \left( T_{ij} \frac{\partial u_i}{\partial x_j} \right)^*, \\
  \Pi^{\delta \omega^2} = & \ 4 \pi^{\delta \omega^2,*}/\rho = \frac{4}{\rho} \left( T_{ij}^{\omega} \frac{\partial \omega_i}{\partial x_j} \right)^*, \\
  \Pi^{\delta h} = & \ 4 \pi^{\delta h,*}/\rho = \frac{4}{\rho} \left( \frac{1}{2} \left( T_{ij} \frac{\partial \omega_i}{\partial x_j} \right) +
                                                                       \frac{1}{2} \left( T_{ij}^{\omega} \frac{\partial u_i}{\partial x_j} \right) \right)^*.
\end{align}
Equation \eqref{eq:gkhe} can be interpreted as a balance equation for the two-point average of the total dissipation rate. At each scale $\bm{r}$ and position $\bm{X}$, the sum of the terms on the left-hand side must balance the combined contribution of $\langle \mathcal{E} \rangle$ and $\langle \Pi \rangle$.

The various terms in the fluxes $\{\bm{\phi},\bm{\psi}\}$ can be interpreted as contributions from mean and turbulent transport and pressure transport. Notably, the pressure does not contribute to $\bm{\psi}^{\delta \omega^2}$, as its effect is eliminated by the curl operator. In addition to the standard nonlinear transfer terms, the fluxes include an additional contribution that reflects a distinct transfer mechanism induced by fluid-polymer interactions, which is absent in purely Newtonian flows. As previously noted by \cite{baj-etal-2022}, the budget equation for $\langle \delta h \rangle$ reveals that helicity transfer in both physical space ($\bm{X}$) and scale space ($\bm{r}$) arises not only from the interaction between $\delta h$ and the velocity increments $\delta u_j$, but also from the coupling between $\delta q^2$ and the vorticity increments $\delta \omega_j$. These two distinct mechanisms give rise to separate pathways for helicity transfer, consistent with the findings of \cite{yan-etal-2020}.  
The right-hand side of equation~\eqref{eq:gkhe} represents the net source terms for $\langle \delta q^2 \rangle$, $\langle \delta \omega^2 \rangle$, and $\langle \delta h \rangle$, accounting for exchanges in both physical space and across scales. The mean production terms $\mathcal{P}$ describe how these quantities are exchanged with the mean flow, analogous to the single-point budget equations. As such, they are directly linked to the gradients of the mean velocity field.
Unlike $\langle \delta q^2 \rangle$, the mean production of $\langle \delta \omega^2 \rangle$ and $\langle \delta h \rangle$ depends not only on the mean velocity gradients $\partial U_i / \partial x_j$ but also on the gradients of the mean vorticity $\partial \Omega_i / \partial x_j$. Moreover, enstrophy production includes additional contributions from vortex stretching induced by turbulent fluctuations $\mathfrak{p}^{\delta \omega^2}$. When present, gradients in the mean vorticity field further enhance this mechanism.
In addition to the fluid dissipation terms $\mathcal{E}$, the right-hand side of equation~\eqref{eq:gkhe} includes a polymeric contribution $\Pi$, which reflects the influence of fluid-polymer interactions. Unlike dissipation, this term is not sign-definite meaning that it can act as either a source or a sink depending on the local configuration of the polymer stress and velocity gradients, see \S\ref{sec:loc-str}. 

\subsubsection{Homogeneous isotropic turbulence}

For homogeneous statistically steady flows, the three structure functions and the terms of the corresponding budget equations lose their dependence on $\bm{X}$ and $t$, and are functions of only the separation vector $\bm{r} \equiv (r_x,r_y,r_z)$. In HIT the mean-flow is absent and $\bm{U} = \bm{0}$ and $\bm{\Omega} = \bm{0}$, implying $\mathcal{P}^{\delta g} = 0 $. Exploiting isotropy, we integrate equation \eqref{eq:gkhe} in the $\bm{r}$ space over spherical shells of radius $r = |\bm{r}|$, with surface $\mathcal{S}(r)$ and volume $\mathcal{V}(r)$, and obtain the generalised K\'{a}rm\'{a}n-Howarth equation for PHIT \citep[see][for further details on the energy equation]{chiarini-singh-rosti-2025}, which in a compact form reads
\begin{align}
  \langle \Phi_f^{\delta g} \rangle(r) + \langle \Phi_p^{\delta g} \rangle (r) - \langle D_f^{\delta g} \rangle(r) - \langle V_{st}^{\delta g} \rangle (r) - \langle F^{\delta g} \rangle(r) = - \frac{4}{3} \langle \varepsilon_f^{\delta g} \rangle r - \frac{4}{3} \langle \pi^{\delta g} \rangle r.
\end{align}
On the left-hand side, $\Phi_f^{\delta g}$ and $\Phi_p^{\delta g}$ represent the contributions of fluid-fluid and fluid-polymer interactions, respectively, to the scale-space transfer of $\delta g$, and are given by:
\begin{align}
  \langle \Phi_f^{\delta q^2} \rangle = & \ \frac{1}{\mathcal{S}(r)} \oint_{\mathcal{S}(r)} \aver{ \delta u_j \delta q^2 } n_j \text{d} \Sigma, \\
  \langle \Phi_f^{\delta \omega^2} \rangle = & \ \frac{1}{\mathcal{S}(r)} \oint_{\mathcal{S}(r)} \aver{ \delta u_j \delta \omega^2 } n_j \text{d} \Sigma, \\
  \langle \Phi_f^{\delta h} \rangle = & \underbrace{ \frac{1}{\mathcal{S}(r)} \oint_{\mathcal{S}(r)}  \aver{ \delta u_j \delta h} n_j \text{d} \Sigma }_{ \aver{ \Phi_{f,a}^{\delta h} } }  +
                                 \underbrace{ \frac{1}{\mathcal{S}(r)} \oint_{\mathcal{S}(r)} -\aver{ \delta \omega_j \delta q^2 } n_j \text{d} \Sigma }_{ \aver{ \Phi_{f,b}^{\delta h} } }
\end{align}
and
\begin{align}
  \langle \Phi_p^{\delta q^2}      \rangle = & \ \frac{1}{\mathcal{S}(r)} \oint_{\mathcal{S}(r)} - 4 \aver{   \delta      u_i T_{ij}^*          } /\rho \ n_j \text{d} \Sigma, \\
  \langle \Phi_p^{\delta \omega^2} \rangle = & \ \frac{1}{\mathcal{S}(r)} \oint_{\mathcal{S}(r)} - 4   \aver{ \delta \omega_i T_{ij}^{\omega,*} } /\rho \ n_j \text{d} \Sigma, \\
  \langle \Phi_p^{\delta h}        \rangle = & \ \frac{1}{\mathcal{S}(r)} \oint_{\mathcal{S}(r)} - 2 ( \aver{ \delta      u_i T_{ij}^{\omega,*} } +
                                                                                   \aver{ \delta \omega_i T_{ij}^*          })/\rho \ n_j \text{d} \Sigma,
\end{align}
where $\bm{n}$ is the outward-pointing unit vector normal to the spherical surface.

Instead, $V_{st}$ denotes the vortex stretching term, which contributes to enstrophy generation across all scales ($V_{st}^{\delta q^2} = V_{st}^{\delta h} = 0$), and is expressed as:
\begin{align}
  \aver{ V_{st}^{\delta \omega^2} } = \frac{1}{\mathcal{S}(r)} \int_{\mathcal{V}(r)} 2 \left( \aver{ \delta \omega_i \omega_j^* \delta \left( \frac{\partial u_i}{\partial x_j} \right)   } +
                                                                          \aver{ \delta \omega_i \delta \omega_j   \left( \frac{\partial u_i}{\partial x_j} \right)^* } \right) \text{d} \Omega.
\end{align}
Viscous diffusion at scale $r$ is expressed as
\begin{align}
  \langle D_f^{\delta q^2}      \rangle = & \ \frac{1}{\mathcal{S}(r)} \oint_{\mathcal{S}(r)} 2 \nu \frac{\partial \aver{\delta q^2}}{\partial r_j} n_j \text{d} \Sigma, \\
  \langle D_f^{\delta \omega^2} \rangle = & \ \frac{1}{\mathcal{S}(r)} \oint_{\mathcal{S}(r)} 2 \nu \frac{\partial \aver{\delta \omega^2}}{\partial r_j} n_j \text{d} \Sigma, \\
  \langle D_f^{\delta h }       \rangle = & \ \frac{1}{\mathcal{S}(r)} \oint_{\mathcal{S}(r)} 2 \nu \frac{\partial \aver{\delta h}  }{\partial r_j} n_j \text{d} \Sigma,
\end{align}
while the forcing term that accounts for the injection of $\aver{\delta q^2}$, $\aver{\delta \omega^2}$ and $\aver{\delta h}$ at scale $r$ reads
\begin{align}
  \langle F^{\delta q^2}      \rangle = & \ \frac{1}{\mathcal{S}(r)} \int_{\mathcal{V}(r)} 2 \aver{ \delta f_i           \delta u_i      } \text{d}\Omega, \\
  \langle F^{\delta \omega^2} \rangle = & \ \frac{1}{\mathcal{S}(r)} \int_{\mathcal{V}(r)} 2 \aver{ \delta f_i^{\omega } \delta \omega_i } \text{d}\Omega, \\
  \langle F^{\delta h}        \rangle = & \ \frac{1}{\mathcal{S}(r)} \int_{\mathcal{V}(r)} ( \aver{ \delta f_i \delta \omega_i} + \aver{ \delta f_i^{\omega} \delta u_i } ) \text{d}\Omega.
\end{align}

These equations highlight the fundamental differences between Newtonian and polymeric turbulence. In Newtonian fluids, dissipation occurs solely through fluid viscosity, represented by $\varepsilon_f$, and the fluxes arise exclusively from the nonlinear term $\Phi_f$. In the intermediate range of scales where the viscous diffusion $D_f$ and forcing $F$ terms are typically negligible, a positive production term associated with vortex stretching appears only in the enstrophy equation.
In contrast, polymeric flows feature an additional flux component, $\Phi_p$, which (also) accounts for the transfer across scales mediated by the polymers. This term appears in equation~\eqref{eq:gkhe} under a divergence in scale-space ($\partial/\partial r_j$), analogous to $\Phi_f$, and its net contribution averages to zero over the entire triperiodic domain. 
Physically, this reflects how stretched polymers facilitate transfers across scales, providing an alternative cascade pathway that ultimately leads to dissipation.
Furthermore, the presence of polymers introduces an additional source/sink term, $\pi$, on the right-hand side of the equations. For the energy equation, $\langle \pi^{\delta q^2} \rangle $ matches the polymeric dissipation rate $\langle \epsilon_p \rangle \equiv \aver{ \mu_p (R_{ii} - 3)/(2 \tau_p^2)} $, and on average acts as a net sink for the fluid phase \citep{deangelis-etal-2005}. For the enstrophy and the helicity the average $\langle \pi^{\delta \omega^2} \rangle $ and $\langle \pi^{\delta h} \rangle$ can be both positive and negative, depending on the small-scale flow topology.

In the case of homogeneous and isotropic flows, the quantities $\mathcal{E}$ and $\Pi$ reduce to the single-point terms $\varepsilon_f$ and $\pi$, which are independent of the scale $r$. The corresponding equations can therefore be interpreted as balance relations for the average total dissipation rates. At each scale $r$, the sum of the scale-dependent terms on the left-hand side must balance the scale-independent average total dissipation rate, which acts as an invariant of the system. This interpretation is fully consistent with the classical K\'{a}rm\'{a}n–Howarth equation for the second-order velocity structure function in HIT. However, as discussed above, a key distinction in the case of PHIT is that the invariant quantity is no longer the fluid dissipation alone, but the combined contribution $\langle \varepsilon_f \rangle + \langle \pi \rangle$ \citep{chiarini-singh-rosti-2025}.
In this context, it is worth noting that the equation for $\langle \delta q^2 \rangle$ closely resembles the classical $-4/5$ law for Newtonian homogeneous and isotropic turbulence. Specifically, in the absence of polymers, in the high Reynolds number limit ($Re \to \infty$), and for scales in the inertial range ($r/L \to 0$), the energy equation reduces to $1/S(r) \oint_{S(r)} \langle \delta u_j \delta q^2 \rangle n_j \text{d} \Sigma = - (4/3) \langle \varepsilon_f^{\delta q^2} \rangle r$, which, under assumptions of incompressibility and isotropy \citep{hill-1997}, simplifies further to the well-known result $S_3(r) \equiv \langle \left(\left( \bm{u}(\bm{x}+\bm{r},t) - \bm{u}(\bm{x},t) \right) \cdot \bm{r}/r\right)^3 \rangle = -(4/5) \langle \varepsilon_f^{\delta q^2} \rangle r$.

To better characterise the influence of the polymers on the transfers of energy and helicity, we decompose $\Phi_p$ into an inertial transfer term $\Phi_{p,i}$ and a dissipative term $D_p$, following the approach introduced by \cite{rosti-perlekar-mitra-2023} and informed by earlier work by \cite{perlekar-mitra-pandit-2006}. 
\cite{perlekar-mitra-pandit-2006} proposed modelling this behaviour via a scale-dependent effective viscosity. Alternatively, \cite{rosti-perlekar-mitra-2023} separated the dissipative part by assuming that it shares the same asymptotic scale dependence as the fluid viscous term $D_f$ and requiring that at vanishing scales it matches the polymeric dissipation $\langle \pi^{\delta q^2} \rangle \equiv \langle \varepsilon_p \rangle$. Here we adopt this second approach for the budgets of energy and helicity. 
We split the polymeric flux $\Phi_p$ into two contributions, such that
\begin{equation*}
  \Phi_{p} =  \Phi_{p,i} + D_{p}, \ \text{with} \ 
  \aver{D_{p}} = \frac{\aver{\pi}}{\aver{\varepsilon_f}} \aver{D_f} \ \text{and} \ \aver{\Phi_{p,i}} = \aver{\Phi_{p}} - \aver{D_{p}}.
\end{equation*}
To simplify the notation, hereafter we use $\Phi_{p}$ instead of $\Phi_{p,i}$ to indicate the inertial contribution to the polymeric flux. 

We do not apply the same decomposition to the budget for $\langle \delta \omega^2 \rangle$, as at large $De$ the fluid–polymer interaction yields, on average, a positive enstrophy source term; that is $\langle \pi^{\delta \omega^2} \rangle <0 $ (see \S\ref{sec:enstrophy}). Unlike in the cases of energy and helicity, $\langle \pi^{\delta \omega^2} \rangle$ cannot be thus interpreted, on average, as a purely dissipative contribution.

\subsection{The numerical database}
\label{sec:database}

We use data from direct numerical simulations (DNS) of homogeneous, isotropic turbulence in a purely elastic Oldroyd-B polymeric fluid. The DNS dataset was first presented by \cite{singh-rosti-2024}, where comprehensive details on the numerical methods and computational procedures can be found. Here, these aspects are only briefly summarised.
The governing equations are numerically integrated using the in-house solver Fujin (\url{https://www.oist.jp/research/research-units/cffu/fujin}), which uses an incremental pressure-correction scheme. The Navier--Stokes equations written in primitive variables are solved on a staggered grid using second-order finite-differences in all the directions. The momentum equation is advanced in time using a second-order Adams-Bashforth time scheme, while the non Newtonian stress tensor is advanced in time with a second-order Crank-Nicolson scheme. A log-conformation formulation \citep{izbassarov-etal-2018} ensures positive-definiteness of the conformation tensor at all times. Turbulence is sustained using the Arnold-Beltrami-Childress (ABC) cellular-flow forcing \citep{podvigina-pouquet-1994}.

The equations are solved within a cubic domain of size $L=2 \pi$ having periodic boundary conditions in all directions. The parameters are chosen to achieve a Taylor-microscale Reynolds number of $Re_\lambda = u' \lambda/ \nu \approx 460$ in the purely Newtonian case ($u'$ is the root mean square of the velocity fluctuations and $\lambda$ is the Taylor length scale). The Deborah number $De = \tau_p/\tau_f $, where $\tau_f = L/u_{rms}$ is the reference large-eddy turnover time of the flow in the purely Newtonian case, is varied in the range $1/9 \le De \le 9$. For all cases, the fluid and polymer viscosities are fixed such that $\mu_f/(\mu_f + \mu_p) = 0.9$. The computational domain is discretised with $N^3 = 1024^3$ grid points, to ensure that all the scales down to the smallest dissipative ones are resolved, i.e. $\aver{\eta}/\Delta x = \mathcal{O}(1)$, where $\Delta x$ is the grid spacing and $\eta$ the Kolmogorov length scale. We have $0.6 \lessapprox \aver{\eta}/\Delta x \lessapprox 0.82$ for all cases. Simulations are advanced with a constant time step of $\Delta t/\aver{\tau_\eta} = 2 \times 10^{-3}$, where $\tau_\eta$ is the Kolmogorov time scale. Details of the numerical simulations with bulk quantities of interest are provided in table \ref{tab:simulations}. 

Concerning the energy budget equation, the model-independence of the results has been demonstrated both in Fourier space by \cite{rosti-perlekar-mitra-2023} and in physical space by \cite{chiarini-singh-rosti-2025}.

\begin{table}
\caption{Details of the numerical simulations considered in the present study. $De$ is the Deborah number, $Re_\lambda$ is the Reynolds number based on $u'=\sqrt{2E/3}$ and on the Taylor length scale $\lambda$, $\eta$ is the Kolmogorov length scale, $R_{ii}$ is the trace of the conformation tensor and mimics the free energy of the polymeric phase, $\varepsilon_f^{\delta q^2}$, $\varepsilon_f^{\delta \omega^2}$ and $\varepsilon_f^{\delta h}$ are the fluid dissipation of energy, enstrophy and helicity, and $\pi^{\delta q^2}$, $\pi^{\delta \omega^2}$, $\pi^{\delta h}$ are the source/sink of energy, enstrophy and helicity due to the fluid-polymer coupling.}
\label{tab:simulations}
\centering
\begin{tabular}{lcccccccccccccccccc}
$De$ & & $Re_\lambda$ & & $\aver{\lambda}$ & & $\aver{\eta}$ & & $\langle \varepsilon_f^{\delta q^2} \rangle$ & & $\langle \varepsilon_f^{\delta \omega^2} \rangle$ & & $\langle \varepsilon_f^{\delta h^2} \rangle$ & & $\langle \pi^{\delta q^2} \rangle$ & & $\langle \pi^{\delta \omega^2} \rangle$ & & $\langle \pi^{\delta h} \rangle$ \\
$-$   & &  $458.6$        & &    $0.157$           & & $0.0037$           & & $59.82$             & & $5.104\times 10^5$                 & & $61.51$   & & $-$ & & $-$  & & $-$              \\
      & &             & &                  & &               & &            & &                 & &                     & &                     \\
$1/9$ & &  $513.6$        & &    $0.174$           & & $0.0039$          & & $50.84$             & & $3.695\times 10^5$                 & & $55.14$   & & $6.50$ & & $\ \ 2.527 \times 10^4$ & & $\ \ 5.51$                \\
$1/3$ & &  $627.0$        & &    $0.203$           & & $0.0041$           & & $41.76$             & & $2.409\times 10^5$                 & & $44.38$   & & $17.29$ & & $\ \ 2.052 \times 10^4$ & & $12.76$                \\
$1$  & &  $795.8$        & &    $0.276$           & & $0.0050$           & & $21.04$             & & $1.126 \times 10^5$                 & & $33.09$   & & $39.94$ & & $-7.083 \times 10^4$  & & $33.84$              \\
$3$   & &  $761.1$        & &    $0.285$           & & $0.0053$           & & $15.50$             & & $9.152 \times 10^4$                 & & $25.21$   & & $33.52$ & & $-7.799 \times 10^4$ & & $22.48$                \\
$9$   & &  $776.3$        & &    $0.273$           & & $0.0050$           & & $22.03$             & & $1.259 \times 10^5$                 & & $33.06$   & & $45.61$ & & $-1.097 \times 10^5$ & & $36.42$                \\
\end{tabular}
\end{table}

\section{Structure functions and dissipation}
\label{sec:strfun}

\begin{figure}
\centering
\includegraphics[width=1\textwidth]{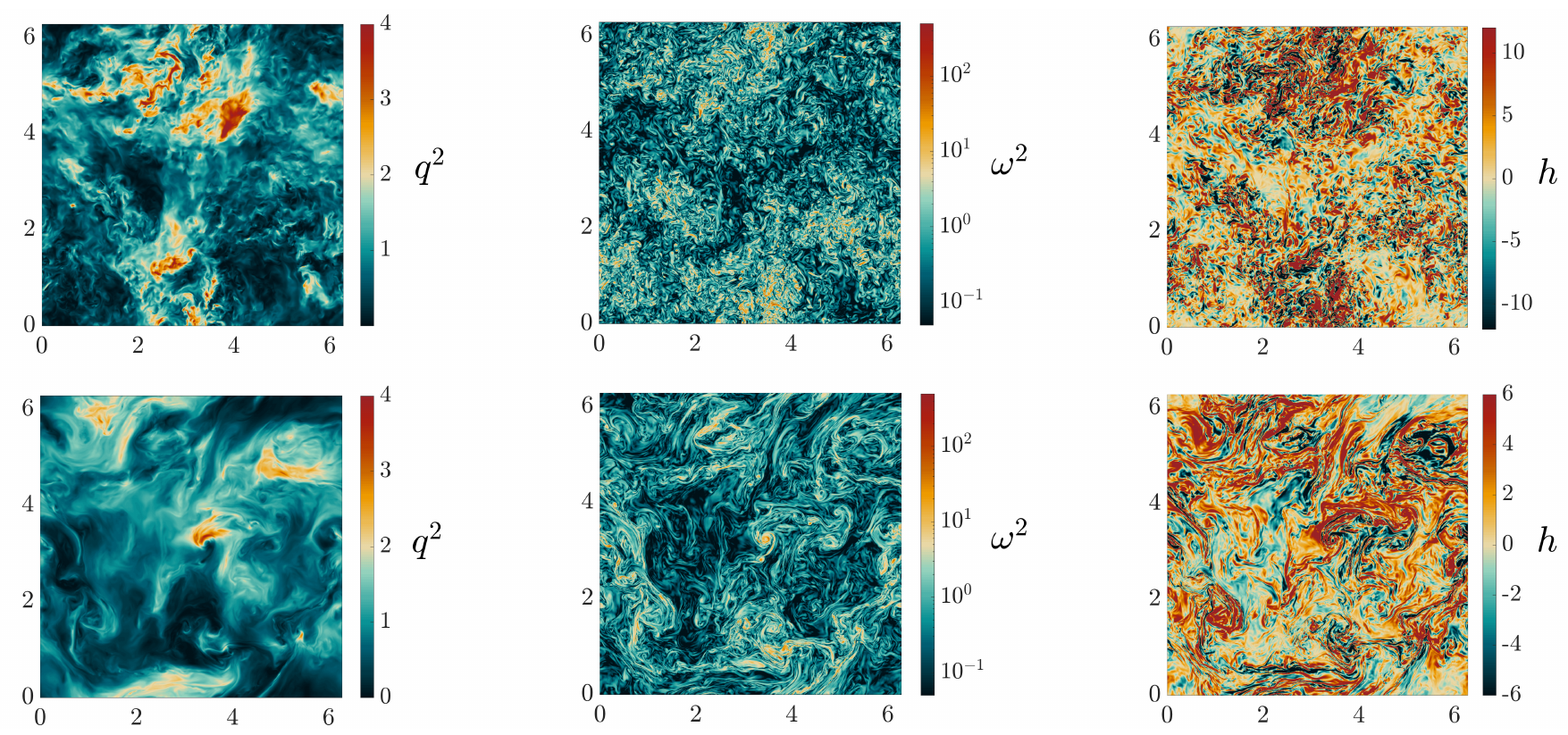}
\caption{Instantaneous visualisations of energy, enstrophy and helicity fields for $De=0$ (top) and $De=1$ (bottom). Left: $q^2 = u_i u_i$. Centre: $\omega^2 = \omega_i \omega_i $. Right: $h = u_i \omega_i$. All quantities are normalised with the average value in the considered slice.}
\label{fig:inst}
\end{figure}
Figure~\ref{fig:inst} presents representative two-dimensional slices of the three key quantities: kinetic energy $q^2 = u_i u_i$ (left), enstrophy $\omega^2 = \omega_i \omega_i$ (centre) and helicity $h=u_i\omega_i$ (right), obtained from both Newtonian $De=0$ (top) and visoelastic $De=1$ (bottom) simulations. Enstrophy and helicity exhibit smaller-scale, more intermittent structures due to their direct dependence on velocity gradients. The addition of polymers significantly modifies the flow structures. In the viscoelastic case, all three fields show noticeably smoother small-scale features, reflecting the polymer-induced suppression of small-scale fluctuations.

\begin{figure}
\centering
\includegraphics[width=0.32\textwidth]{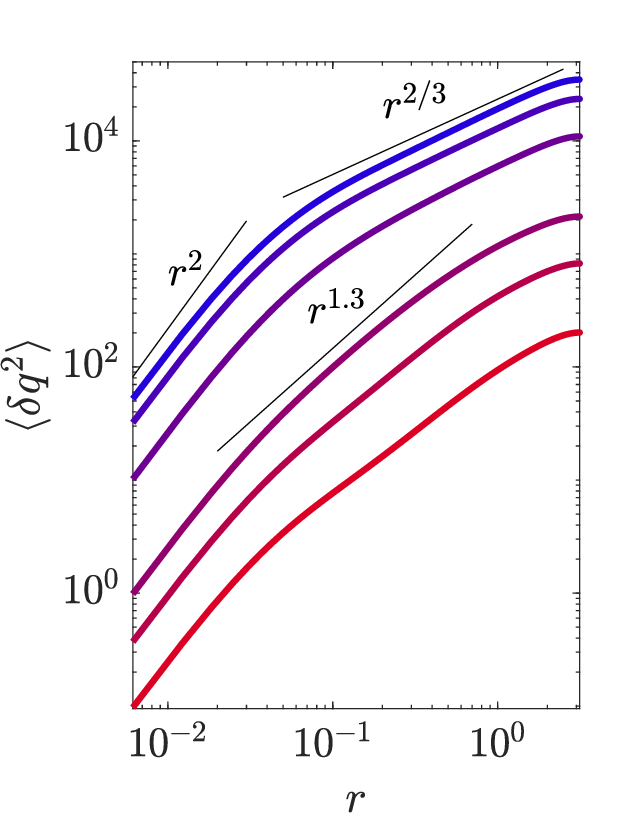}
\includegraphics[width=0.32\textwidth]{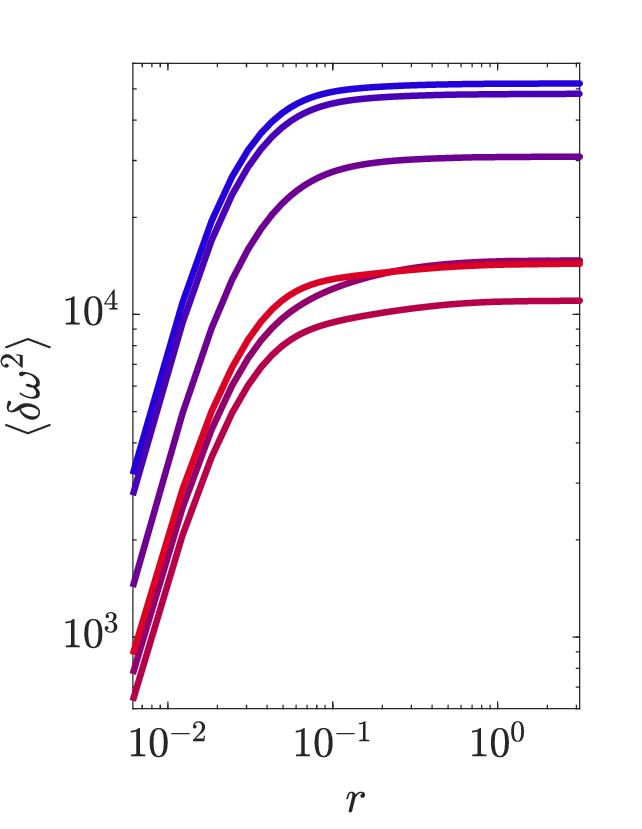}
\includegraphics[width=0.32\textwidth]{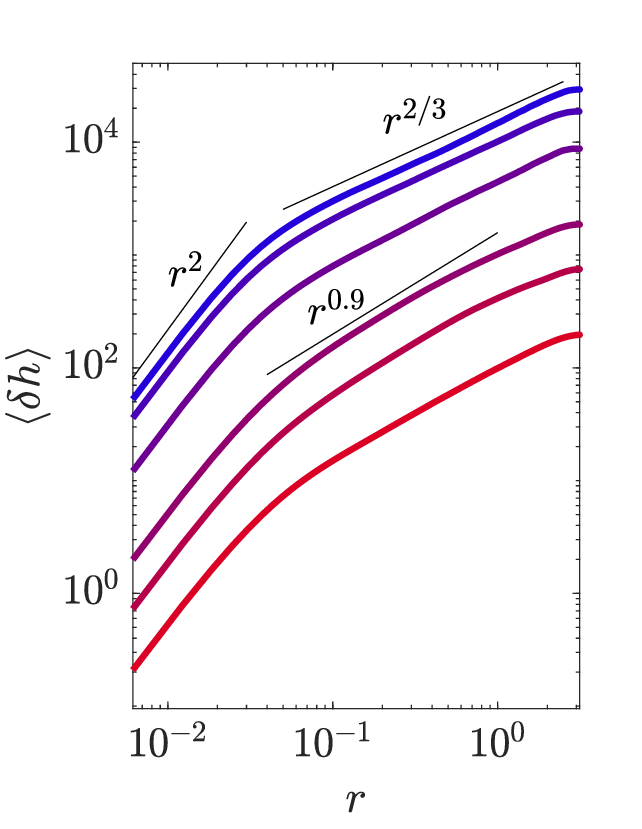}
\includegraphics[trim={0 50 0 0},clip,width=1.0\textwidth]{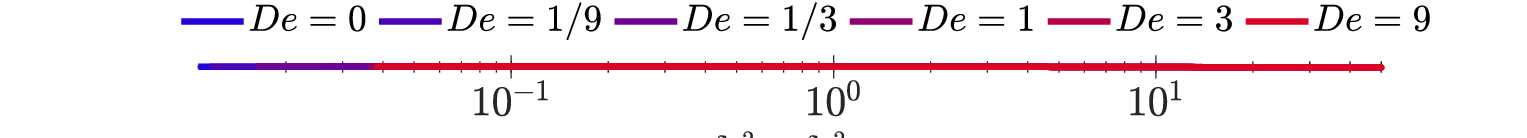}
\caption{Dependence of the velocity structure function $\aver{\delta q^2}$ (left), vorticity structure function $\aver{\delta \omega^2}$ (centre) and helicity structure function $\aver{\delta h}$ (right) on the Deborah number. For $\aver{\delta q^2}$ and $\aver{\delta h}$ the lines for the different $De$ are vertically shifted for increase the clarity.}
\label{fig:duidui}
\end{figure}

We start probing the velocity, vorticity and helicity structure functions $\aver{\delta q^2}$, $\aver{\delta \omega^2}$ and $\aver{\delta h}$ (see figure \ref{fig:duidui}), and discuss how they are influenced by polymeric additives.

The velocity structure function $\aver{\delta q^2}$ clearly shows the multiscaling behaviour of polymeric flows \citep{zhang-etal-2021,rosti-perlekar-mitra-2023,singh-rosti-2024}. In the Newtonian case and for $De \le 1/3$ we observe $\aver{\delta q^2} \sim r^{2/3}$ in the intermediate range of scales in agreement with the Kolmogorov predictions \citep{kolmogorov-1941}. For small $De$, i.e. in the limit of $\tau_p \rightarrow 0$, the polymers do not stretch, and they only marginally influence the velocity structure function. For $De = \mathcal{O}(1)$ the polymeric relaxation time is comparable with the large time scale of the flow and the polymers interact with the energy cascade, inducing a steeper depletion of $\aver{\delta q^2}$ with $r$ in the so-called elastic range of scales \citep{deangelis-etal-2005,perlekar-mitra-pandit-2010}.  For $De \ge 1$ we observe $\aver{\delta q^2} \sim r^{1.3}$ that matches the energy spectrum scaling $E(k) \sim k^{-2.3}$ \citep{rosti-perlekar-mitra-2023}, $k$ being the wavenumber. At large Deborah numbers ($De \gg 1$), polymers possess long relaxation times and tend to remain stretched, progressively decoupling from the carrier flow. As a result, their influence diminishes and the scaling of $\aver{\delta q^2}$ reverts toward the classical Kolmogorov prediction $\aver{\delta q^2} \sim r^{2/3}$ as $De \rightarrow \infty$ \citep{rosti-perlekar-mitra-2023}. Note that Kolmogorov scaling is not fully recovered for the largest $De$ considered in the current parameter set (see \S\ref{sec:spec}).

The scenario is thus the following. At low Deborah numbers $De \ll 1$, the polymers behave essentially as stiff chains and undergo only limited stretching. Consequently, the fluctuations in the polymer conformation tensor $R_{ij}$ and the polymer stress tensor $T_{ij}$ are rather small, resulting in negligible non-Newtonian stresses $\partial T_{ij}/\partial x_j \approx 0$. Under these conditions, the flow statistics closely resemble those of a purely Newtonian fluid, with the energy spectrum and structure functions exhibiting the classical $E(k) \sim k^{-5/3}$ and $S_p(r) \sim r^{p/3}$ scalings, where $S_p \equiv \aver{ \left(\left( \bm{u}(\bm{x}+\bm{r},t) - \bm{u}(\bm{x},t) \right) \cdot \bm{r}/r\right)^p }$, with $r = |\bm{r}|$; in HIT $S_2 \sim \aver{\delta q^2}$ due to the incompressibility constraint \citep{hill-1997}.
As $De$ increases towards $\mathcal{O}(1)$, the scaling behaviour changes. In this intermediate regime, the polymer relaxation time becomes comparable to the characteristic time scale of the fluid. Here, the energy transfer is mediated by the polymer elasticity rather than solely by the nonlinear cascade (see \S\ref{sec:energy}), leading to the modified scalings $E(k) \sim k^{-2.3}$ for the energy spectrum and $S_2(r) \sim r^{1.3}$ for the second-order structure function. However, this regime cannot persist in the limit $De \rightarrow \infty$. For $De \gg 1$, indeed, the polymers effectively decouple from the fluid dynamics. Due to their long relaxation times, the polymers respond to flow changes with significant delay, and remain stretched for extended periods. This suppresses fluctuations in polymer lengths, weakens the non-Newtonian stresses $\partial T_{ij}/\partial x_j$, and causes the flow to revert towards a Kolmogorov-like behaviour. The progressive narrowing of the scale range associated with the elastic scaling for $De \gtrsim 1$ is more clearly observed in Fourier space, as discussed in appendix~\S\ref{sec:spec}.

The exact value of $De$ at which the flow transitions between different regimes is influenced by the method used to sustain turbulence. For instance, \citet{rosti-perlekar-mitra-2023} conducted simulations at the same Reynolds number as in the present study but adopted the forcing scheme of \citet{eswaran-pope-1988}---which injects energy randomly at low wavenumbers---, rather than the ABC forcing used here---which generates a more coherent large-scale flow structure \citep{chiarini-cannon-rosti-2024}. With their approach, \citet{rosti-perlekar-mitra-2023} observed the onset of Kolmogorov-like scaling already at $De = 9$, a behaviour not seen in our dataset. Distinct forcing mechanisms adopted to sustain turbulence indeed affect differently the large-scale dynamics and, therefore, the characteristic time scale used to define $De$.
As a result, the precise values and range of $De$ over which polymers dominate, and the $S_2 \sim r^{1.3}$ scaling emerges, varies slightly depending on the external forcing. Nevertheless, the overall qualitative behaviour remains unchanged, with only modest $\mathcal{O}(1)$ shifts in the critical $De$ values delineating the different regimes.

It is worth noting that $\aver{\delta q^2} \sim r^2$ as $r\rightarrow 0$ across all cases, consistent with the expected smoothness of the velocity field at small scales~\citep{schumacher-etal-2007}. In real space, this smooth behaviour precludes the observation of the rise in energy at deep dissipative scales that has been reported in the Fourier-space energy spectra by \cite{perlekar-mitra-pandit-2010}, \cite{rosti-perlekar-mitra-2023}, and by \cite{singh-rosti-2024} using the present dataset.

The vorticity structure function $\aver{\delta \omega^2}$ is shown in the central panel of figure \ref{fig:duidui}. It is clear that enstrophy growth (generation) is confined to small scales for all $De$, as already observed by other authors for Newtonian HIT \citep{jimenez-etal-1993,ishihara-kaneda-hunt-2013,elsinga-etal-2017}. Indeed, $\aver{\delta \omega^2}$ rapidly increases with $r$ at the small scales and saturates in the inertial range. The polymers reduce the vorticity fluctuations at the small scales \citep{liberzon-etal-2005,liberzon-etal-2006,perlekar-mitra-pandit-2010} which results in a decrease of $\aver{\delta \omega^2}$ at all $r$. Note that, like for $\aver{\delta q^2}$, the effect of the polymers on the vorticity structure function is non-monotonous, with $\aver{\delta \omega^2}$ being minimum for $De=3$ at all scales.

The dependence of the helicity structure functions $\aver{\delta h}$ on $De$ is shown in the right panel of figure \ref{fig:duidui}. In the Newtonian case, we find $\aver{\delta h} \sim r^{2/3}$ in the inertial range, in agreement with the existence of the cascade of helicity from larger to the smaller scales envisaged by \cite{brissaud-etal-1973}; see also \cite{polifke-shtilman-1989}, \cite{borue-orszag-1997} and \cite{baj-etal-2022} and \S\ref{sec:helicity}. Note also that $\aver{\delta h}$ and $\aver{\delta q^2}$ have an almost identical $r^{2/3}$ range, indicating that the inertial range for helicity spans nearly the same range of scales as that for energy, confirming the findings of \cite{chen-etal-2003b}. The effect of the polymers on $\aver{\delta h}$ resembles what has been observed for $\aver{\delta q^2}$, suggesting that the two cascades are influenced in a similar fashion. For small and large $De$, the influence of the polymers is rather small: the $r^{2/3}$ scaling is recovered for $De \rightarrow 0$ and $De \rightarrow \infty$. For intermediate $De$, instead, $\aver{\delta h}$ shows a steeper dependence on $r$ in the elastic range of scales. For $De = 1$, we find $ \aver{\delta h} \sim r^{0.9}$, which is shallower compared to the $\aver{\delta q^2} \sim r^{1.3}$ scaling for the energy.
We have $\aver{\delta h} \sim r^2$ as $r \rightarrow 0$, similar to the velocity structure function. Overall, the results show that helicity is mainly concentrated at the large scales for all $De$: turbulence cascade tends to restore the mirror symmetry at the small scales in both Newtonian and polymeric turbulence~\citep{ditlevsen-giuliani-2001,chen-etal-2003b,baj-etal-2022}; see \S\ref{sec:helicity} for additional discussion.

\begin{figure}
  \centering
  \includegraphics[width=\textwidth]{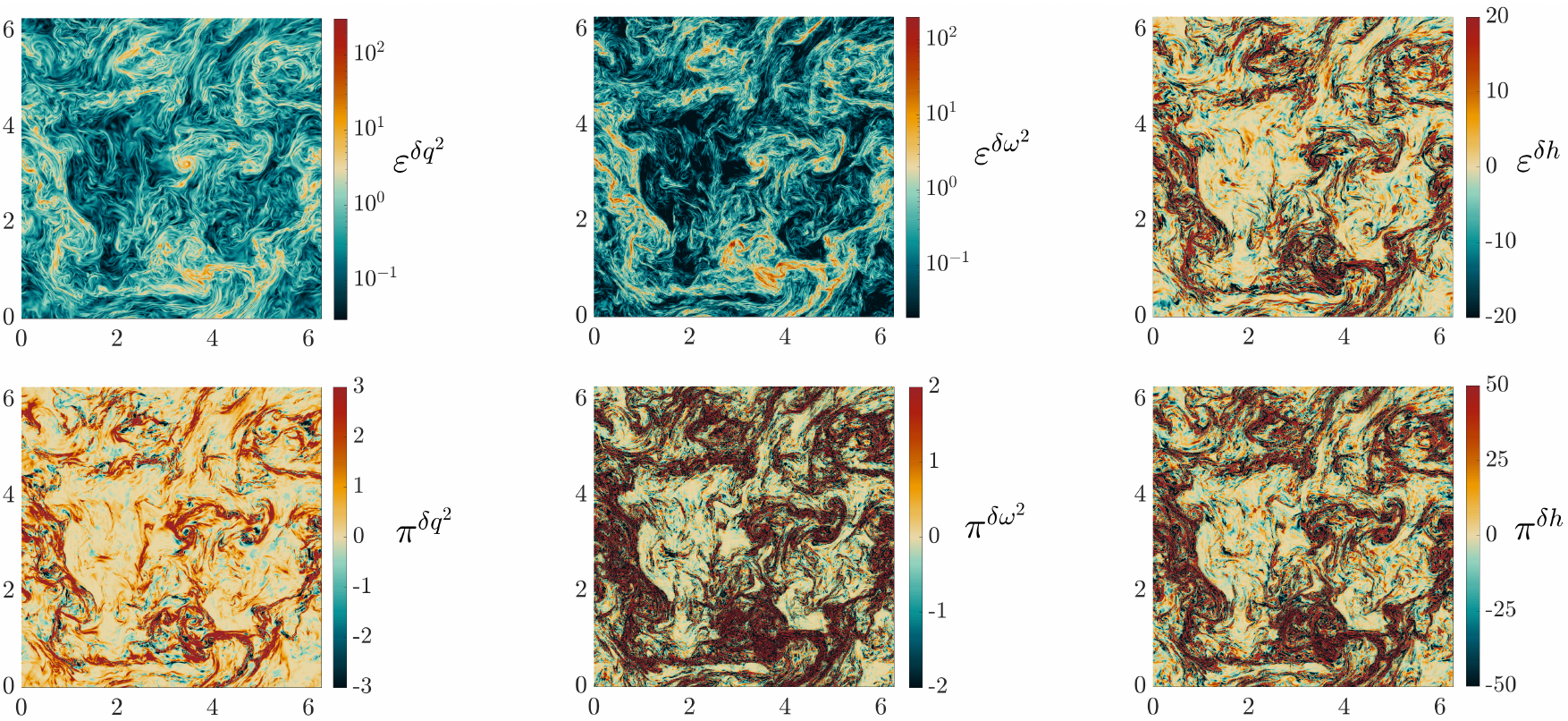}
  \caption{Instantaneous visualisations of the dissipation fields and the polymeric source/sink terms for $De=1$. Top: Dissipation fields $\varepsilon_f^{\delta q^2}$ (left), $\varepsilon_f^{\delta \omega^2}$ (centre) and $\varepsilon_f^{\delta h}$ (right). Bottom: Polymeric source/sink terms $\pi^{\delta q^2}$ (left), $\pi^{\delta \omega^2}$ (centre) and $\pi^{\delta h}$ (right). All the quantities are normalised with the modulus of their averaged value.}
  \label{fig:diss_snap}
\end{figure}
We now turn our attention to the dissipation rate fields and the polymeric source/sink terms. As illustrated in figure~\ref{fig:diss_snap} through instantaneous visualisations, the fluid and polymer contributions exhibit markedly different behaviours. For both energy and enstrophy, the fluid dissipation fields $\varepsilon_f^{\delta q^2}$ and $\varepsilon_f^{\delta \omega^2}$ are strictly positive, reflecting the fact that viscous effects irreversibly dissipate these quantities at all spatial locations.
In contrast, the polymeric terms $\pi^{\delta q^2}$ and $\pi^{\delta \omega^2}$ can take on both positive and negative values, indicating that polymers may locally act either as sinks or sources of energy and enstrophy. In the case of energy, this behaviour is consistent with the appearance of the same term (but with opposite sign) in the evolution equation for the free energy of the polymers \citep{deangelis-etal-2005}: it represents a bidirectional exchange of energy between the fluid and the polymeric phases. On average, however, $\langle \pi^{\delta q^2} \rangle >0 $, indicating a net transfer of energy from the fluid to the polymeric phase, where it is eventually dissipated. 
Figure \ref{fig:diss_snap} shows that both positive and negative intense events of $\pi^{\delta \omega^2}$ are concentrated in regions of intense shear.
Unlike energy and enstrophy dissipation, both $\varepsilon_f^{\delta h}$ and $\pi^{\delta h}$ associated with helicity are not sign-definite, meaning they can contribute positively or negatively depending on the local flow configuration.

\begin{figure}
\centering
\includegraphics[width=0.32\textwidth]{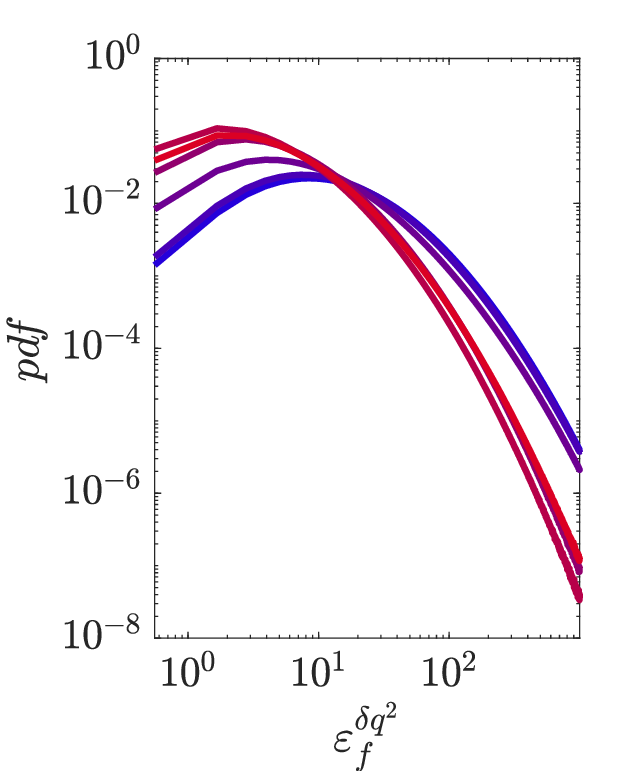}
\includegraphics[width=0.32\textwidth]{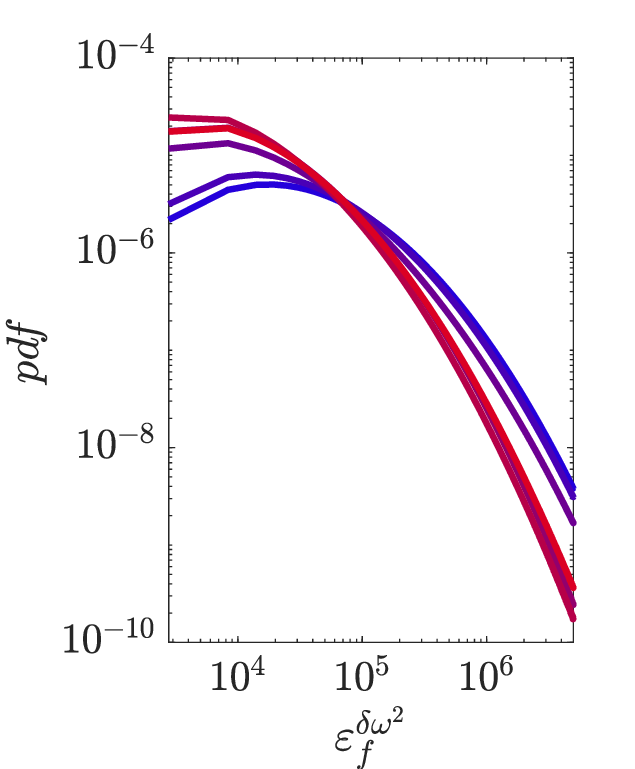}
\includegraphics[width=0.32\textwidth]{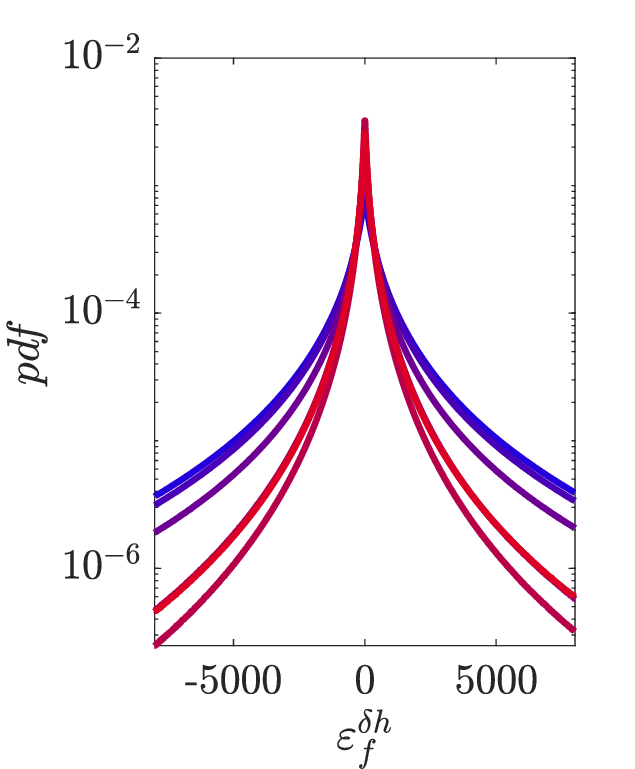}
\includegraphics[width=0.32\textwidth]{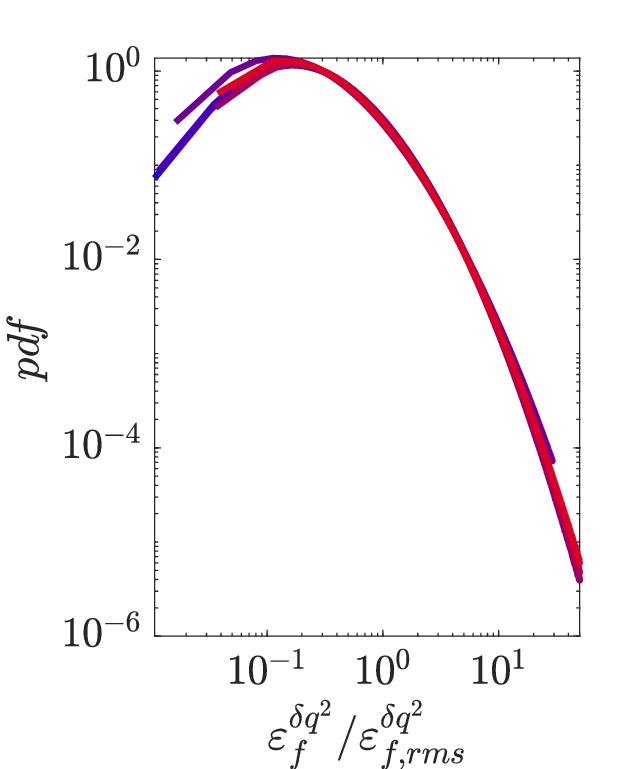}
\includegraphics[width=0.32\textwidth]{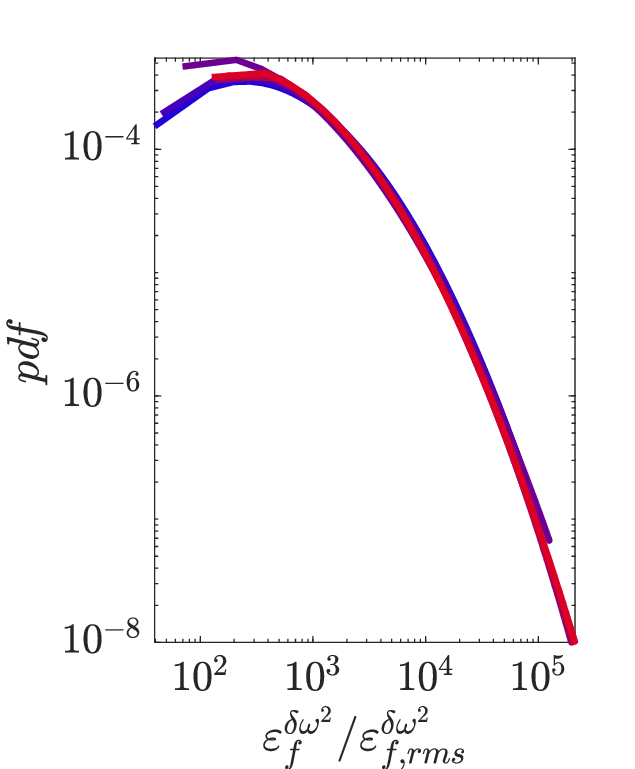}
\includegraphics[width=0.32\textwidth]{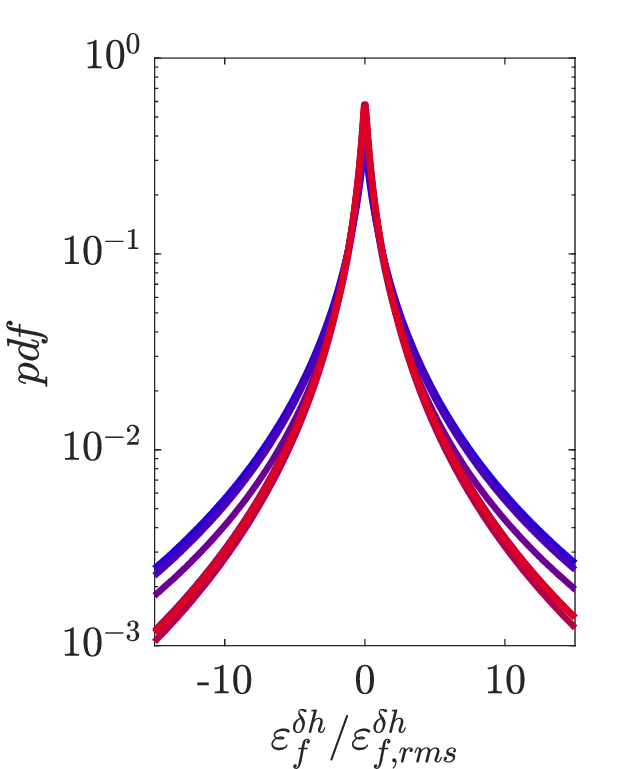}
\includegraphics[trim={0 50 0 0},clip,width=1.0\textwidth]{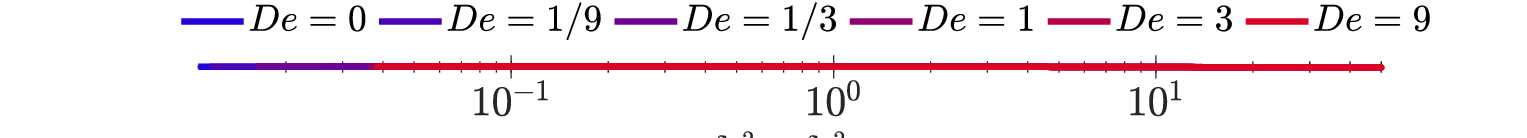}
\caption{Distribution of the dissipation of (left) energy, (centre) enstrophy, and (right) helicity for different Deborah numbers. The bottom panels plot the distribution of the three quantities normalised with their root-mean-square value.}
\label{fig:diss}
\end{figure}
We now present a statistical quantification of the fluid dissipation associated with all three quantities in figure~\ref{fig:diss}. These quantities play a key role in the scale-by-scale cascades. In Newtonian HIT, for the two averaged, inviscid invariants (energy and helicity) the dissipation rate matches the injection rate at the large scales and the averaged total flux in the inertial range of scales. The distributions of the dissipations show large tails, which are indicative of small-scale intermittency \citep{sreenivasan-antonia-1997}, see figure \ref{fig:diss_snap}. As expected, the distributions narrow for the polymeric case: polymers reduce the magnitude of the velocity derivatives, thus largely reducing the regions of large dissipations \citep{liberzon-etal-2005,liberzon-etal-2006}. The distributions of the energy and enstrophy dissipation collapse nicely for all $De$ once the quantities are normalised with the respective standard deviation, similarly to what was shown by \cite{perlekar-mitra-pandit-2010} and is consistent with the $De$ invariance of intermittency corrections observed in~\citet{rosti-perlekar-mitra-2023}.

\begin{figure}
\centering
\includegraphics[width=0.32\textwidth]{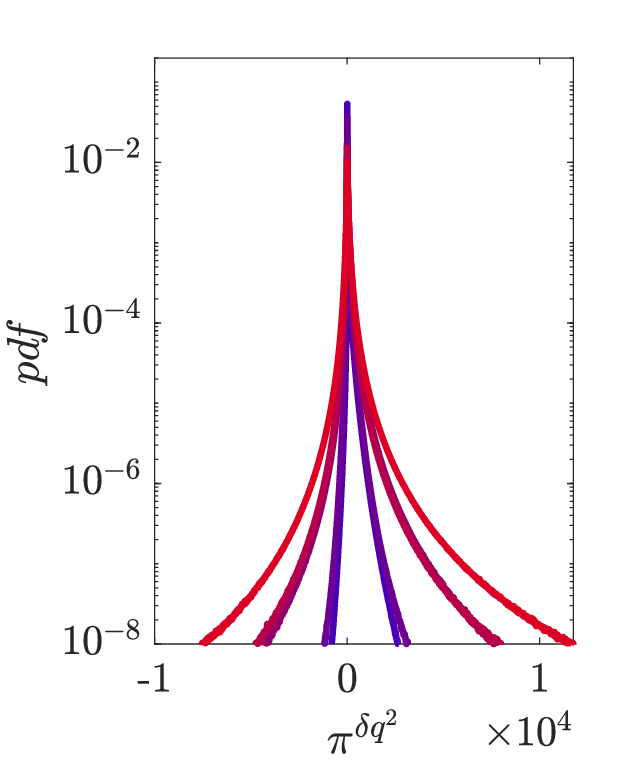}
\includegraphics[width=0.32\textwidth]{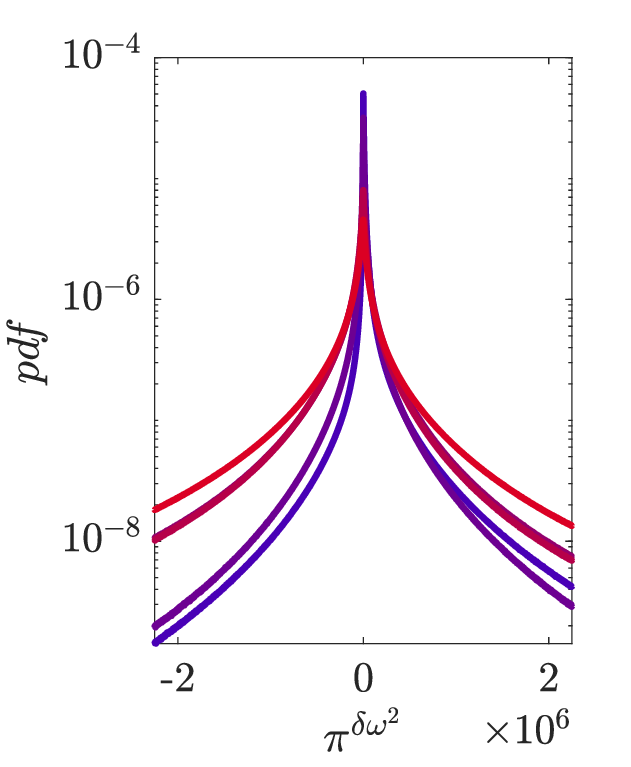}
\includegraphics[width=0.32\textwidth]{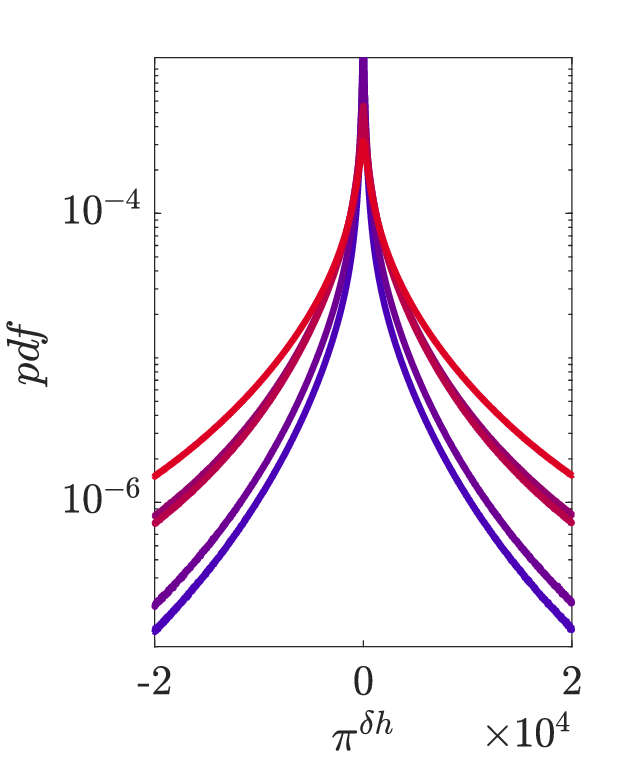}
\includegraphics[width=0.32\textwidth]{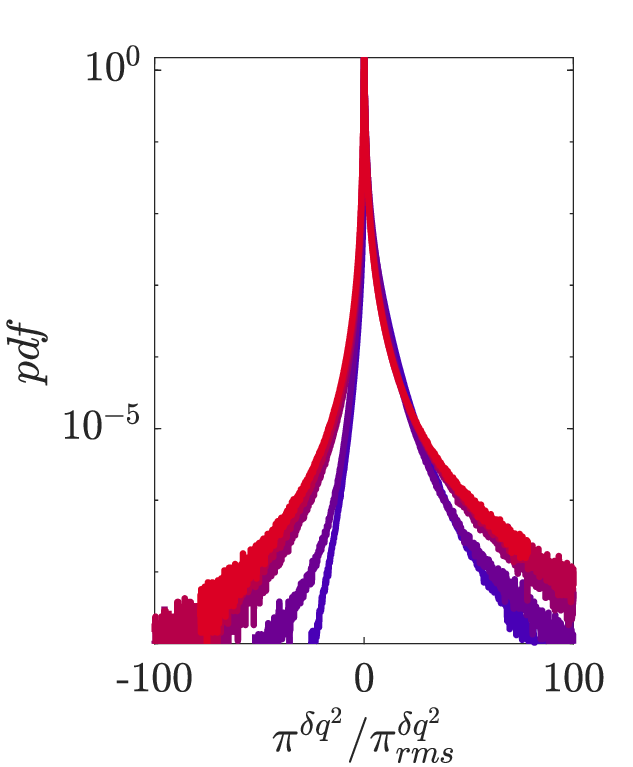}
\includegraphics[width=0.32\textwidth]{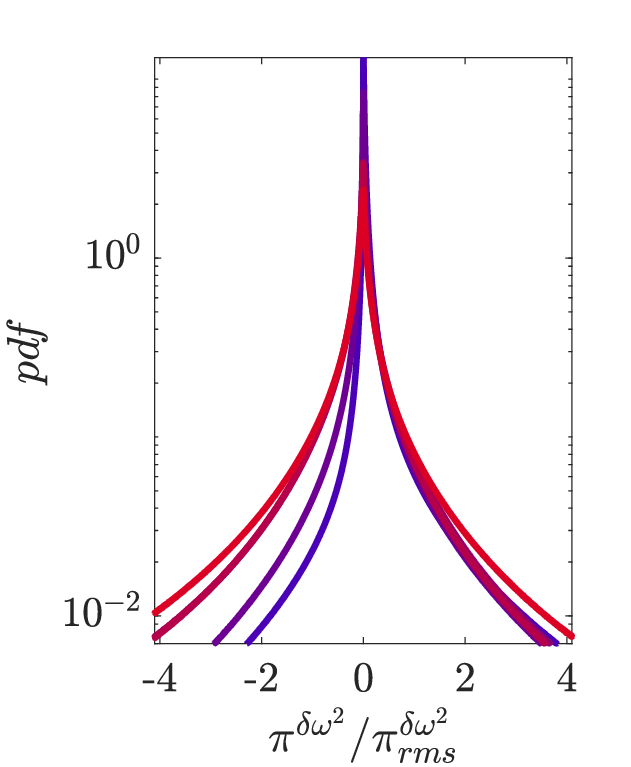}
\includegraphics[width=0.32\textwidth]{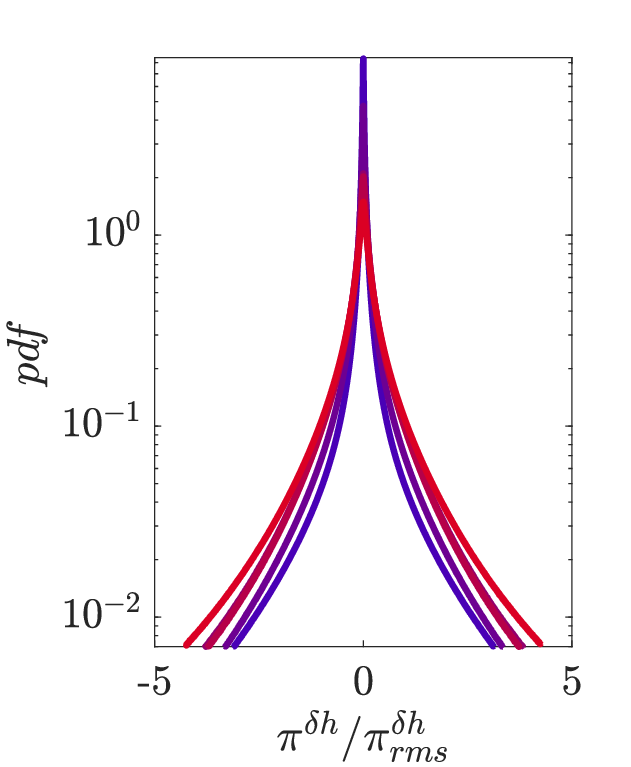}
\includegraphics[trim={0 50 0 0},clip,width=1.0\textwidth]{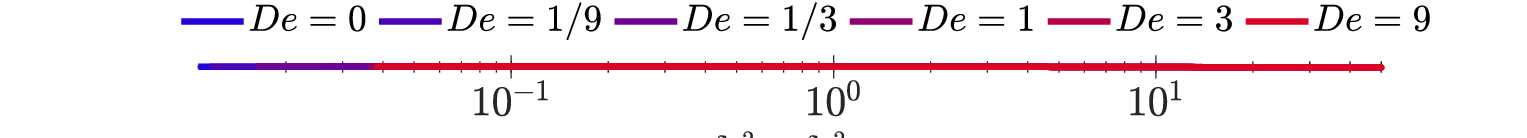}
\caption{Distribution of the the $\pi$ terms for $\delta q^2$, $\delta \omega^2$ and $\delta h$ for different Deborah numbers. Left: $\pi^{\delta q^2}$, centre: $\pi^{\delta \omega^2}$ and right: $\pi^{\delta h}$. The bottom panels plot the distribution of the three quantities normalised with their root-mean-square value.}
\label{fig:Pi}
\end{figure}
We conclude this section with the distributions of $\pi^{\delta q^2}$, $\pi^{\delta \omega^2}$ and $\pi^{\delta h}$.  
As mentioned above, these quantities are not positive definite, meaning that locally polymers may act as either a sink or a source for the velocity/vorticity fluctuations. The distribution of $\pi^{\delta q^2}$ is right-skewed for all $De$, confirming the positive average value $\langle \pi^{\delta q^2} \rangle = \langle \epsilon_p \rangle>0$.
As $De$ increases, both the right and left tails widen, meaning that sink and source intense events are equally promoted by the larger stretching of the polymers. The effect of $De$ is much more significant on $\pi^{\delta \omega^2}$, whose distributions are shown in the central panels of figure \ref{fig:Pi}, see also table \ref{tab:simulations}.
At small $De < 1$, the distributions are right skewed and, on average, the polymers act as a sink for the enstrophy $\langle \pi^{\delta \omega^2} \rangle >0$. For $De \ge 1$, instead, the distribution of $\pi^{\delta \omega^2}$ is left skewed. In this case, due to the larger stretching of the polymers, the fluid-polymer interaction facilitates intense production of enstrophy $\langle \pi^{\delta \omega^2} \rangle<0$ and for large enough $De$ it emerges as the dominant source of enstrophy generation; this will be further discussed in \S\ref{sec:enstrophy} and \S\ref{sec:loc-str}. Eventually, the right panels in figure \ref{fig:Pi} shows that the distribution of $\pi^{\delta h}$ is symmetric, and that both tails become longer as $De$ increases.

\section{Scale-by-scale balance equations}
\label{sec:resbud}
\subsection{Scale energy $\aver{\delta q^2}$}
\label{sec:energy}

We now examine the influence of polymeric additives on the energy transfers. We begin by analysing the budget equation for $\aver{\delta q^2}$, focusing on how the average energy cascade is altered. This real-space analysis complements the Fourier-space study conducted by \cite{rosti-perlekar-mitra-2023}. Subsequently, we explore the nonlinear and polymeric fluxes in detail, highlighting the role of polymers in modifying the intermittency of energy transfers and the localised backscatter.

\subsubsection{Scale-by-scale budget}

\begin{figure}
\centering
\includegraphics[width=0.49\textwidth]{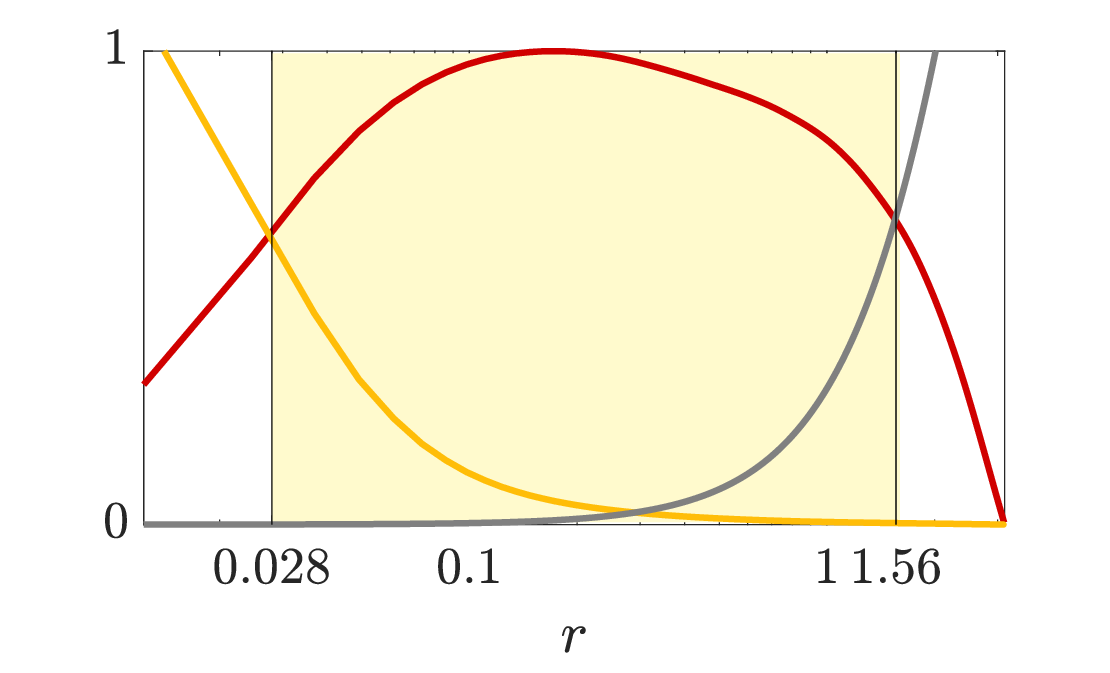}
\includegraphics[width=0.49\textwidth]{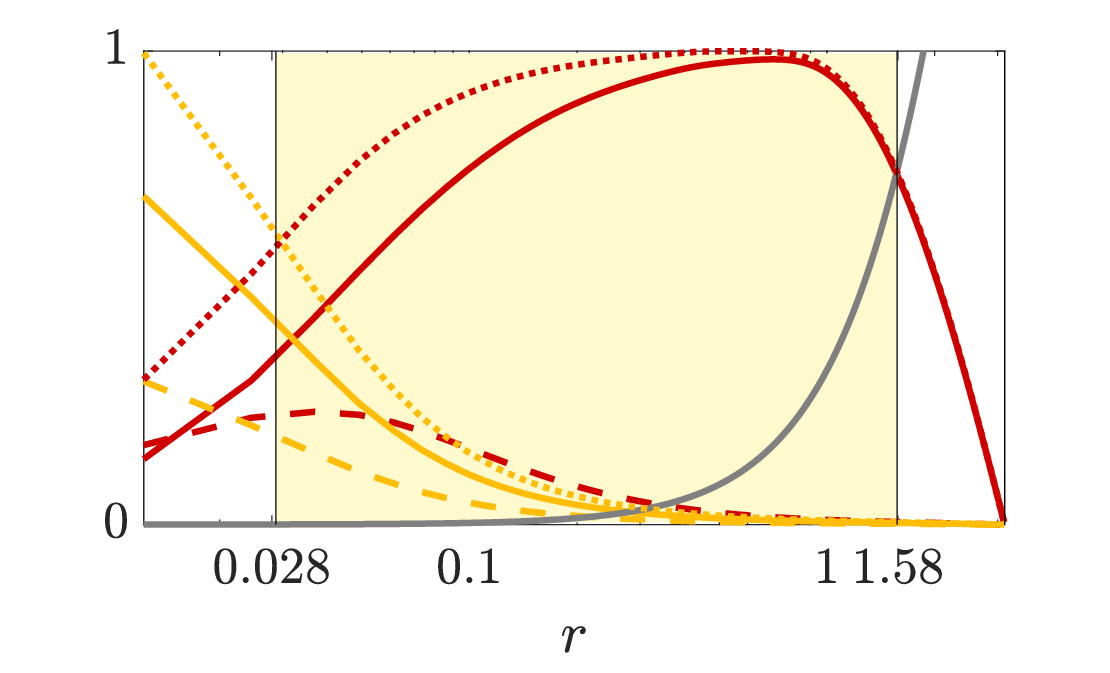}
\includegraphics[width=0.49\textwidth]{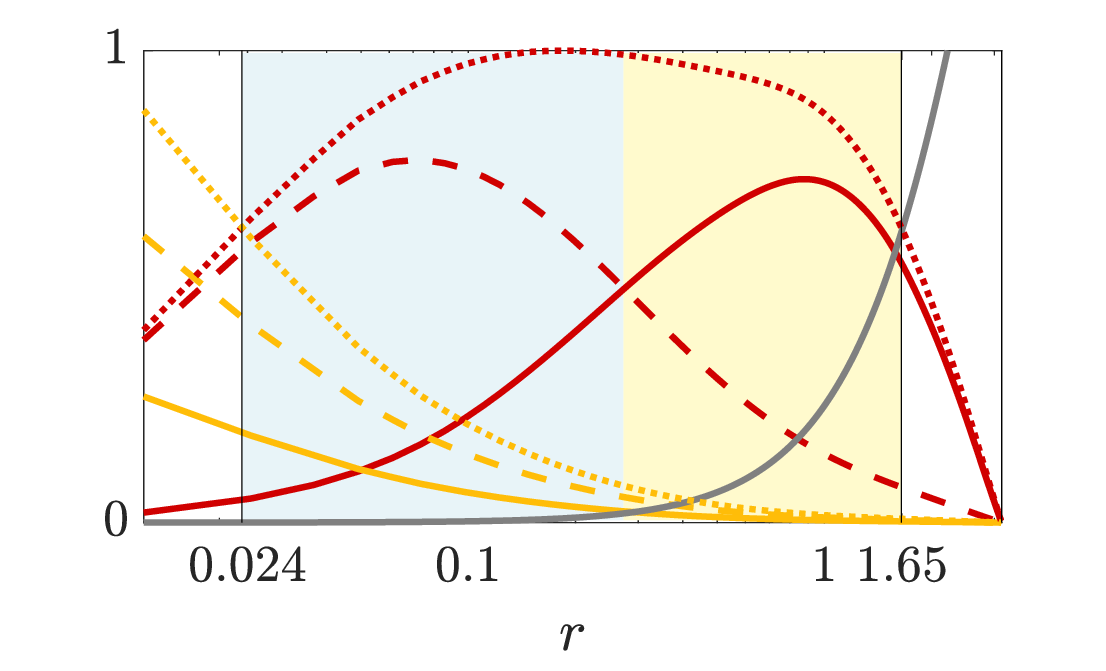}
\includegraphics[width=0.49\textwidth]{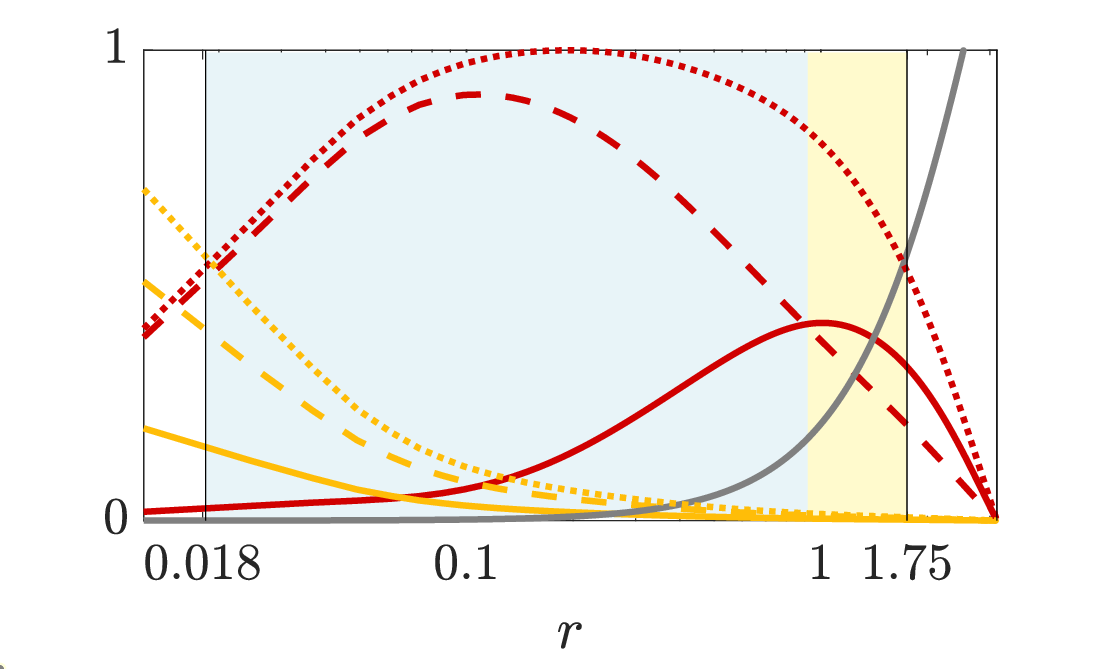}
\includegraphics[trim={0 10 0 10},clip,width=0.7\textwidth]{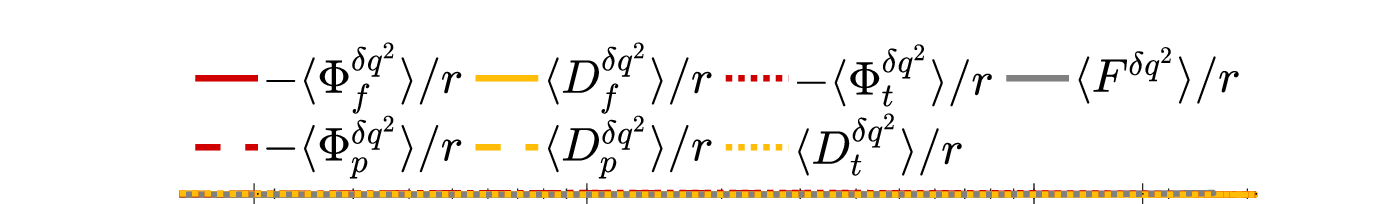}
\caption{All contributions in budget equation for $\aver{\delta q^2}$. The panels are for $De=0$, $De=1/3$, $De=1$, and $De=9$, from top-left to bottom-right. The yellow and blue shaded regions identify the inertial and elastic ranges of scales. Here $\Phi_t = \Phi_f + \Phi_p$ and $D_t = D_f + D_p$.}
\label{fig:ener-bud}
\end{figure}

Figure \ref{fig:ener-bud} represents the terms in the budget equation for $\aver{\delta q^2}$ for both Newtonian HIT and PHIT. 

We begin by examining the $De=0$ case, which represents the classical Richardson-Kolmogorov turbulence scenario for Newtonian fluids. Energy is injected at the largest scales, $r>r_{IE} \approx 1.56$, through external forcing, with $\langle F^{\delta q^2} \rangle \approx (4/3) \langle \varepsilon_f^{\delta q^2} \rangle r$. From there, energy is transferred from larger to smaller scales within the inertial range of scales $0.028 \approx r_{DI} \le r \le r_{IE}$ via the classical inertia-driven cascade. Here, $r_{IE}$ marks the scale above which forcing dominates, separating the inertial and energetic ranges, while $r_{DI}$ denotes the scale below which viscous dissipation dominates, separating the dissipative and inertial ranges. Consistent with the Kármán-Howarth equation, the energy transfer rate remains constant within the inertial range and equal in magnitude to the dissipation rate as evidenced by the plateau in the top-left panel of figure \ref{fig:ener-bud}, here $\langle \Phi_f^{\delta q^2} \rangle \approx -(4/3) \langle \varepsilon_f \rangle r$,. At scales smaller than $r_{DI}$, viscous effects $\langle D_f^{\delta q^2} \rangle$ dominate, and $\langle \delta q^2 \rangle $ is dissipated into heat through viscous friction.

Polymeric additives introduce an additional dissipative process $D_p^{\delta q^2}$ and an alternative route for energy transfer among scales $\Phi_p^{\delta q^2}$. For small $De \ll 1$, the global picture resembles what observed in Newtonian HIT. In this case, indeed, the polymers are only marginally stretched as their relaxation time is small compared to the characteristic time scales of the flow, and they quickly relax back to their equilibrium lengths; here, $\Phi_{p}^{\delta q^2} \approx D_{p}^{\delta q^2} \approx \pi^{\delta q^2} \approx 0$. 

As $De$ increases towards $\mathcal{O}(1)$ the relevance of the non-Newtonian contribution increases, and the polymeric additives modify the energy cascade in a non trivial way. Energy enters the system at the large scales ($r>r_{IE}$) and is dissipated away at the same rate, via both the fluid dissipation and the polymeric sink term, i.e.
$ \langle F^{\delta q^2} \rangle = (4/3) (\langle \varepsilon_f^{\delta q^2} \rangle + \langle \pi^{\delta q^2} \rangle ) r$.
In the intermediate range of scales ($r_{DI} < r < r_{IE}$) energy is then transferred from larger to smaller scales by two alternative routes, respectively associated with (i) the classic inertia-driven term $\Phi_{f}^{\delta q^2}$ and (ii) the fluid-polymer interaction term $\Phi_{p}^{\delta q^2}$. At these scales the viscous effects are negligible, and the total energy transfer rate is an invariant and matches the total dissipation rate \citep{chiarini-singh-rosti-2025}; here
 $\langle \Phi_f^{\delta q^2} \rangle + \langle \Phi_f^{\delta q^2} \rangle \approx - (4/3) ( \langle \varepsilon_f^{\delta q^2} \rangle + \langle \pi^{\delta q^2} \rangle ) r$.
Eventually, at the smallest scales ($r < r_{DI}$) energy is dissipated away by both the fluid and the polymers as 
$\text{d} \langle D_f^{\delta q^2} \rangle/\text{d}r \rightarrow (4/3) \langle \varepsilon_f^{\delta q^2} \rangle$
and 
$\text{d} \langle D_p^{\delta q^2} \rangle/\text{d}r \rightarrow (4/3) \langle \pi^{\delta q^2} \rangle$
 for $r \rightarrow 0$.

The range of scales over which the energy cascade is significantely altered by the presence of polymers varies with $De$, as indicated by the blue shaded region in figure \ref{fig:ener-bud}. To quantify this, we define for each $De$ a characteristic scale $r_p^*$ below which elastic effects are expected to dominate. This scale is estimated by comparing the local eddy turnover time at scale $r$, i.e. $\tau_f(r) \sim  r/\aver{\delta V^2}^{1/2}$, where $\delta V(r) = ( u_i(\bm{x}+\bm{r}) - u_i(\bm{x}) ) \cdot r_i/r)$, with the polymeric relaxation time $\tau_p$. Thus, $r_p^*$ is estimated as
\begin{equation}
\frac{r_p^*}{\sqrt{\aver{\delta V^{2}}^{*}}} = \tau_p, \quad \text{where} \quad \aver{\delta V^2}^* = \aver{\delta V^2}(r_p^*).
\end{equation}
For $r>r_p^*$ the time scale of the fluid fluctuations is larger than $\tau_p$ ($\tau_p/\tau_f(r)<1$): the polymers are only marginally stretched and thus their influence on the flow is weak. On the contrary, when $r<r_p^*$ the time scale of the fluid is smaller than the polymeric relaxation time ($\tau_p/\tau_f(r)>1$): the polymers interact more effectively with the fluid and thus modify the organisation of the corresponding fluctuations. For the present cases, we measure $r_p^* \approx 0.04, 0.32, 1.34$ and $2.81$ for $De=1/9,1/3,1$ and $3$: the range of scales where elastic effects significantly modify the energy cascade increases with $De$, in agreement with the widening of the elastic range of scales shown in figure \ref{fig:ener-bud}. For $De=9$, $\tau_p> r/\aver{\delta V^2}^{1/2}$ for all $r$, suggesting that the polymeric additives effectively modify the energy cascade in the whole range of scales. The scaling behavior observed in figure \ref{fig:duidui} aligns with $r_p^*$ lying well within the inertial range $r_{DI} < r \le r_{IE}$ when $De$ and $Re$ are sufficiently large. For small $De$ and/or low $Re$, the $\langle \delta q^2 \rangle \sim r^{1.3}$ scaling disappears as $r_p^*$ falls within the dissipative range \citep{deangelis-etal-2005,perlekar-mitra-pandit-2010}. We reiterate that when $De$ is very large and $\tau_p \gg r/\aver{\delta V^2}^{1/2}$ the polymeric chains decouple from the carrier phase and their influence on the carrier flow is rather low.

This suggests that the relevant control parameter is not the global Deborah number, but rather the scale-dependent or local Deborah number, defined as $De(r) = \tau_p / \tau_f(r)$, where $\tau_f(r)$ is indeed the characteristic timescale of eddies of size $r$.
This perspective is consistent with the observation that different bulk flow quantities exhibit a change in behaviour at different values of (the global) $De$; see table \ref{tab:simulations}. Specifically, the inversion in the behaviour of a given quantity due to the change of regime is expected to occur when $De(r)$ becomes sufficiently large at the scale $r$ where that quantity is most active or sensitive.

A last comment regards the influence of $De$ on the dissipative range of scales. Our data show that $r_{DI}$ (i.e. scale below which the total dissipative term $D_t^{\delta q^2} = D_f^{\delta q^2} + D_p^{\delta q^2}$ dominates) decreases with $De$, thus indicating that the dissipative range of scales progressively shrinks as the polymeric relaxation time increases.

\subsubsection{The nonlinear and non Newtonian fluxes}

\begin{figure}
\centering
\includegraphics[width=0.32\textwidth]{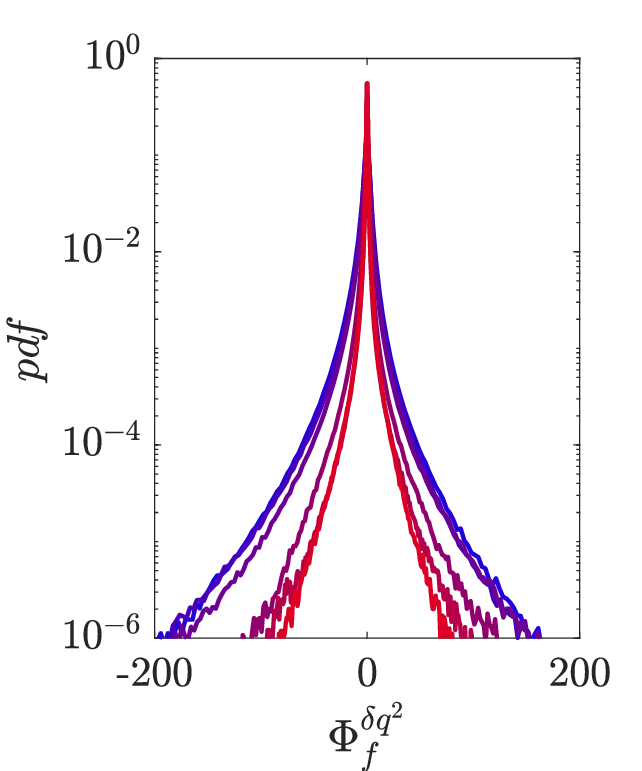}
\includegraphics[width=0.32\textwidth]{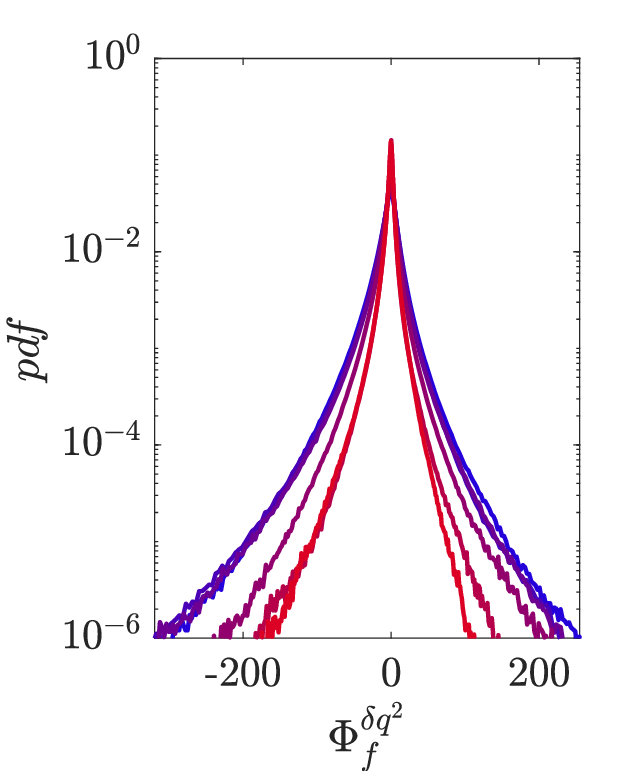}
\includegraphics[width=0.32\textwidth]{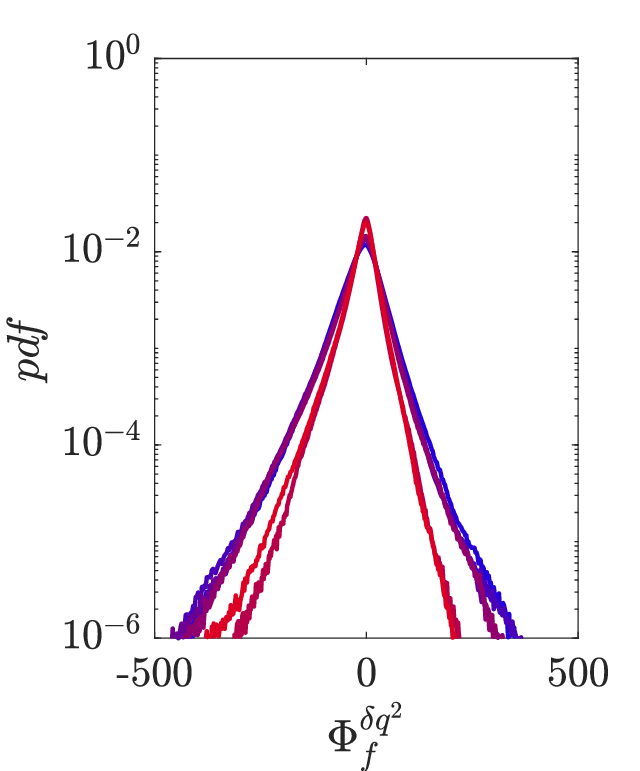}
\includegraphics[width=0.32\textwidth]{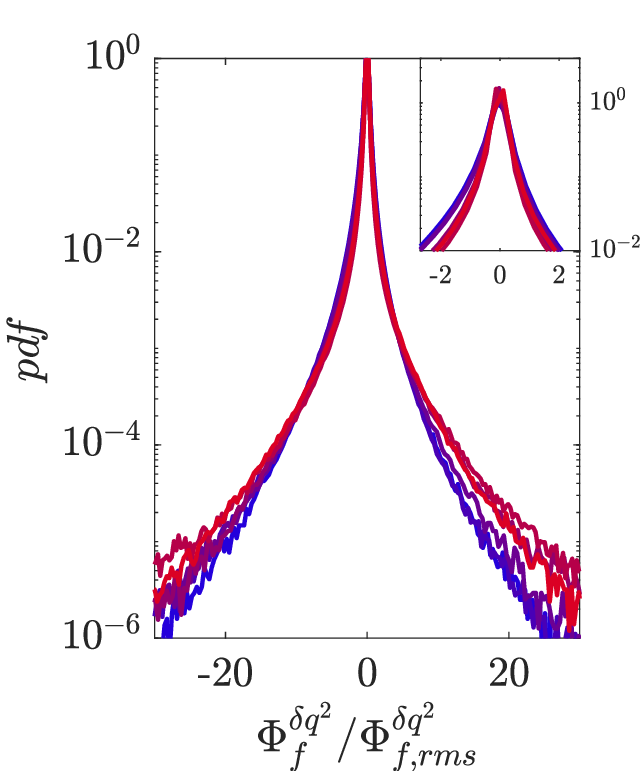}
\includegraphics[width=0.32\textwidth]{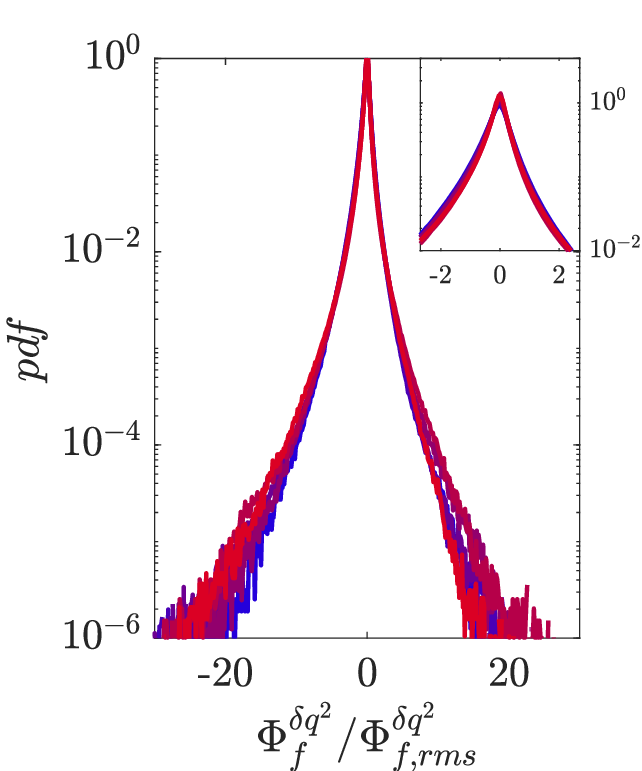}
\includegraphics[width=0.32\textwidth]{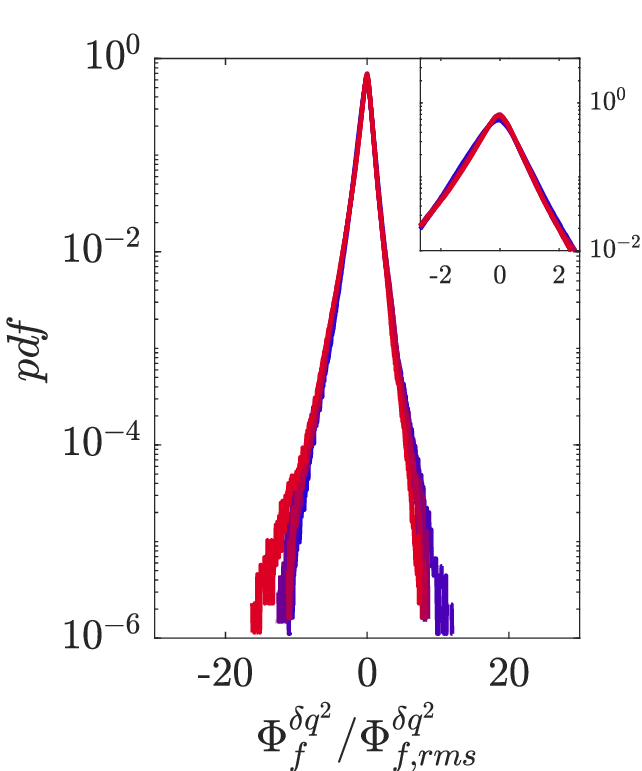}
\includegraphics[trim={0 50 0 0},clip,width=1.0\textwidth]{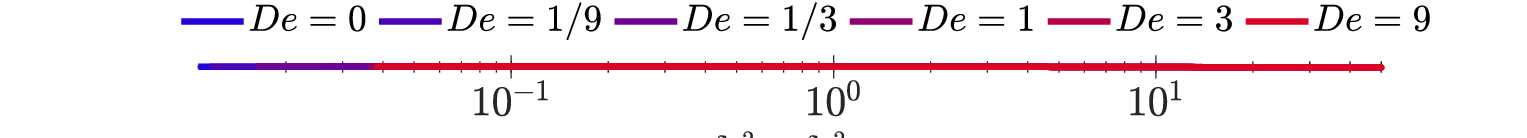}
\caption{Dependence of the distribution of the nonlinear flux $\Phi_{f}$ on the Deborah number, for (left) $r=0.1$, (centre) $r=0.3$, and (right) $r=0.9$. Top: distribution of $\Phi_f^{\delta q^2}$. Bottom: distribution of $\Phi_f^{\delta q^2}/\Phi_{f,rms}^{\delta q^2}$. In the bottom panels the inset provide a zoom of the normalised distributions.}
\label{fig:phinl_ener}
\end{figure}

\begin{figure}
\centering
\includegraphics[width=0.32\textwidth]{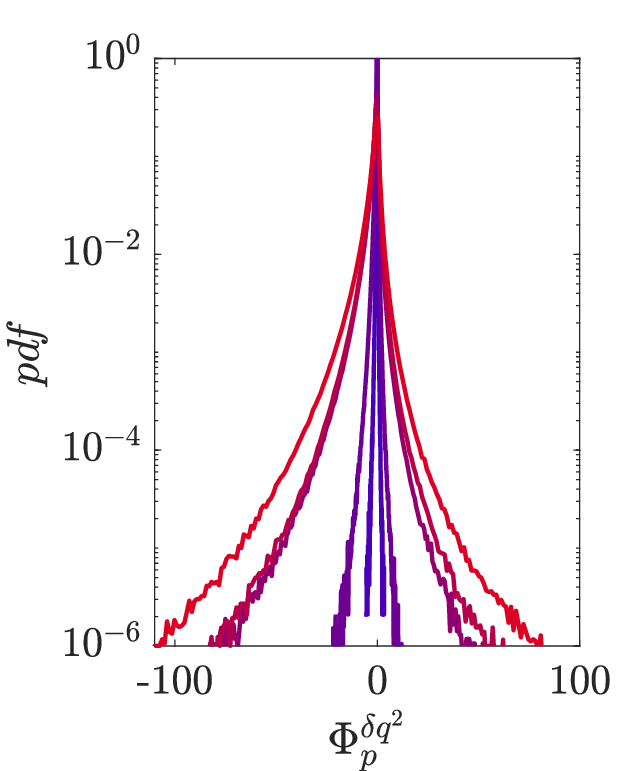}
\includegraphics[width=0.32\textwidth]{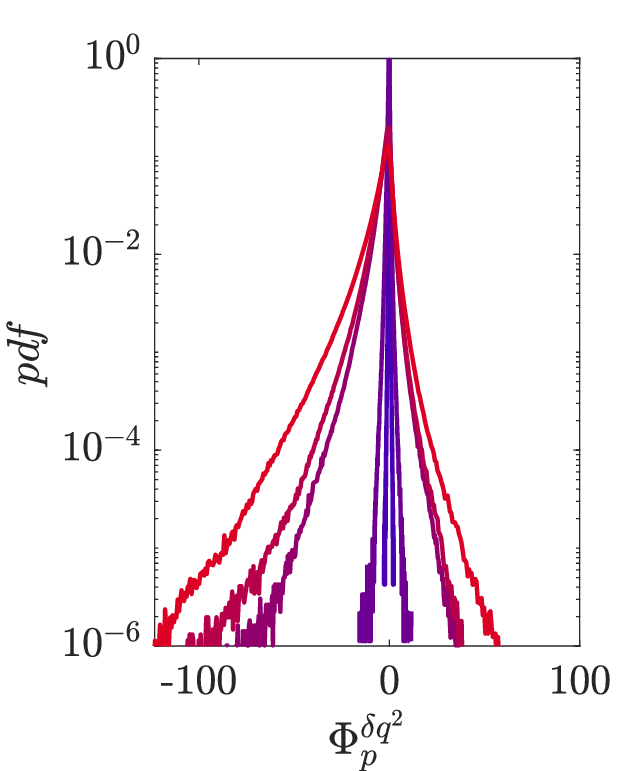}
\includegraphics[width=0.32\textwidth]{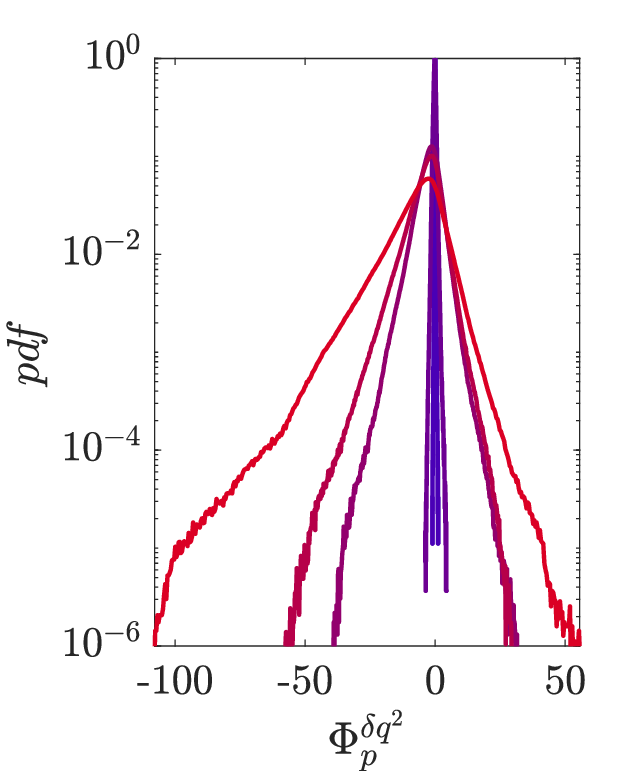}
\includegraphics[width=0.32\textwidth]{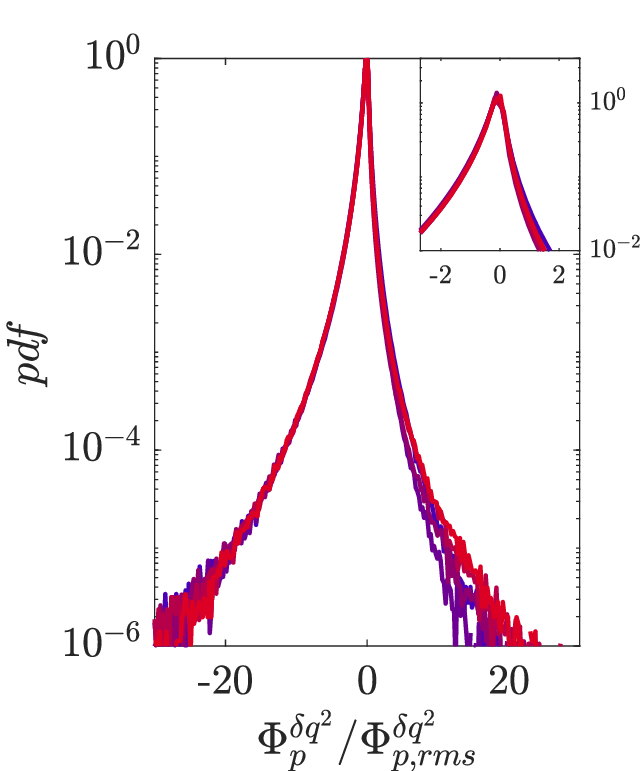}
\includegraphics[width=0.32\textwidth]{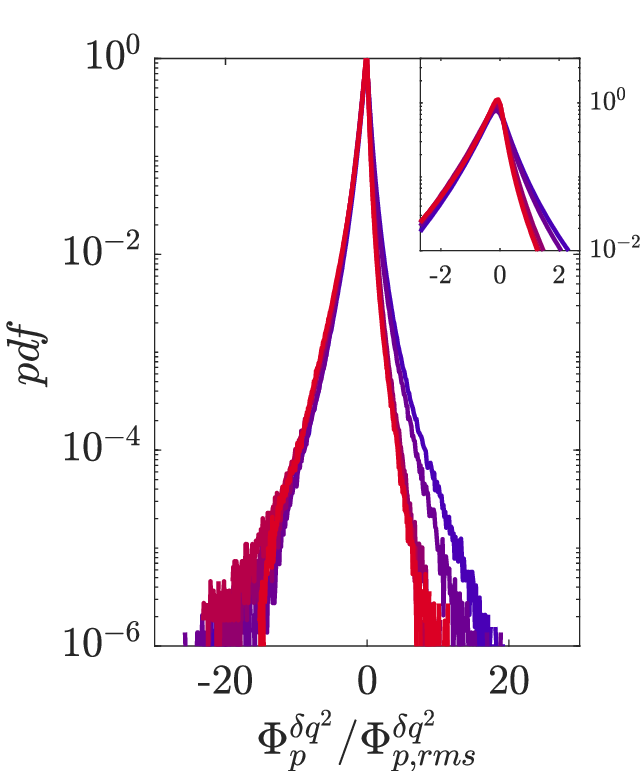}
\includegraphics[width=0.32\textwidth]{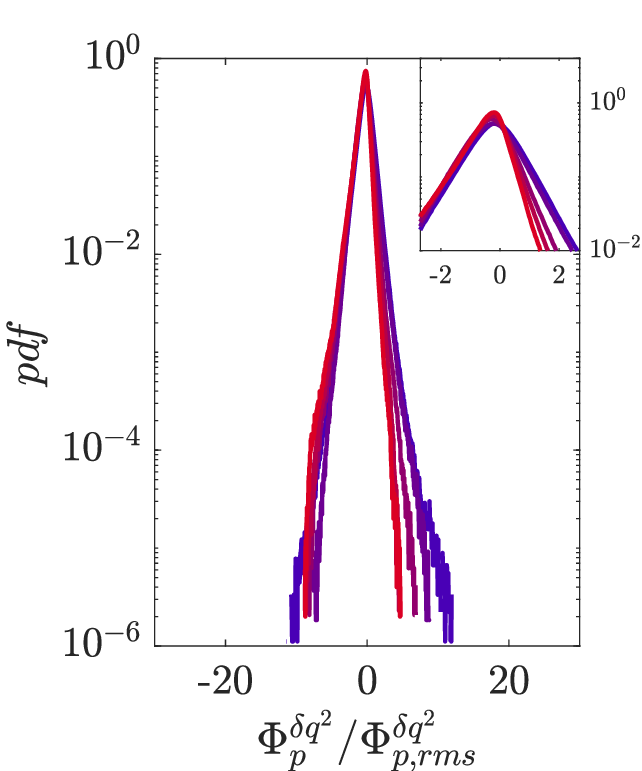}
\includegraphics[trim={0 50 0 0},clip,width=1.0\textwidth]{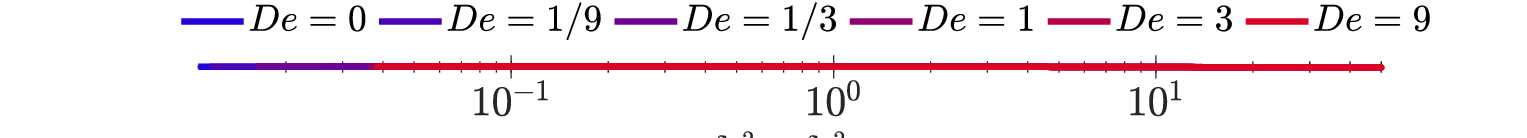}
\caption{Dependence of the distribution of the non Newtonian flux $\Phi_{p}^{\delta q^2}$ on the Deborah number, for (left) $r=0.1$, (centre) $r=0.3$, and (right) $r=0.9$.  Top: distribution of $\Phi_p^{\delta q^2}$. Bottom: distribution of $\Phi_p^{\delta q^2}/\Phi_{p,rms}^{\delta q^2}$. In the bottom panels the inset provide a zoom of the normalised distributions.}
\label{fig:phinn_ener}
\end{figure}

The average picture of the cascade and interscale exchanges in the $r$-space is not representative of the actual physical processes. In the previous section, we have shown the influence of the polymers on the energy cascade in an average sense only, even though it is known that the interscale energy transfer is highly intermittent \citep[see for example][]{piomelli-etal-1991,domaradzki-liu-brachet-1993,cerutti-meneveau-1998}. Moreover, even though energy cascade in HIT is on an average from larger to smaller scales, there exist localised regions in the flow where energy actually cascades from smaller to larger scales, opposite to the average sense. We now investigate how the presence of polymeric additives modifies this picture, and whether this localised inverse energy transfer is also detected in the transfer mechanism driven by the fluid-polymer interaction.

Figures \ref{fig:phinl_ener} and \ref{fig:phinn_ener} show the probability distribution functions of the nonlinear energy fluxes $\Phi_f^{\delta q^2}$ and the polymeric energy fluxes $\Phi_p^{\delta q^2}$ for various values of $De$. We analyse three different separations: $r=0.1$, where $\Phi_p^{\delta q^2}$ dominates for $De > 1$, and $r=0.3$ and $r=0.9$ where $\Phi_f^{\delta q^2}$ dominates for all $De$.
In our convention, a positive flux $\Phi>0$ corresponds to backscatter events, i.e., local transfers of energy from smaller to larger scales, whereas a negative flux $\Phi<0$ indicates forward cascade events, transferring energy from larger to smaller scales. The distributions reveal the highly intermittent and strongly non-Gaussian nature of both $\Phi_f^{\delta q^2}$ and $\Phi_p^{\delta q^2}$, being characterised by heavy tails \citep{ishihara-etal-2009,yasuda-vassilicos-2018}. Both extreme forward transfers and backscatter events occur with significantly higher probability than predicted by a normal distribution.
The distributions are skewed towards negative values, confirming that on average both fluxes transfer energy forward across scales. Quantitatively, for $De=1$ at $r=0.9$ ($r=0.1$) the probability that $\Phi_f^{\delta q^2}$ exceeds its root-mean-square value is $0.0672$ ($0.0274$), while the probability that it falls below minus its root-mean-square value is $0.1640$ ($0.0767$). Similarly, for $\Phi_p^{\delta q^2}$, the probabilities of exceeding and being less than minus its root-mean-suqare value are $0.0323$ ($0.0105$) and $0.2402$ ($0.1545$), respectively.

We now examine how polymers affect the tails of the distributions, focusing on extreme events. Starting with the nonlinear flux $\Phi_f^{\delta q^2}$ shown in figure \ref{fig:phinl_ener}, we observe that across all three scales, the presence of polymers narrows the tails of the distribution. This indicates a suppression of extreme forward and backward energy transfer events along the classical cascade route. This attenuation becomes more pronounced as $De$ increases, consistent with a smoother velocity field and the diversion of energy towards the polymer-driven transfer mechanism.
When examining the fluxes normalised by their root-mean-square values (bottom panels), we find that the distributions at $r=0.3$ and $r=0.9$ collapse reasonably well across all $De$. Some deviations appear at the largest values, likely due to insufficient statistical convergence. This suggests that, in the $\Phi_f^{\delta q^2}$-dominated range, polymers modify the intensity of extreme nonlinear cascade events while having little effect on their intermittency. In other words, in the inertial range of scales where the polymeric contribution is weak, the self-similarity of the classical cascade process appears to be largely maintained. 
On the contrary, the normalised distributions do not overlap at the small scales $r=0.1$.
In the elastic range of scales the nonlinear flux $\Phi_f^{\delta q^2}$ has a small average contribution, and its intermittency increases with $De$.
We now move to the polymeric flux $\Phi_p^{\delta q^2}$, shown in figure \ref{fig:phinn_ener}. As $De$ increases, the distribution tails widen, and the asymmetry becomes more pronounced consistent with the rise in the average value. When examining the normalised fluxes (bottom panels), the behaviour differs from that of $\Phi_f^{\delta q^2}$. Specifically, the distributions collapse reasonably well at the small scale $r=0.1$ (with some deviations at extreme values), but not at the larger scales $r=0.3$ and $r=0.9$. An increase in $De$ appears to alter the space-time intermittency of the interscale energy transfer driven by the polymeric microstructure primarily at larger scales, while self-similarity with respect to $De$ appears to hold at smaller scales within the elastic range of scales. Consistent with the increased asymmetry, at large scales higher $De$ corresponds to narrower positive tails and slightly heavier negative tails.

\begin{figure}
\centering
\includegraphics[width=0.49\textwidth]{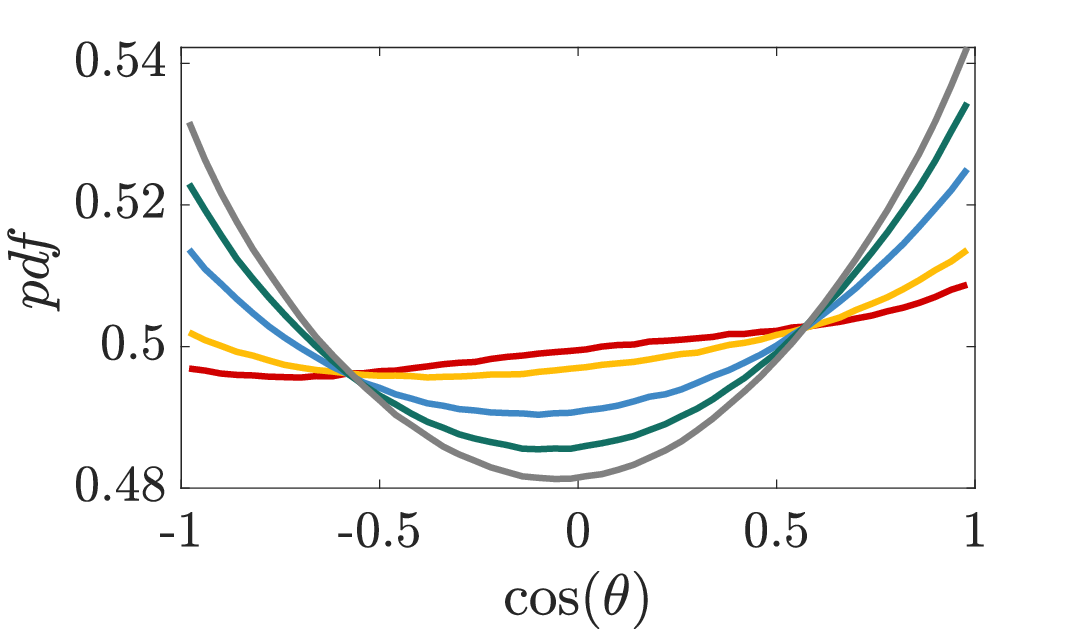}
\includegraphics[width=0.49\textwidth]{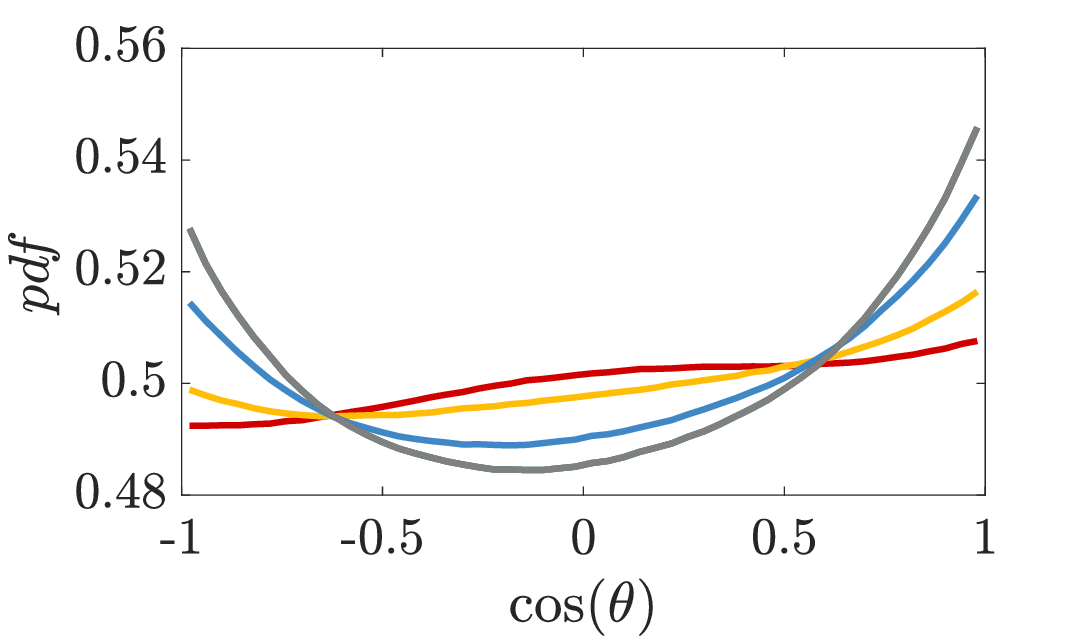}
\includegraphics[width=0.49\textwidth]{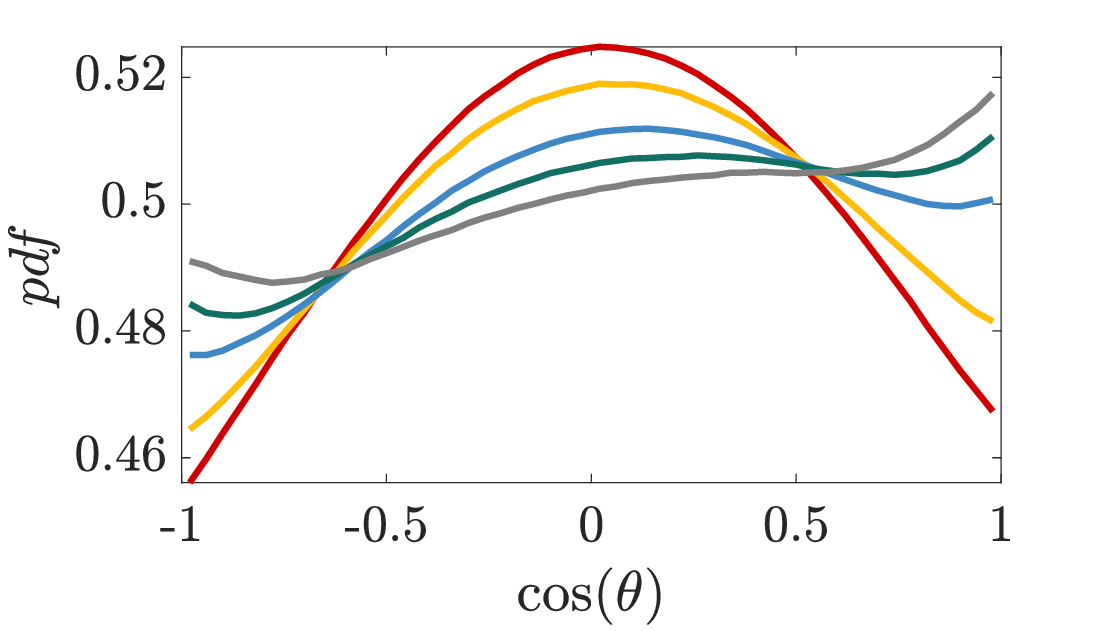}
\includegraphics[width=0.49\textwidth]{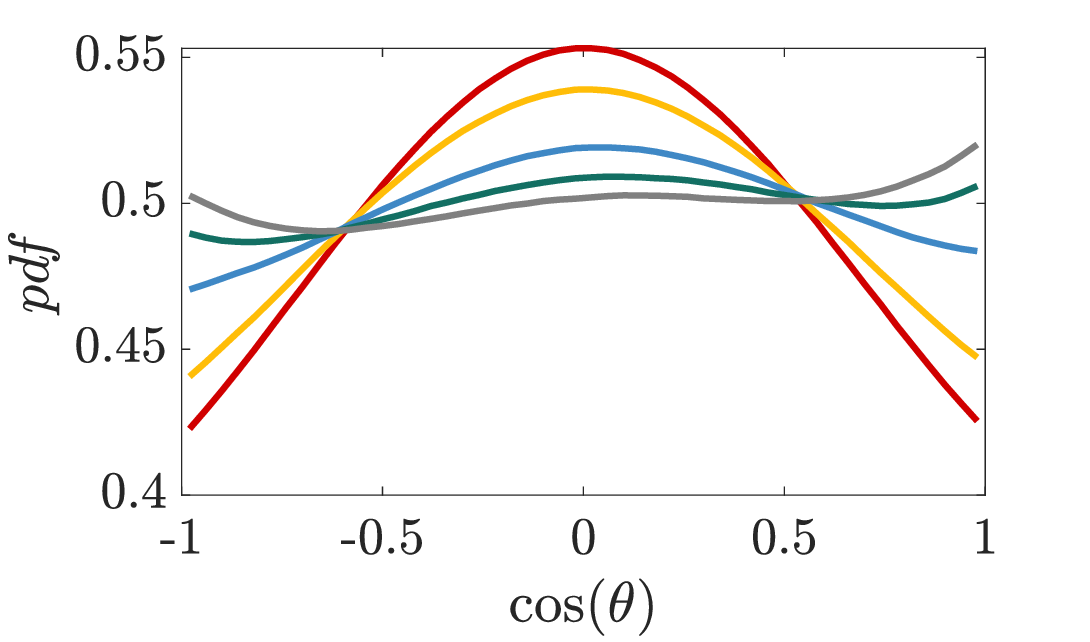}
\includegraphics[trim={0 15 0 0},clip,width=0.8\textwidth]{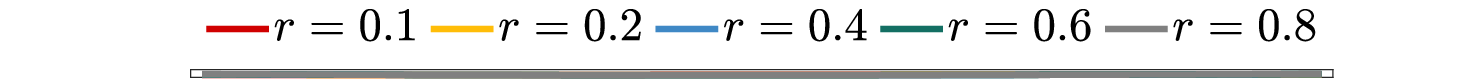}
\caption{Distribution of $\cos(\theta)= \bm{u}^* \cdot ( \delta \bm{\omega} \times \delta \bm{u} )/( ||\bm{u}^*|| || \delta \bm{u} \times \delta \bm{\omega} ||)$ for different scales. In order from top-left to bottom-right, the panels are for $De=0$, $De=1/3$, $De=1$ and $De=9$.}
\label{fig:costeta}
\end{figure}

We now provide a further insight on the influence of the polymers on the inertia-driven energy cascade process by looking at the simple relation introduced by \cite{baj-etal-2022}, that links the nonlinear flux with the Lamb vector $\delta \bm{\omega} \times \delta \bm{u}$. 
Exploiting the Lamb decomposition and manipulating the equations, it is indeed possible to show that 
\begin{equation}
-\frac{\partial}{\partial r_j} \aver{ \delta u_j \delta q^2} = 2 \aver{\bm{u}^* \cdot ( \delta \bm{\omega} \times \delta \bm{u} ) }.
\label{eq:lamb}
\end{equation}
Since $|| \delta \bm{u} \times \delta \bm{\omega}||^2 = || \delta \bm{u} ||^2 || \delta \bm{\omega} ||^2 - || \delta h ||^2$, it is clear that there is an implicit connection between the scale-by-scale helicity $\delta h$ and the interscale energy transfer. Indeed,
a large $\delta h = \delta \bm{\omega} \cdot \delta \bm{u}$ corresponds to a small $\delta \bm{\omega} \times \delta \bm{u}$, and therefore to a weaker local transfer of energy. This agrees with the observation of previous authors which found that the magnitude of the helicity has an impact on the local transfer of energy \citep[see for example][]{pelz-etal-1985,stepanov-etal-2015}. To gain further insights on how polymers influence the interscale energy transfer, we look at the angle $\theta$ between $\bm{u}^*$ and $\delta \bm{\omega} \times \delta \bm{u}$ such that
\begin{equation}
  \cos(\theta) = \frac{ \bm{u}^* \cdot \left( \delta \bm{\omega} \times \delta \bm{u} \right) }
                      { || \bm{u}^* || \ || \delta \bm{\omega} \times \delta \bm{u}|| }.
\end{equation}
Figure \ref{fig:costeta} plots the distributions of $\cos(\theta)$ at different scales $r$ for $0 \le De \le 9$. We start our discussion with the Newtonian case, where the distribution is positively skewed for all $r$. This is consistent with equation \eqref{eq:lamb}, which implies that on average the left-hand side ($|| \bm{u}^* || || \delta \bm{\omega} \times \delta \bm{u} || \cos(\theta)$) is positive. Similar to what found by \cite{baj-etal-2022}, as the separation increases the probability of parallel ($\cos(\theta) = 1$) and antiparallel ($\cos(\theta) = -1$) alignment increases, with a larger probability of the parallel events. For PHIT ($De>0$), the picture changes in a scale-dependent manner. At small $De (\ll1 )$, as expected, the distributions remain similar to Newtonian HIT. For large $De (\gtrsim 1)$, instead, at small scales $r$ where the polymeric flux $\Phi_p^{\delta q^2}$ dominates the Lamb vector has a larger likelihood of being normal to the local advecting velocity, with the distribution still being right skewed. This agrees with the fact that in this range of scales, only a small fraction of energy is transferred to smaller scales by the inertia-driven transfer route. 
The weakening of the inertia-driven interscale-energy transfer within the elastic range of scales is thus accompanied by an increase of the events where $\bm{u}^*(r)$ and $\delta \bm{\omega}(r) \times \delta \bm{u}(r)$ are perpendicular. 
When considering large $r$, where the influence of the polymers is weaker and $\Phi_f^{\delta q^2}$ dominates, the distribution of $\cos(\theta)$ flattens and progressively recovers a distribution similar to that in Newtonian HIT.

\subsection{Scale enstrophy $\aver{\delta \omega^2}$}
\label{sec:enstrophy}

\subsubsection{Scale-by-scale budget}

\begin{figure}
\centering
\includegraphics[width=0.49\textwidth]{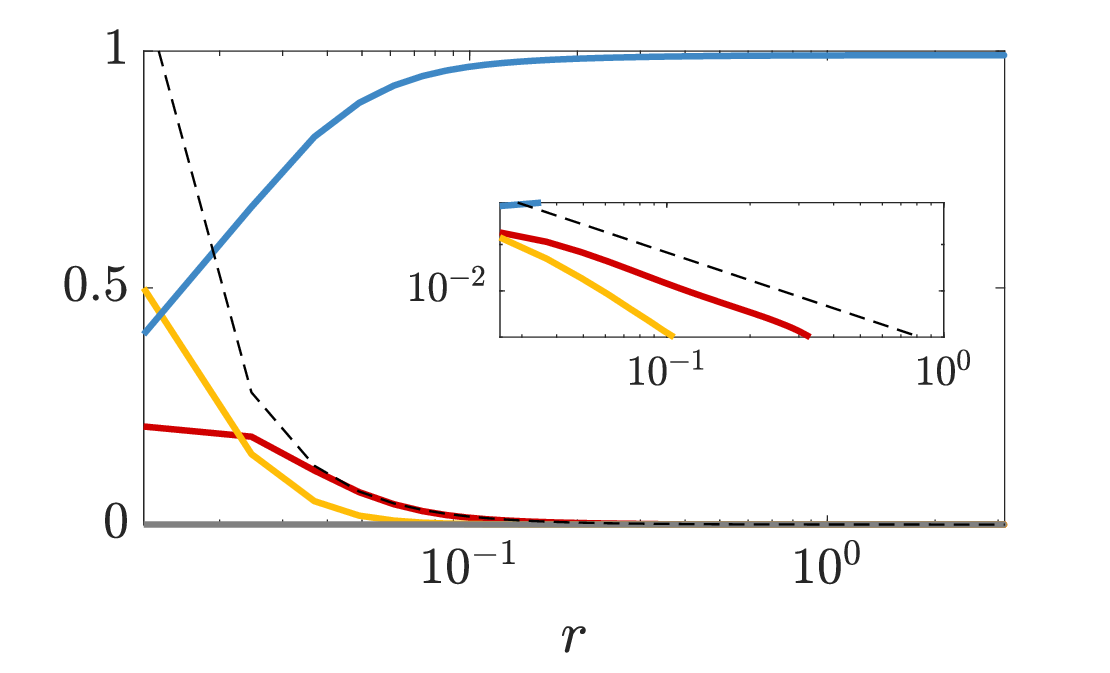}
\includegraphics[width=0.49\textwidth]{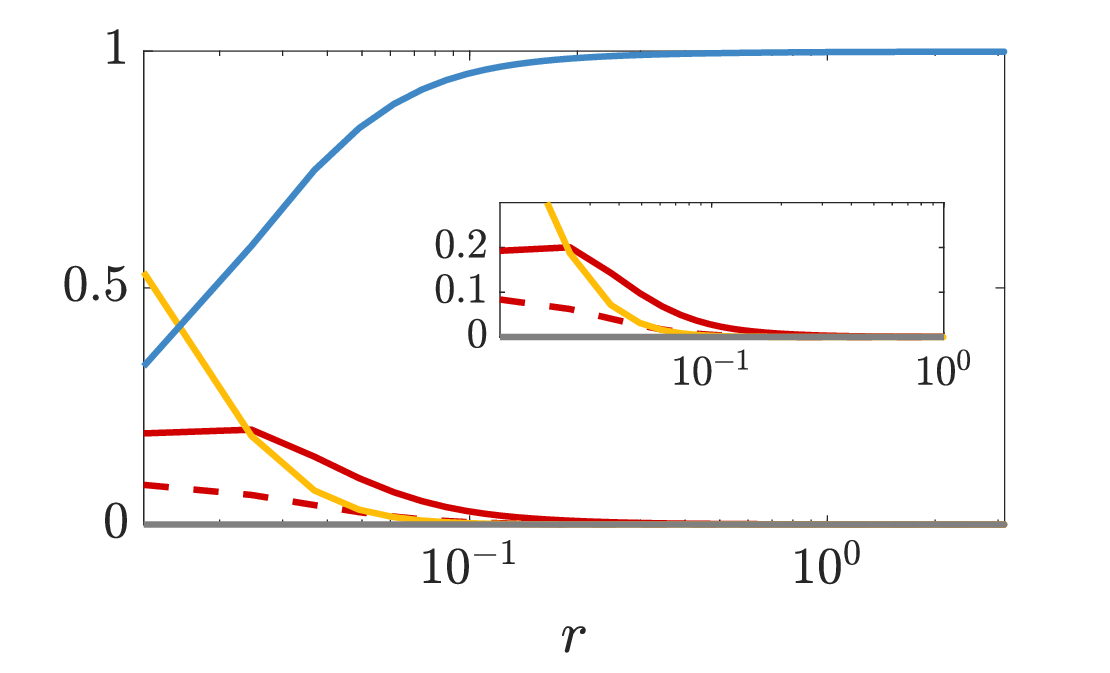}
\includegraphics[width=0.49\textwidth]{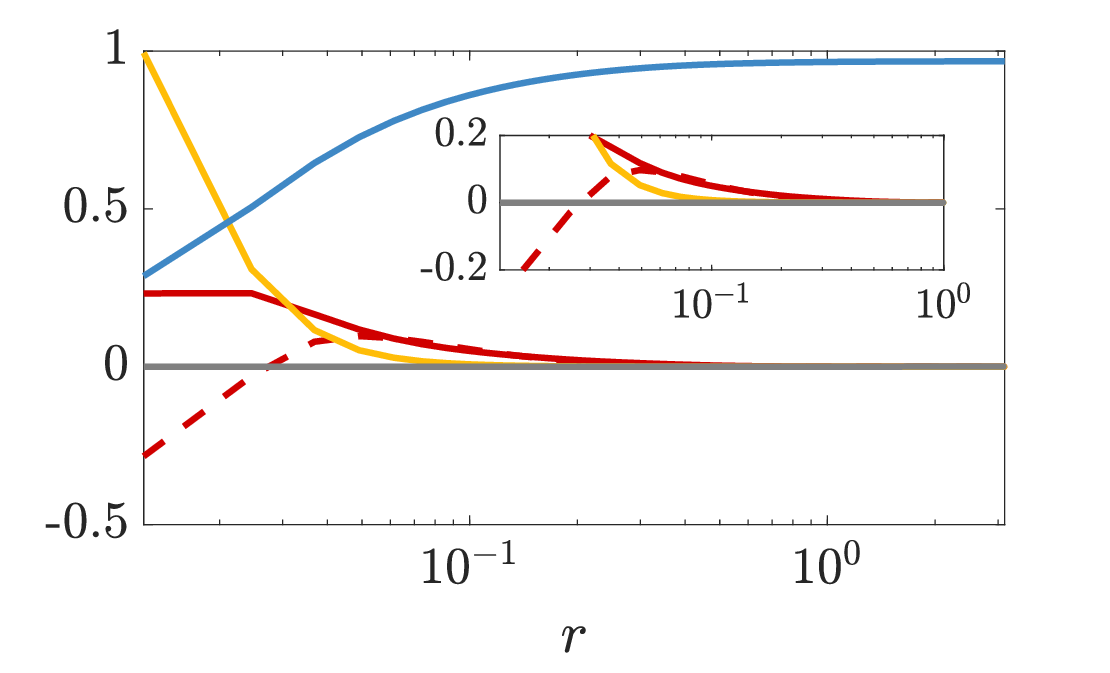}
\includegraphics[width=0.49\textwidth]{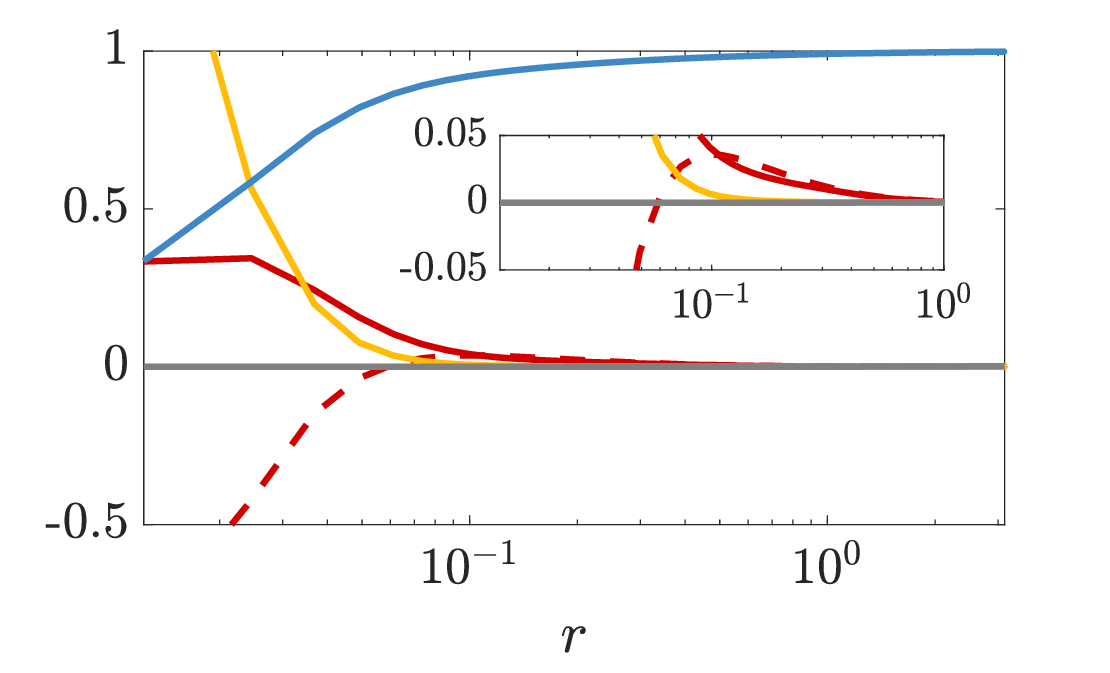}
\includegraphics[trim={0 30 0 0},clip,width=0.8\textwidth]{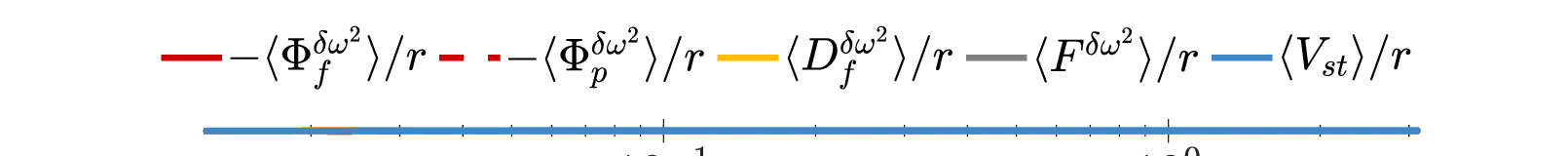}
\caption{All contributions in budget equation for $\aver{\delta \omega^2}$. In order from left to right and from top to bottom, the panels are for $De=0$, $De=1/3$, $De=1$ and $De=9$. For $De=0$, the inset shows the budget with the $y$ axis is logarithmic scale. The black dashed line denotes $r^{-2}$.}
\label{fig:enstr-bud}
\end{figure}

This section considers the enstrophy structure function. Unlike energy and helicity, enstrophy is not an inviscid invariant of the Navier--Stokes equations. The budget equation for $\aver{\delta \omega^2}$, indeed, features a source term $\langle V_{st} \rangle$, that quantifies the amount of enstrophy that is created up to the scale $r$ by means of vortex stretching-like processes. This means that even in Newtonian HIT, the notion of cascade used for energy transfer does not apply here, and most of the enstrophy is generated directly at the small scales. However, as stated by several authors \citep[see for example][]{davidson-etal-2008}, vortex stretching does occur in the inertial range of HIT and hence a net transfer of enstrophy to smaller scales is expected at these $r$. 

We plot the terms in the budget equation for $\aver{\delta \omega^2}$ in figure \ref{fig:enstr-bud}.
We start looking at the Newtonian case ($De=0$). At intermediate and large scales, vortex stretching substantially matches the dissipation of enstrophy, i.e., $\langle V_{st} \rangle (r) \approx (4/3) \langle \varepsilon_f^{\delta \omega^2} \rangle r$, resulting in a $\langle V_{st} \rangle (r) \sim r^{1}$ scaling. 
In this intermediate range of scales the forcing $\langle F^{\delta \omega^2} \rangle$ and dissipation $\langle D_f^{\delta \omega^2} \rangle$ contributions are negligible, and the flux is subdominant $\langle \Phi_f^{\delta \omega^2} \rangle  \ll \langle V_{st}^{\delta \omega^2} \rangle$. 

An insightful interpretation of the linear scaling $\langle V_{st}^{\delta \omega^2} \rangle \sim r^1$ arises from the observation that vortex stretching is predominantly contributed by the smallest scales. This can be seen by the dimensional estimate of the scale-by-scale vorticity $\omega_r$
\begin{align}
	\omega_r \sim \frac{u_r}{r} \sim \langle \varepsilon_f^{\delta q^2} \rangle^{1/3} r^{1/3} r^{-1} =  \langle \varepsilon_f^{\delta q^2} \rangle^{1/3}  r^{-2/3},
\end{align}
which becomes maximum as $r \to \eta$, so that $\omega_\eta \sim \langle \varepsilon_f^{\delta q^2} \rangle^{1/3}  \eta^{-2/3}$; here we have used the $u_r \sim \langle \varepsilon_f^{\delta q^2} \rangle^{1/3} r^{1/3}$ Kolmogorov prediction. Now, to estimate the cumulative vortex stretching $\aver{V_{st}}$ up to scale $r$,
\begin{align}
	\aver{V_{st}}(r) &=  \frac{1}{\mathcal{S}(r)} \int_{\mathcal{V}(r)} \ \underbrace{ 2 \left( \aver{ \delta \omega_i \omega_j^* \delta \left( \frac{\partial u_i}{\partial x_j} \right) } 	+  \aver{ \delta \omega_i \delta \omega_j \left( \frac{\partial u_i}{\partial x_j} \right)^* } \right) }_{v_{st,r}} \text{d}\Omega ,
	\label{eq:Vst}
\end{align}
we begin by estimating the scale-by-scale vortex stretching $v_{st,r}$, as
\begin{align}
	v_{st,r} \sim \omega_r \omega_r  \frac{u_\eta}{\eta}.
\end{align}
We now use the notion that $v_{st,r}$ gets a maximal contribution from the smallest scales \citep{davidson-etal-2008} and write
\begin{align}
	v_{st,\eta} \sim \omega^2_{\eta} \ \frac{u_\eta}{\eta} \sim \langle \varepsilon_f^{\delta q^2} \rangle^{2/3} \eta^{-4/3} 
	                                                            \langle \varepsilon_f^{\delta q^2} \rangle^{1/3} \eta^{1/3} \eta^{-1} =
	                                                            \langle \varepsilon_f^{\delta q^2} \rangle \eta^{-2} 
	                                                       \sim \langle \varepsilon_f^{\delta \omega^2} \rangle,
\end{align}
where we have used the $\langle \varepsilon_f^{\delta \omega^2} \rangle \sim \langle \varepsilon_f^{\delta q^2} \rangle\eta^{-2}$ dimensional estimate. By neglecting the contribution to vortex stretching from larger scales, we thus estimate the cumulative vortex stretching up to a scale $r$ as $\langle V_{st} \rangle (r) \approx 1/\mathcal{S}(r)\int_{\mathcal{V}(r)} v^{st}_\eta \text{d}\Omega  \sim r^{-2}r^{3} \langle \varepsilon_f^{\delta \omega^2} \rangle \sim \langle \varepsilon_f^{\delta \omega^2} \rangle r$, which is consistent with the data plotted in figure~\ref{fig:enstr-bud}. 

Moreover, figure~\ref{fig:enstr-bud} shows that the nonlinear flux exhibits a $\langle \Phi_f^{\delta \omega^2} \rangle(r) \sim r^{-1}$ scaling behaviour at the intermediate scales, with its magnitude increasing as $r$ decreases. Since enstrophy is predominantly generated at the smallest scales, the inter-scale transfer rate intensifies with decreasing $r$, reaching a maximum near the transition between the inertial and dissipative ranges. Following the dimensional analysis approach of \cite{davidson-etal-2008}, this scaling can be interpreted by analogy with the vortex stretching mechanism. Consider, for example, a vortex tube of length $\ell$ and cross-sectional area $A \sim r^2$. When subjected to external shear, the tube is stretched, reducing its cross-section and thereby increasing its local vorticity. This illustrates how enstrophy at a larger scale is transferred to more intense enstrophy at smaller scales. Based on this picture, the enstrophy transfer in the inertial range is driven by eddies of size $r$, with characteristic velocity $\delta u(r) \sim u_r \sim \langle \varepsilon_f^{\delta q^2} \rangle^{1/3} r^{1/3}$ and vorticity $\delta \omega(r) \sim \omega_r \sim \langle \varepsilon_f^{\delta q^2} \rangle^{1/3} r^{-2/3}$. These scalings naturally lead to the observed behaviour:
\begin{align}
  \aver{\Phi_f^{\delta \omega^2}}(r)  \sim \frac{1}{\mathcal{S}(r)} \int_{\mathcal{S}(r)}  \ u_r \omega^2_r \text{d} \Sigma
                                      \sim r^{-2} r^2 \ \aver{\varepsilon_f^{\delta q^2}} r^{1/3} r^{-4/3}
                                      = \aver{\varepsilon_f^{\delta q^2}} r^{-1}.
\end{align}

\begin{figure}
\centering
\includegraphics[width=0.9\textwidth]{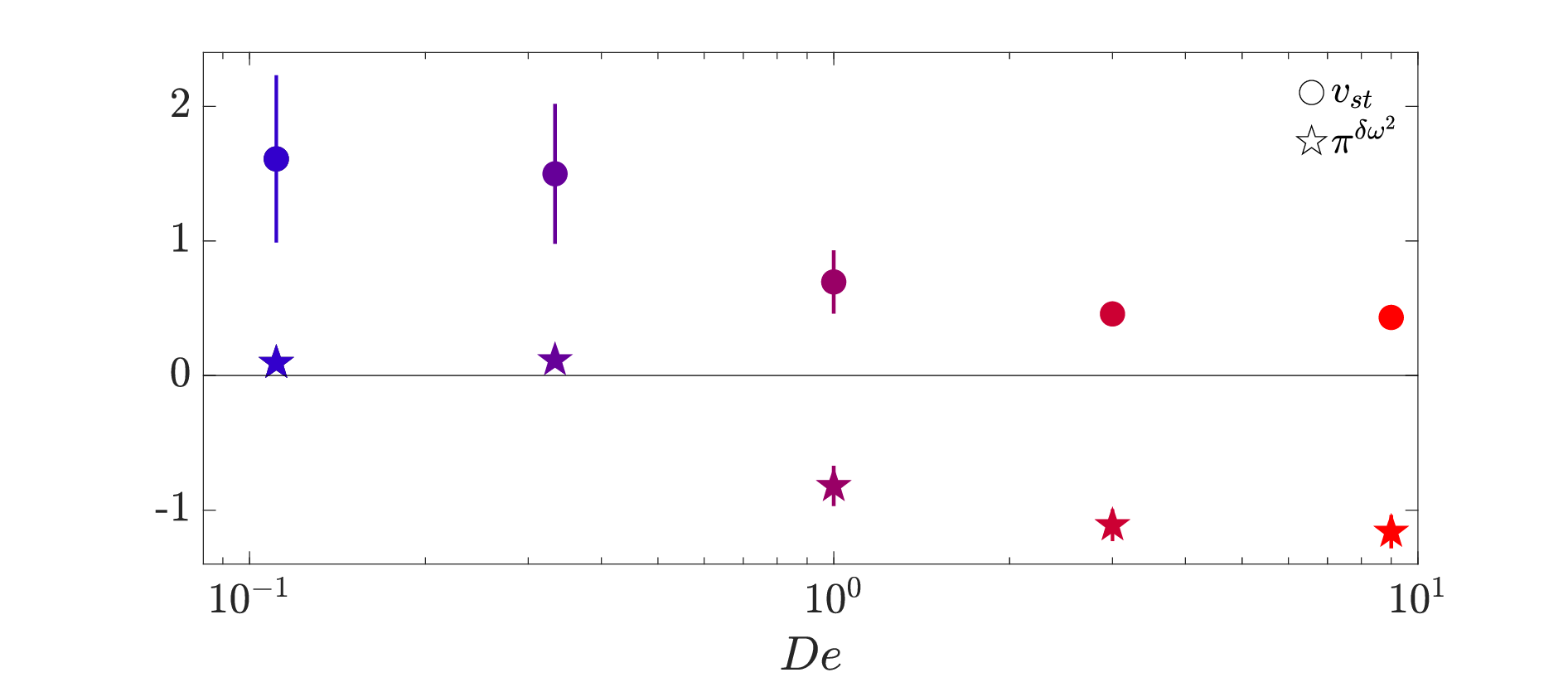}
\caption{Dependence of $\langle v_{st} \rangle$ and $\langle \pi^{\delta \omega^2} \rangle$ on $De$. All quantities are made dimensionless with $\langle \varepsilon_f^{\delta \omega^2} \rangle$. Here $v_{st} = \omega_i \omega_j S_{ij}$ is the vortex-stretching production term that appears in the budget equation for $\langle \omega^2 \rangle$. The bars show the standard deviation computed from six different snapshots.}
\label{fig:VstEpsp}
\end{figure}

We now examine the influence of polymers on the budget of $\langle \delta \omega^2 \rangle$. Their presence introduces two additional terms into the enstrophy budget: a polymeric source/sink term, $\pi^{\delta \omega^2}$, and the associated flux term, $\Phi_p^{\delta \omega^2}$. In other words, analogous to the case of $\langle \delta q^2 \rangle$, polymers contribute with a source/sink mechanism and an alternative pathway for interscale enstrophy transfer. Recall that for enstrophy we have not split $\Phi_p^{\delta \omega^2}$ into the inertial and dissipative contributions.

For small Deborah numbers ($De < 1$), the polymeric flux $\langle \Phi_{p}^{\delta \omega^2} \rangle$ and source term $\langle \pi^{\delta \omega^2} \rangle$ remain weak, and the classical enstrophy production $\langle V_{st} \rangle$ and transfer $\langle \Phi_f^{\delta \omega^2} \rangle$ are only marginally affected by the polymers; see the top right panel of figure~\ref{fig:enstr-bud}. In this regime $\langle \pi^{\delta \omega^2} \rangle < 0$, indicating that polymers act, on average, as a sink of enstrophy for the fluid phase (see also figure~\ref{fig:VstEpsp}). Moreover, $\langle \Phi_p^{\delta \omega^2} \rangle<0$ at all scales, implying that enstrophy is transferred from larger to smaller scales even along the polymer-induced interscale transfer.

At large Deborah numbers, the dynamics of the enstrophy transfer undergoes a marked transformation. As $De$ increases, the average fluid contribution $\langle \Phi_f^{\delta \omega^2} \rangle $ remains nearly unchanged, indicating that polymers exert limited influence on the classical inertia-driven enstrophy interscale transfer. However, once $De \ge 1$, the polymeric contribution $\langle \Phi_p^{\delta \omega^2} \rangle $ becomes comparable in magnitude to $\langle \Phi_f^{\delta \omega^2} \rangle $, and displays a different behaviour across scales. At intermediate scales, both fluid- and polymer-driven transfer processes contribute to the forward enstrophy transfer, from larger to smaller scales, though both remain subdominant compared to the vortex-stretching term.
At sufficiently small scales, the scenario changes. For $De \ge 1$, the polymeric flux $\langle \Phi_p^{\delta \omega^2} \rangle $ becomes negative as $r \rightarrow 0$, suggesting that the fluid–polymer interaction facilitates an inverse enstrophy transfer from smaller to larger scales. Notably, this shift is accompanied by a sign reversal in $\langle \pi^{\delta \omega^2} \rangle$ (see \S\S\ref{sec:strfun} and \ref{sec:loc-str}). The data suggest that, at these values of $De$, polymers generate enstrophy at the smallest scales, which is then transferred towards slightly larger scales before being ultimately dissipated by viscosity.
This behaviour stands in contrast to that observed for $\langle \delta q^2 \rangle $, where the nonlinear flux dominates the energy cascade at large scales and the polymeric flux becomes dominant at small scales, but both consistently drive a forward energy transfer across all $r$.

A further key effect at high $De$ is the strong modulation of the vortex-stretching term. Figure~\ref{fig:VstEpsp} illustrates this by showing the dependence of $\langle v_{st} \rangle = \langle \omega_i \omega_j S_{ij} \rangle $ on $De$, where $S_{ij} = ( \partial u_i/\partial x_j + \partial u_j/\partial x_i)/2$ is the rate-of-strain tensor. The quantity $v_{st}$ represents the one-point vortex-stretching production of enstrophy $\omega^2$ \citep{tsinober-2001}, and satisfies $\langle V_{st}^{\delta \omega^2} \rangle \rightarrow 2 \langle v_{st} \rangle $ as $r \rightarrow \infty$.
As shown in figure~\ref{fig:VstEpsp}, $\langle v_{st} \rangle$ decreases significantly with increasing $De$, in agreement with earlier findings by \citet{liberzon-etal-2005, liberzon-etal-2006, cai-etal-2010, rehman-etal-2022}. Remarkably, this attenuation of vortex stretching is closely accompanied by the striking sign reversal in $\langle \pi^{\delta \omega^2} \rangle$, which becomes negative and approximately balances the viscous dissipation $\langle \pi^{\delta \omega^2} \rangle \approx \langle \varepsilon_f^{\delta \omega^2} \rangle $ for $De \ge 1$. In other words, at high elasticity, polymers not only suppress classical enstrophy production via vortex stretching, but also introduce a different dynamical regime, in which polymer stresses act as a net source of enstrophy.

As discussed in \S\ref{sec:loc-str}, the suppression of vortex stretching and the sign reversal in $\langle \pi^{\delta \omega^2} \rangle$ with increasing $De$ are closely linked to changes in the local flow topology and alignments, a transformation that redefines enstrophy dynamics in polymer-laden turbulence.

\subsection{Scale helicity}
\label{sec:helicity}

\subsubsection{Scale-by-scale budget}

\begin{figure}
\centering
\includegraphics[width=0.49\textwidth]{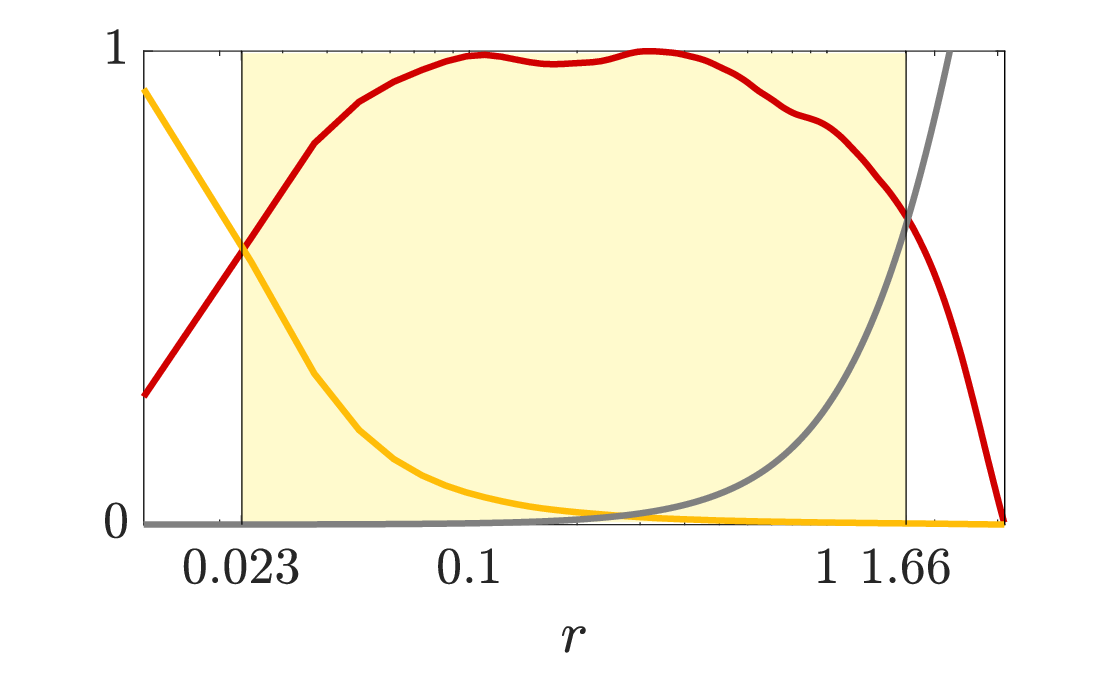}
\includegraphics[width=0.49\textwidth]{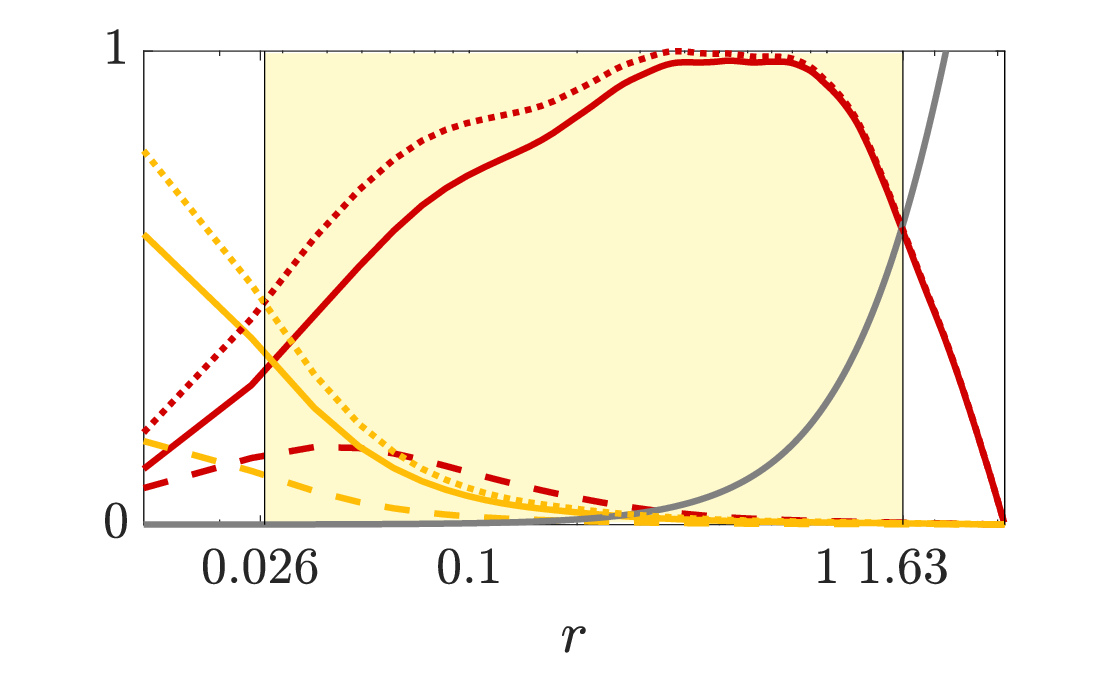}
\includegraphics[width=0.49\textwidth]{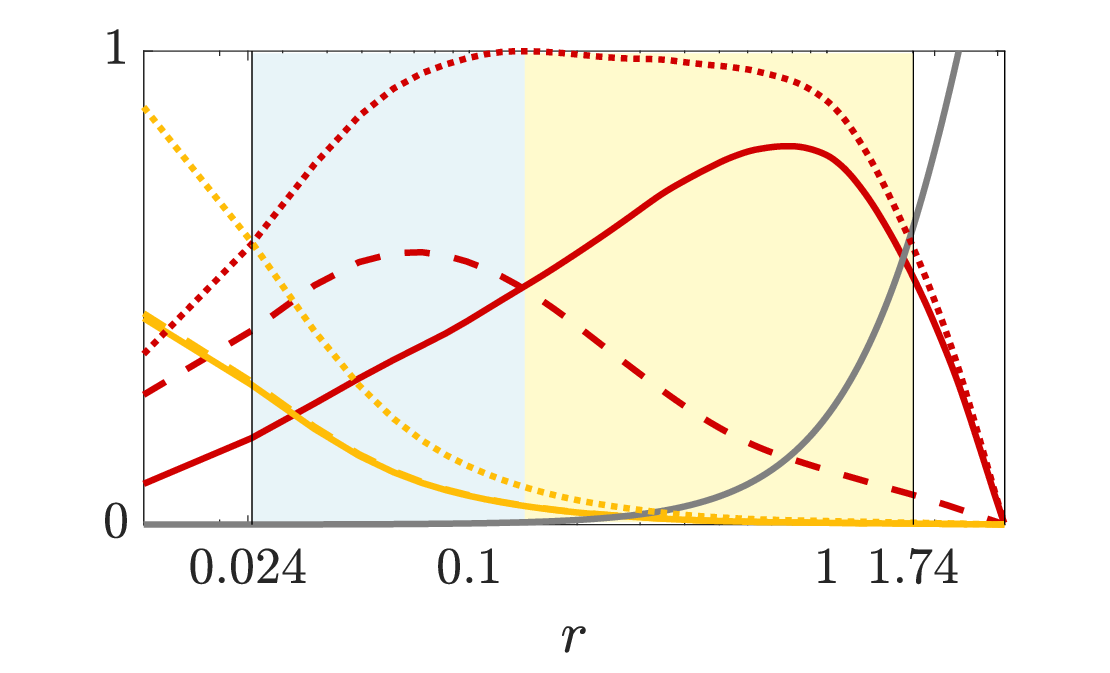}
\includegraphics[width=0.49\textwidth]{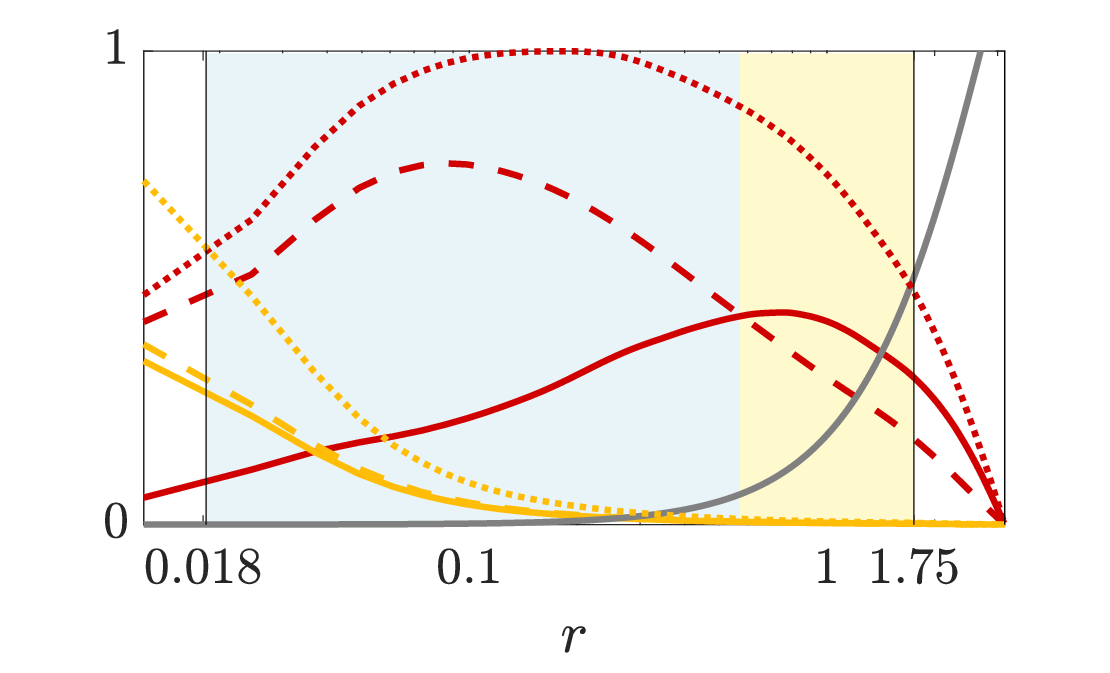}
\includegraphics[trim={0 15 0 10},clip,width=0.7\textwidth]{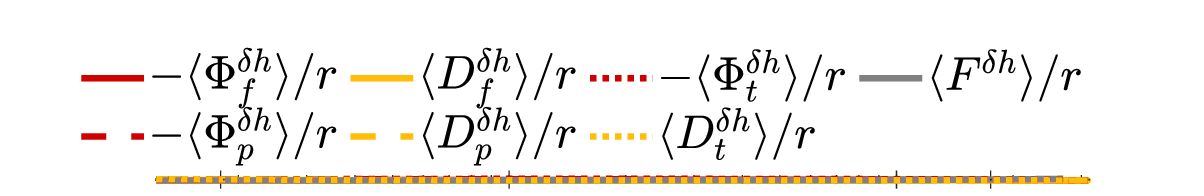}
\caption{All contributions in budget equation for $\aver{\delta h}$. In order from top-left to bottom-right, the panels are for $De=0$, $De=1/3$, $De=1$ and $De=9$. The yellow and blue shaded regions identify the inertial and elastic ranges of scales. Here $\Phi_t = \Phi_f + \Phi_p$ and $D_t = D_f + D_p$.}
\label{fig:heli-bud}
\end{figure}
We now move to the helicity structure function $\delta h$. In the present set up, turbulence is sustained with the ABC forcing that injects both energy and helicity at the largest scales. As such, like energy, helicity is expected to cascade from large to small scales in the inertial range of scales, and to eventually vanish at the small scales where the mirror symmetry $\bm{u} \cdot \bm{\omega} = 0$ is restored. 

Our results, presented in figure~\ref{fig:heli-bud}, provide clear evidence of a dual direct cascade of energy and helicity across all values of $De$, consistent with the theoretical predictions and numerical observations by \cite{kraichnan-1973}, \cite{polifke-shtilman-1989}, and \cite{borue-orszag-1997}.

In the purely Newtonian case, helicity is injected at the largest scales ($r > r_{IE}$) at a rate that matches its dissipation, i.e.\ $ \langle F^{\delta h} \rangle = (4/3) \langle \varepsilon_f^{\delta h} \rangle r$. It is then transferred from large to small scales within the inertial range ($r_{DI} < r < r_{IE}$), where $\langle \Phi_f^{\delta h} \rangle \sim - (4/3) \langle \varepsilon_f^{\delta h} \rangle r$, and is ultimately dissipated by viscous effects at the smallest scales.

As with the energy $\langle \delta q^2 \rangle$, polymers introduce additional helicity transfer and dissipation mechanisms, whose significance increases with $De$. Specifically, the helicity cascade is governed predominantly by the nonlinear flux $\langle \Phi_f^{\delta h} \rangle$ in the range $r_p^* \lessapprox r < r_{IE}$, and by the polymeric flux $\langle \Phi_p^{\delta h} \rangle$ in the smaller-scale range $r_{DI} < r \lessapprox r_p^*$. As $De$ increases and the polymers stretch more, the inertia-driven helicity cascade $\langle \Phi_f^{\delta h} \rangle$ is progressively suppressed, mirroring the behaviour observed for $\langle \Phi_f^{\delta q^2} \rangle$.

The emergence of two distinct cascade mechanisms dominating in separate scale ranges is consistent with the multiscaling behaviour of $\langle \delta h \rangle$ reported in figure~\ref{fig:duidui}.

\begin{figure}
\centering
\includegraphics[width=0.49\textwidth]{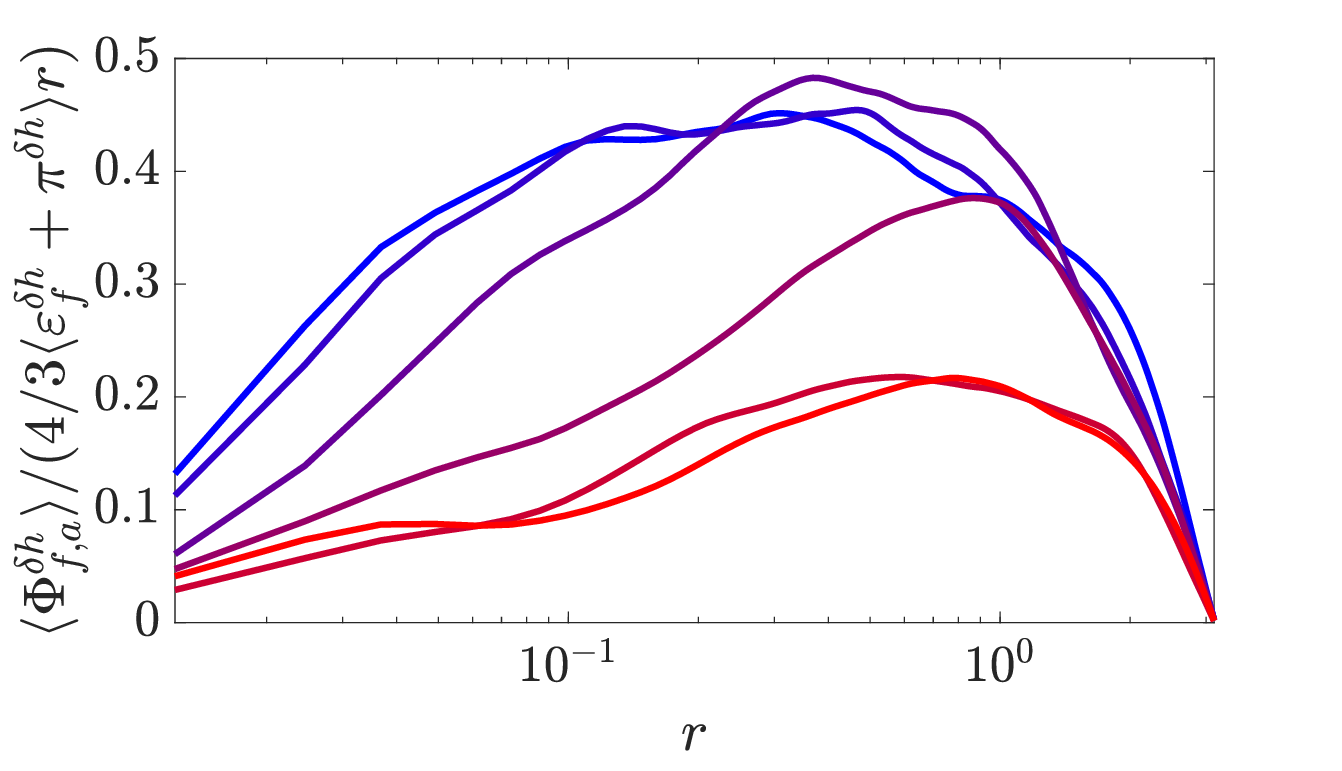}
\includegraphics[width=0.49\textwidth]{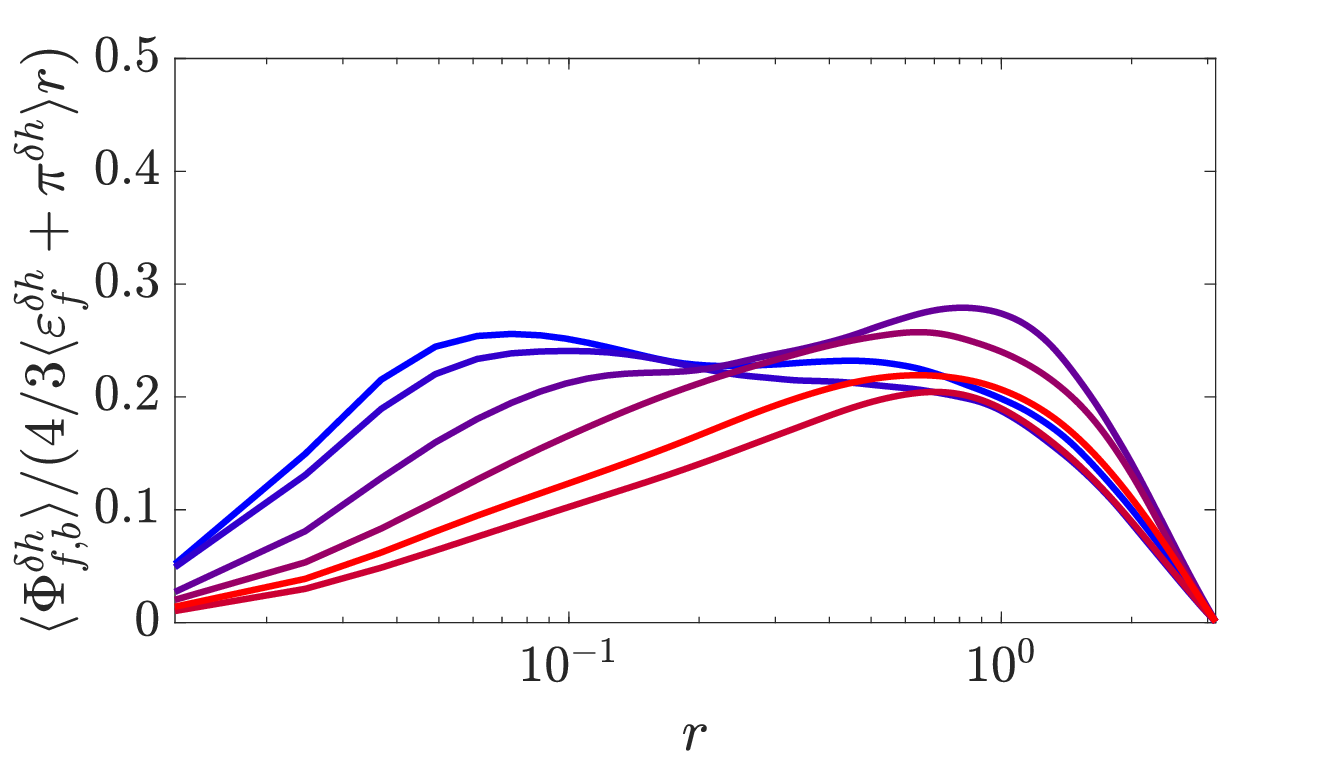}
\includegraphics[trim={0 50 0 0},clip,width=1.0\textwidth]{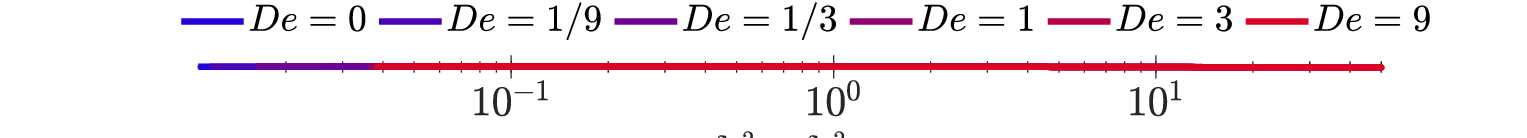}
\caption{Dependence of $\langle \Phi_{f,a}^{\delta h} \rangle$ (left) and $\langle \Phi_{f,b}^{\delta h} \rangle$ (right) on $r$ for different $De$.}
\label{fig:phinlab}
\end{figure}
Figure~\ref{fig:phinlab} offers a more detailed view of how polymer additives influence the nonlinear helicity flux $\langle \Phi_{f}^{\delta h} \rangle$. As discussed in \S\ref{sec:formulation}, this flux comprises two distinct contributions. The first is the classical convective term, $\langle \Phi_{f,a}^{\delta h} \rangle$, associated with the correlation $\langle \delta h  \delta u_j \rangle$. The second, $\langle \Phi_{f,b}^{\delta h} \rangle$, captures a transfer mechanism involving the correlation $\langle \delta \omega_j  \delta q^2 \rangle$.
In the Newtonian case, figure~\ref{fig:phinlab} shows that $\langle \Phi_{f,a}^{\delta h} \rangle$ clearly dominates, being nearly twice as large as $\langle \Phi_{f,b}^{\delta h} \rangle$ across all scales. This indicates that, in Newtonian HIT, helicity is primarily transferred through the convective mechanism. As $De$ increases, however, both components of the flux are progressively suppressed in the elastic range of scales, reflecting the polymers' ability to attenuate both transfer pathways while redirecting a portion of the helicity flux through the polymeric channel.
Notably, polymers have a stronger damping effect on $\langle \Phi_{f,a}^{\delta h} \rangle$ than on $\langle \Phi_{f,b}^{\delta h} \rangle$. As a result, at $De = 9$, the two contributions become comparable on average, with $\langle \Phi_{f,a}^{\delta h} \rangle \approx \langle \Phi_{f,b}^{\delta h} \rangle$.

\subsubsection{Relative helicity}

\begin{figure}
\centering
\includegraphics[width=0.49\textwidth]{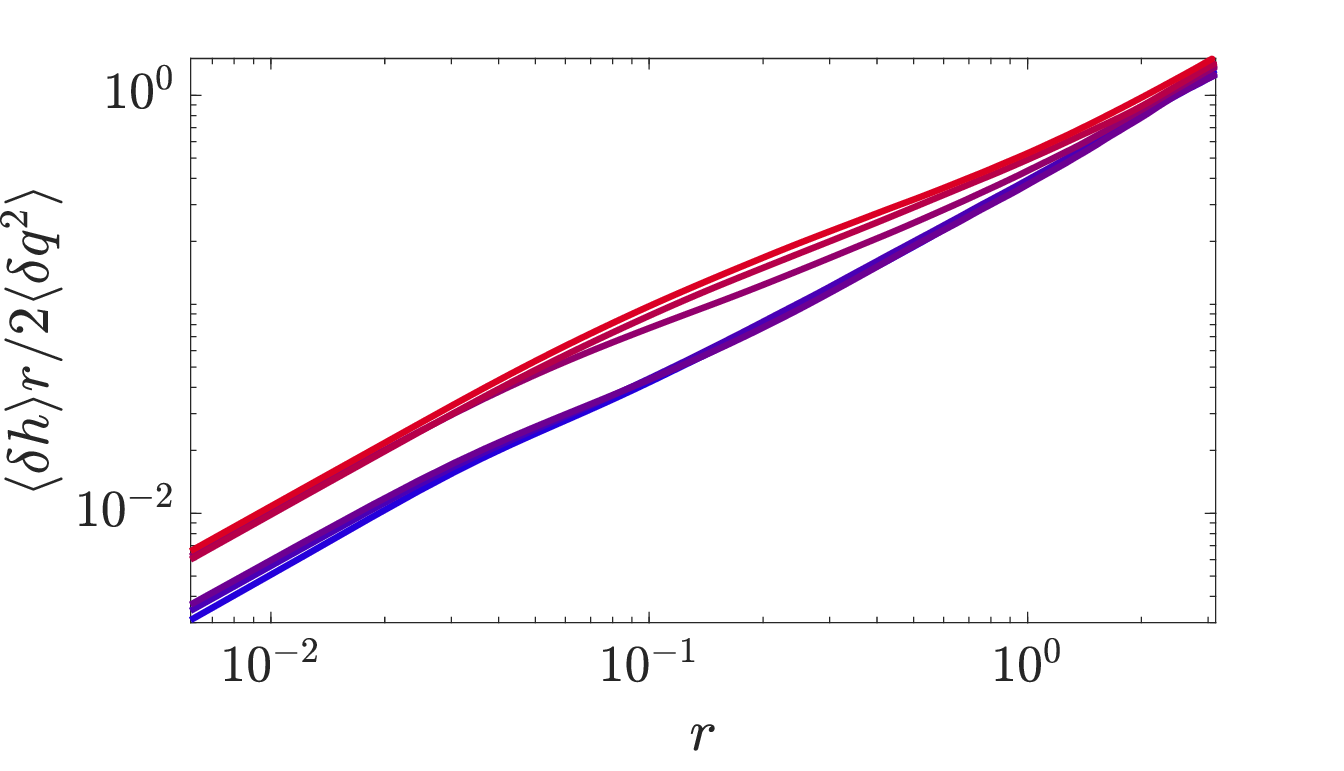}
\includegraphics[width=0.49\textwidth]{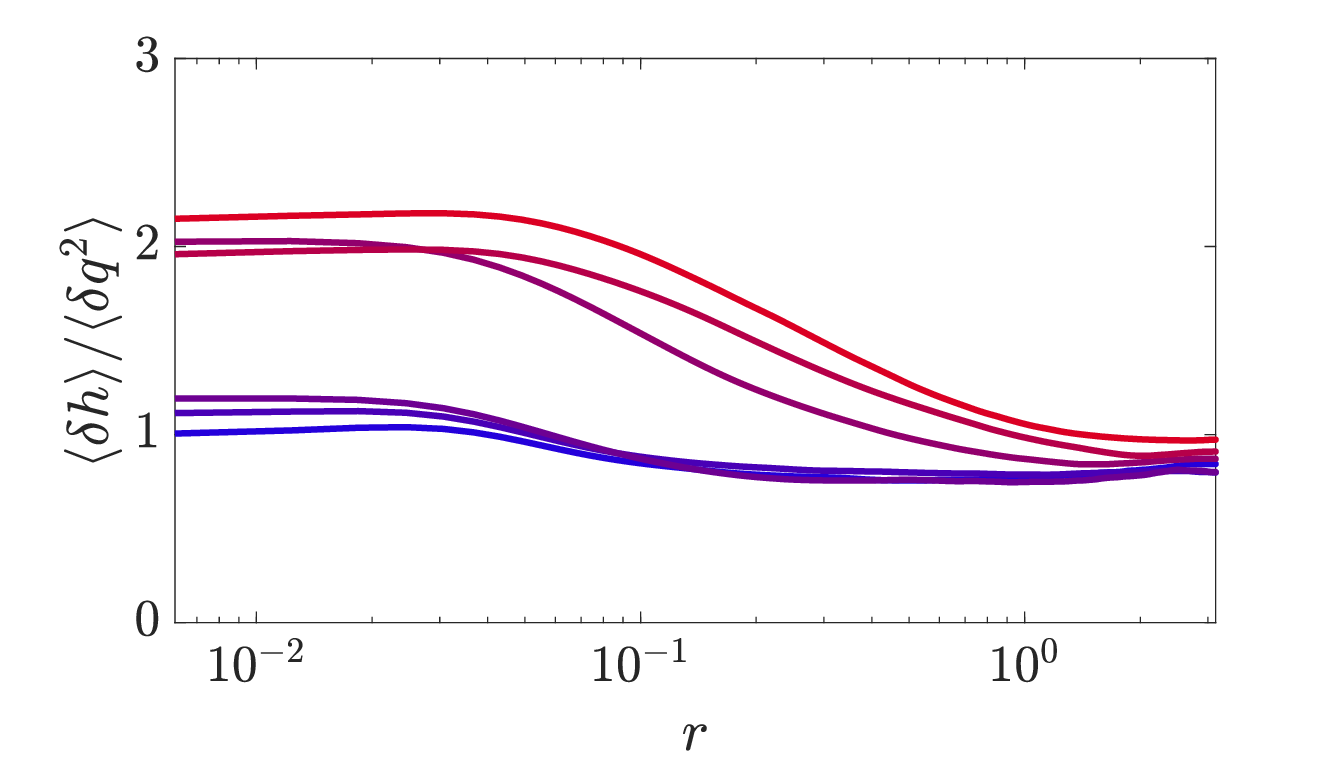}
\includegraphics[trim={0 50 0 0},clip,width=1.0\textwidth]{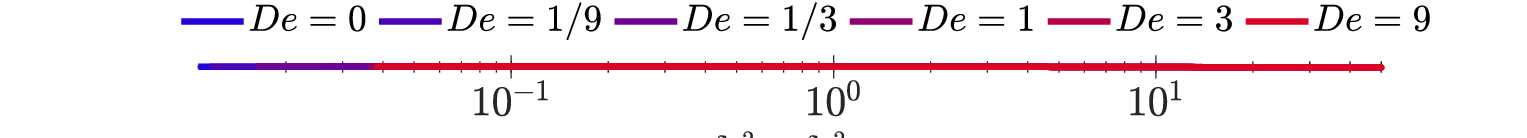}
\caption{(left) Dependence of the relative helicity $r\aver{\delta h}/(2 \aver{\delta q^2})$ on $r$ for different $De$. (right) Dependence of $\aver{\delta h}/\aver{\delta q^2}$ on $r$ for different $De$.}
\label{fig:heliDivener}
\end{figure}
We now examine the relative helicity, defined as $h_r \equiv r \langle \delta h \rangle / (2 \langle \delta q^2 \rangle)$ following \citet{borue-orszag-1997}, to assess how polymers influence the flow's tendency to restore mirror symmetry at small scales. As shown in figure~\ref{fig:heliDivener}, the flow exhibits maximal helicity at large scales, with $h_r$ decaying monotonically as $r$ decreases, indicating that helicity becomes dynamically negligible at small scales.

In the Newtonian case, we recover the classical scaling $h_r \sim r^1$ within the inertial range, consistent with the findings of \citet{borue-orszag-1997}. This behaviour is illustrated in the right panel of figure~\ref{fig:heliDivener}, where the ratio $\langle \delta h \rangle / \langle \delta q^2 \rangle \approx 0.75$ is observed, comparable to that found for passive scalar transport in isotropic turbulence \citep{borue-orszag-1997}.

In PHIT, however, the decay of $h_r$ with $r$ is noticeably slower in the intermediate range, and the Newtonian $r^1$ scaling in the inertial range is lost. This reflects the fact that $\langle \delta h \rangle$ and $\langle \delta q^2 \rangle$ scale differently for $De > 0$, as previously shown in figure~\ref{fig:duidui}. Furthermore, we find that $h_r$ increases systematically with $De$ across all scales, suggesting a growing asymmetry with increasing elasticity.

At the smallest scales, the ratio $\langle \delta h \rangle / \langle \delta q^2 \rangle$ tends to a plateau for all cases, which reflects the smoothness of the velocity and vorticity fields and is consistent with a Taylor expansion representation (see also figure~\ref{fig:duidui}). Overall, the results in figure~\ref{fig:heliDivener} indicate that polymers act to break mirror symmetry, and that this symmetry-breaking effect becomes more pronounced with increasing elasticity.

\section{Local structure of the flow}
\label{sec:loc-str}

In this section, we examine how polymer additives modify the local flow topology. While the preceding section has focused on the budgets of energy, enstrophy, and helicity, the local structure of the flow provides a complementary perspective, that connects small-scale dynamics more directly to the mechanisms of production, transfer and dissipation of the three quantities. Specifically, we relate the polymer-induced changes in energy transfer and enstrophy production to alterations in local strain and rotation dynamics.

The flow topology is characterised using the three principal invariants of the velocity gradient tensor $A_{ij} = \partial u_j/\partial x_i$ \citep{davidson-2004, meneveau-2011}, which offer compact yet insightful descriptors of the local kinematics.

\subsection{Local straining state and alignments}

The velocity gradient tensor $A_{ij} = \partial u_j / \partial x_i$ can be decomposed into its symmetric and antisymmetric components, namely $S_{ij}$ and the rate-of-rotation tensor $W_{ij}$:
\begin{equation}
A_{ij} = \frac{\partial u_j}{\partial x_i} =
\underbrace{ \frac{1}{2} \left( \frac{\partial u_j}{\partial x_i} + \frac{\partial u_i}{\partial x_j} \right) }_{S_{ij}} +
\underbrace{ \frac{1}{2} \left( \frac{\partial u_j}{\partial x_i} - \frac{\partial u_i}{\partial x_j} \right) }_{W_{ij}}.
\end{equation}
The tensor $S_{ij}$ admits a set of real eigenvalues $\alpha \geq \beta \geq \gamma$, corresponding to the principal strain rates, with associated orthonormal eigenvectors. The local kinematics of the flow can thus be characterised by the triplet $(\alpha, \beta, \gamma)$, by the orientation of the vorticity vector $\bm{\omega}$ with respect to the eigenvectors of $S_{ij}$, and by the magnitude of the vorticity vector, i.e. the enstrophy.
In incompressible flows, the trace of $S_{ij}$ vanishes, yielding the constraint $\alpha + \beta + \gamma = 0$. This implies that at least one eigenvalue must be non-positive ($\gamma \leq 0$), and at least one must be non-negative ($\alpha \geq 0$).

The vortex stretching term is related to the eigenvalues of $S_{ij}$ through the identity \citep[see][]{davidson-2004}:
\begin{equation}
\langle v_{st} \rangle = \langle \omega_i \omega_j S_{ij} \rangle = -4 \langle \alpha \beta \gamma \rangle.
\end{equation}
In Newtonian HIT this term is positive, i.e. $\langle \omega_i \omega_j S_{ij} \rangle > 0$, which implies that $\langle \alpha \beta \gamma \rangle < 0$. This reflects a characteristic flow configuration in which one large compressive strain is accompanied by two weaker extensional ones. On average, the eigenvalues of $S_{ij}$ follow the approximate ratios $(\langle \alpha \rangle, \langle \beta \rangle, \langle \gamma \rangle) \approx (3, 1, -4)|\beta|$ \citep{davidson-2004, meneveau-2011}. Note that, although this seems to be consistent with the
generation of sheet-like structures, it is also consistent with the stretching of vortex tubes, once the self-induced strain is considered \citep{davidson-2004}.

We start looking at the influence of the polymers on the local state of strain of the flow, by analysing the distribution of $\alpha$, $\beta$ and $\gamma$.
\begin{figure}
\centering
\includegraphics[width=0.9\textwidth]{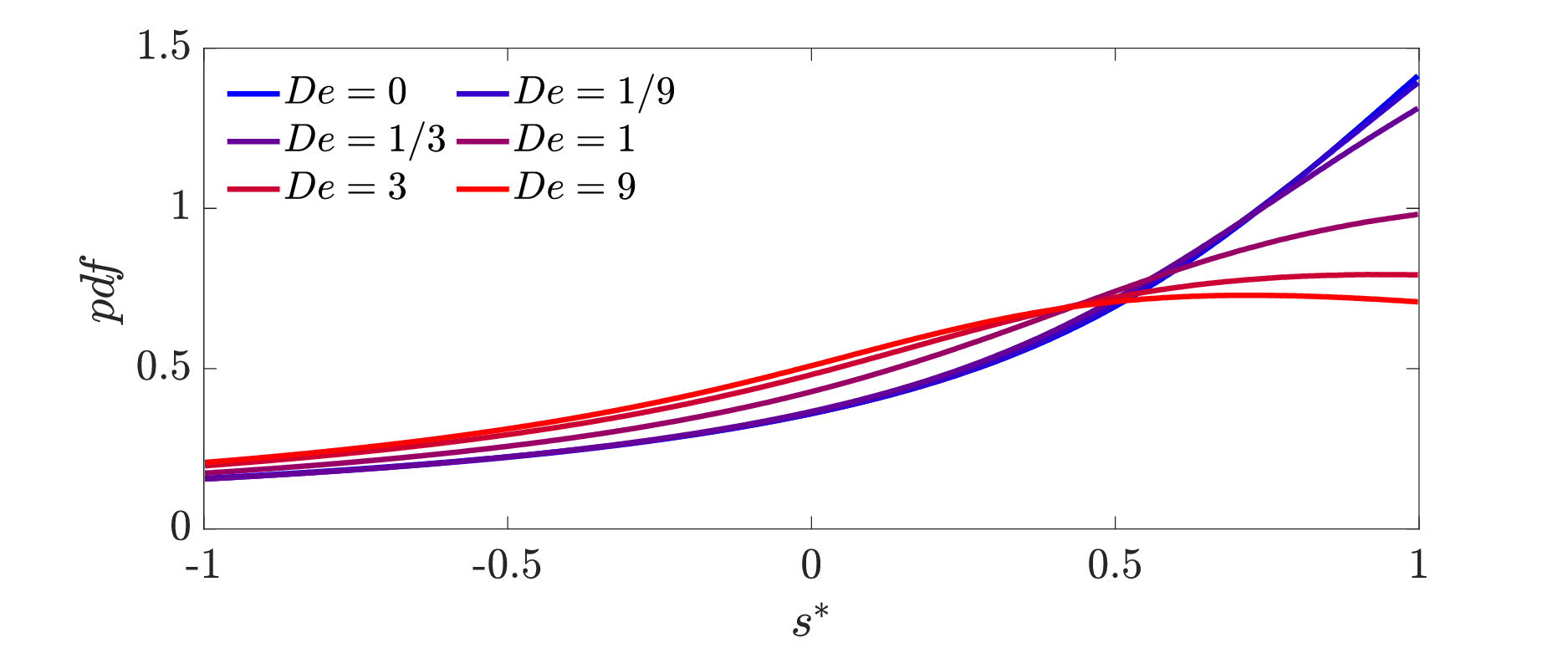}
\caption{Eigenvalues of $S_{ij}$. Distribution of $s^*= - 3 \sqrt{6} \alpha \beta \gamma/(\alpha^2 + \beta^2 + \gamma^2)^{3/2}$ for different $De$. }
\label{fig:sstar_abc}
\end{figure}
To quantify the local strain state, we consider the normalised parameter
\begin{equation}
s^* = - \frac{3 \sqrt{6} \alpha \beta \gamma}{(\alpha^2 + \beta^2 + \gamma^2)^{3/2}},
\end{equation}
originally introduced by \citet{lund-rogers-1994}. For a velocity gradient field with no preferred structure, the distribution of $s^*$ is uniform. The limiting values of $s^*$ correspond to distinct flow topologies. The value $s^* = 1$ occurs when $\alpha = \beta = -\gamma/2 > 0$, representing axisymmetric extension. In this state, a small fluid element symmetrically stretches along two principal directions while contracting along the third. Conversely, $s^* = 0$ corresponds to $\beta = 0$, indicating a two-dimensional straining state where the fluid element experiences equal stretching and compression in two directions. Here, the product $\alpha \beta \gamma = 0$ vanishes, and consequently the vortex stretching production term is zero, $v_{st} = \omega_i \omega_j S_{ij} = 0$.

\begin{figure}
  \centering
  \includegraphics[width=0.9\textwidth]{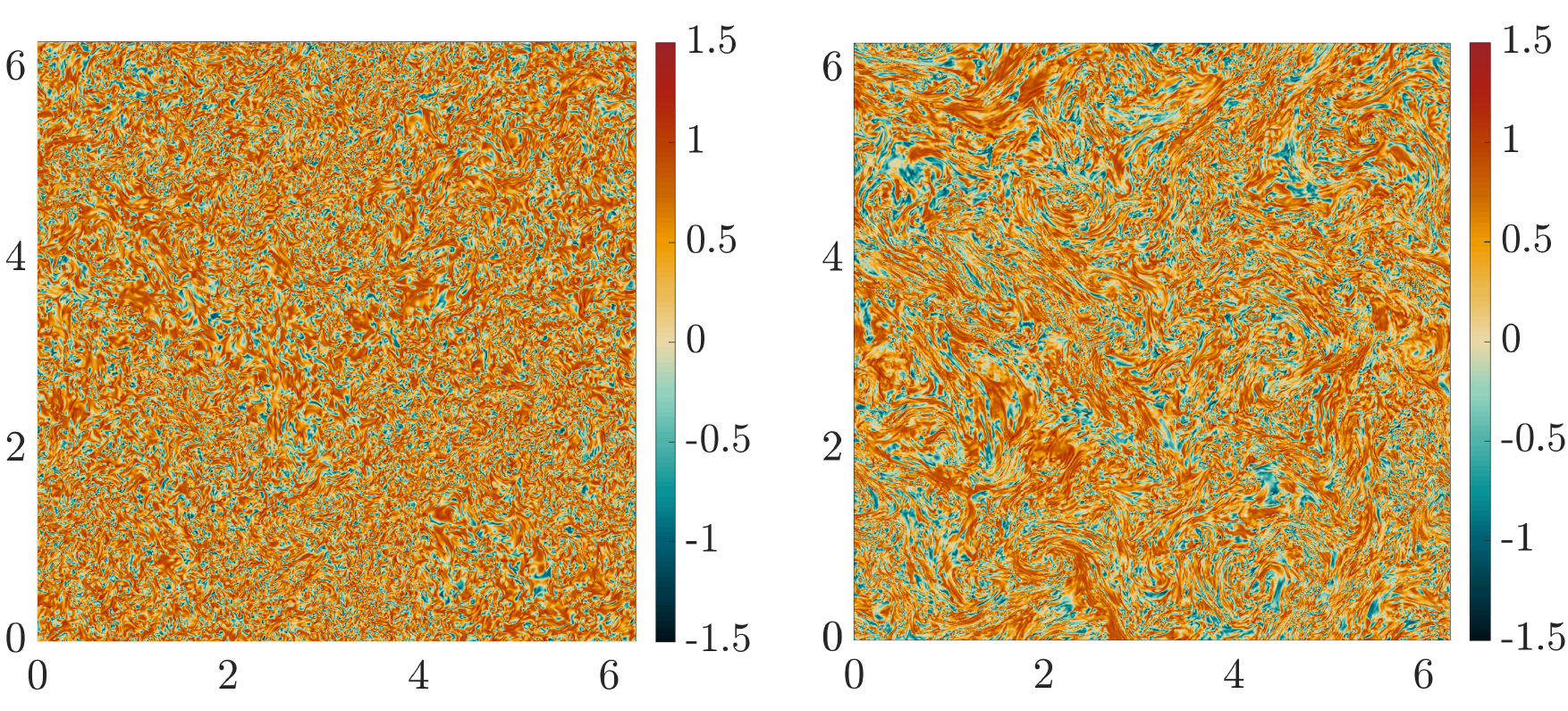}
  \caption{Instantaneous $s^*$ field on a two-dimensional slice for $De=0$ (left) and $De=1$ (right). Regions with $s^* \le 0$ (highlighted in blue and white) occur more frequently at $De=1$.}
  \label{fig:snap_sstar}
\end{figure}

Figure \ref{fig:sstar_abc} depicts the distribution of $s^*$ for $0 \le De \le 9$. Consistent with \cite{lund-rogers-1994}, the Newtonian HIT case shows a pronounced peak at $s^*=1$, indicating that axisymmetric extension is the most probable straining state. In contrast, as $De$ increases in PHIT, the distribution progressively flattens and the mode shifts toward $s^*=0$, reflecting a growing tendency toward quasi-two-dimensional strain states. This trend is further supported by direct inspection of the instantaneous flow fields illustrated in figure \ref{fig:snap_sstar}.

\begin{table}
\centering
\begin{tabular}{lcccccccccc}
	$De$ & & & $\aver{\alpha}/\aver{\beta}$ & & & $\aver{\gamma}/\aver{\beta}$ & & & $\aver{\beta}$ \\
	$1/9$   & & &  $4.3$      & & &    $-5.3$        & & &    $11$     \\
	$1/3$   & & & $4.5$       & & &    $-5.5$   		 & & &    $12$	  \\
	$1$     & & &  $5.9$      & & &    $-6.9$  		   & & &    $5$	   \\
	$3$     & & & $7.5$       & & &    $-8.5$    	   & & &    $4$	   \\
	$9$     & & &  $9.5$      & & &    $-10.5$   		 & & &    $3.5$	  \\
\end{tabular}
\caption{Dependence of the average eigenvalues of $S_{ij}$ on $De$.}
\label{tab:eigen}		
\end{table}

This shift toward two-dimensionality is further supported by the separate eigenvalue distributions of $\alpha$, $\beta$, and $\gamma$ (not shown for brevity). For the Newtonian case, $\beta$ is predominantly positive, with mean values $(\langle \alpha \rangle, \langle \beta \rangle, \langle \gamma \rangle) = (4.3, 1, -5.3) |\langle \beta \rangle|$, confirming the dominance of axisymmetric extension. For PHIT, the eigenvalue distributions contract and their modal values decrease monotonically with $De$, indicating a greater prevalence of shear and planar extensional regions (we quote in table \ref{tab:eigen} the average values).

The increased occurrence of two-dimensional straining explains the observed reduction in the average vortex stretching term $v_{st}$ with increasing $De$ (figure \ref{fig:VstEpsp}). This trend aligns with the results of \cite{warwaruk-ghaemi-2024}, who reported similar modifications in the local flow topology within polymer-laden turbulent boundary layers. Furthermore, these findings support the classic mechanism of polymer drag reduction proposed by \cite{lumley-1973} and \cite{roy-etal-2006}, whereby polymer extensional viscosity suppresses uniaxial and biaxial extensional flows, promoting two-dimensional strain states and thereby inhibiting the formation of quasi-streamwise vortices.

\begin{figure}
\centering
\includegraphics[width=0.325\textwidth]{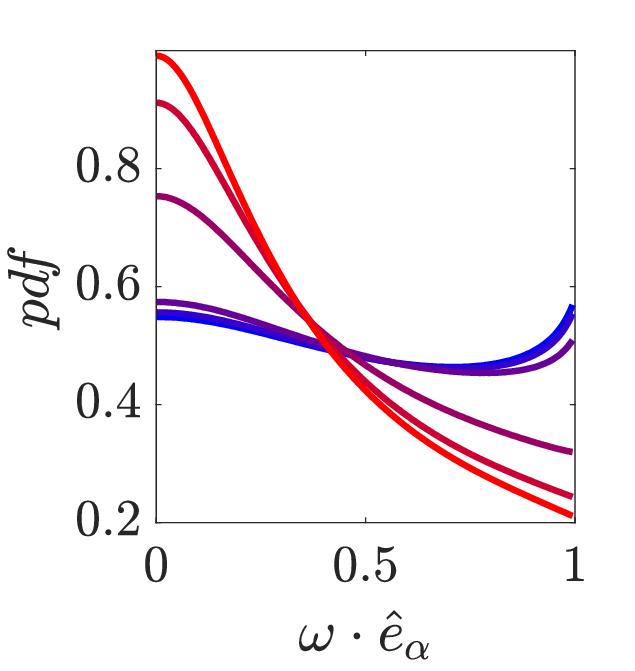}
\includegraphics[width=0.325\textwidth]{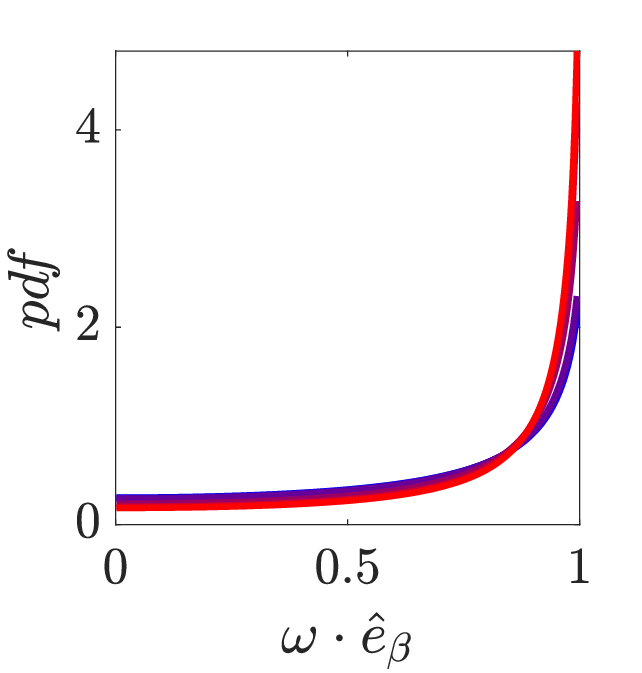}
\includegraphics[width=0.325\textwidth]{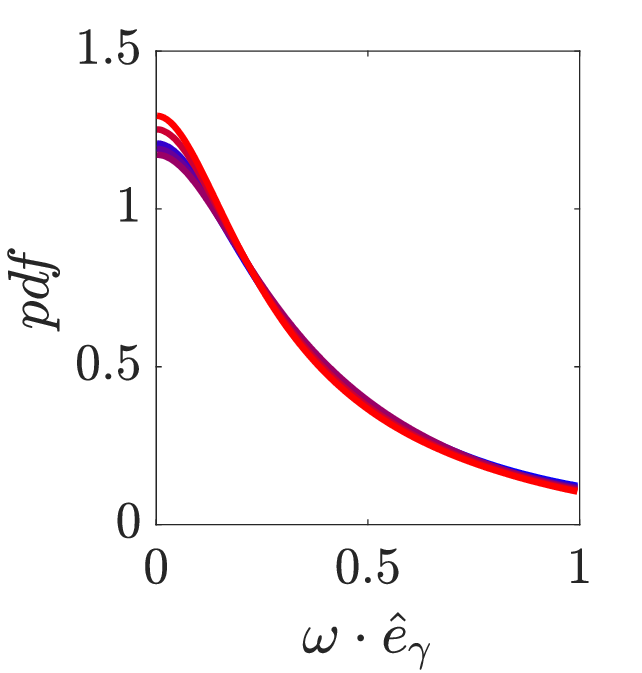}
\includegraphics[trim={0 50 0 0},clip,width=1.0\textwidth]{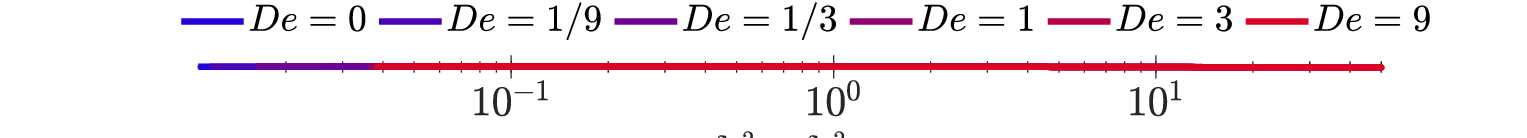}
\caption{Distribution of the cosine of the angle between the vorticity and the eigenvectors of the strain-rate tensor, for different $De$.}
\label{fig:vort-angle}
\end{figure}
To further characterise the influence of the polymers on the local flow topology, it is essential to examine the orientation of the vorticity vector $\bm{\omega}$ relative to the principal axes of the strain-rate tensor $S_{ij}$ ---  $\hat{\bm{e}}_\alpha$, $\hat{\bm{e}}_\beta$, and $\hat{\bm{e}}_\gamma$ --- which correspond to the eigenvectors associated with the eigenvalues $\alpha$, $\beta$ and $\gamma$ \citep{meneveau-2011}.  This alignment is closely linked to the process of vortex stretching, as expressed by the identity \citep{tsinober-2001}:
\begin{equation}
v_{st} = \omega_i \omega_j S_{ij} = \omega^2 \alpha \cos^2(\bm{\omega}, \hat{\bm{e}}_\alpha) +
\omega^2 \beta \cos^2(\bm{\omega}, \hat{\bm{e}}_\beta) +
\omega^2 \gamma \cos^2(\bm{\omega}, \hat{\bm{e}}_\gamma),
\end{equation}
which decomposes the vortex-stretching term into contributions from each principal strain direction. In Newtonian HIT, the dominant contribution typically arises from the extensional eigenvalue $\alpha$ \citep{ashurst-etal-1987}.
Figure \ref{fig:vort-angle} shows how this alignment distribution changes with increasing $De$. The results indicate that polymer additives progressively alter the orientation statistics of $\bm{\omega}$. Specifically, as $De$ increases, the probability of $\bm{\omega}$ being orthogonal to $\hat{\bm{e}}_\alpha$ increases, as seen from the growing peak at $\bm{\omega} \cdot \hat{\bm{e}}_\alpha = 0$. At the same time, alignment with the intermediate strain direction $\hat{\bm{e}}_\beta$ becomes more pronounced, with peaks at $\bm{\omega} \cdot \hat{\bm{e}}_\beta = 1$ strengthening with $De$.

These trends suggest that polymers suppress the classical vortex stretching mechanism via two complementary effects: (i) reducing the alignment between vorticity and the most extensional direction, thereby diminishing the dominant $\alpha$-term in $v_{st}$, and (ii) promoting alignment with the intermediate strain direction $\hat{\bm{e}}_\beta$, where stretching is weaker, while also increasing the frequency of configurations where $\beta \approx 0$, in which case the associated contribution to vortex stretching vanishes. Together, these modifications lead to the observed reduction in enstrophy production at high $De$.

\begin{figure}
  \centering
  \includegraphics[width=0.49\textwidth]{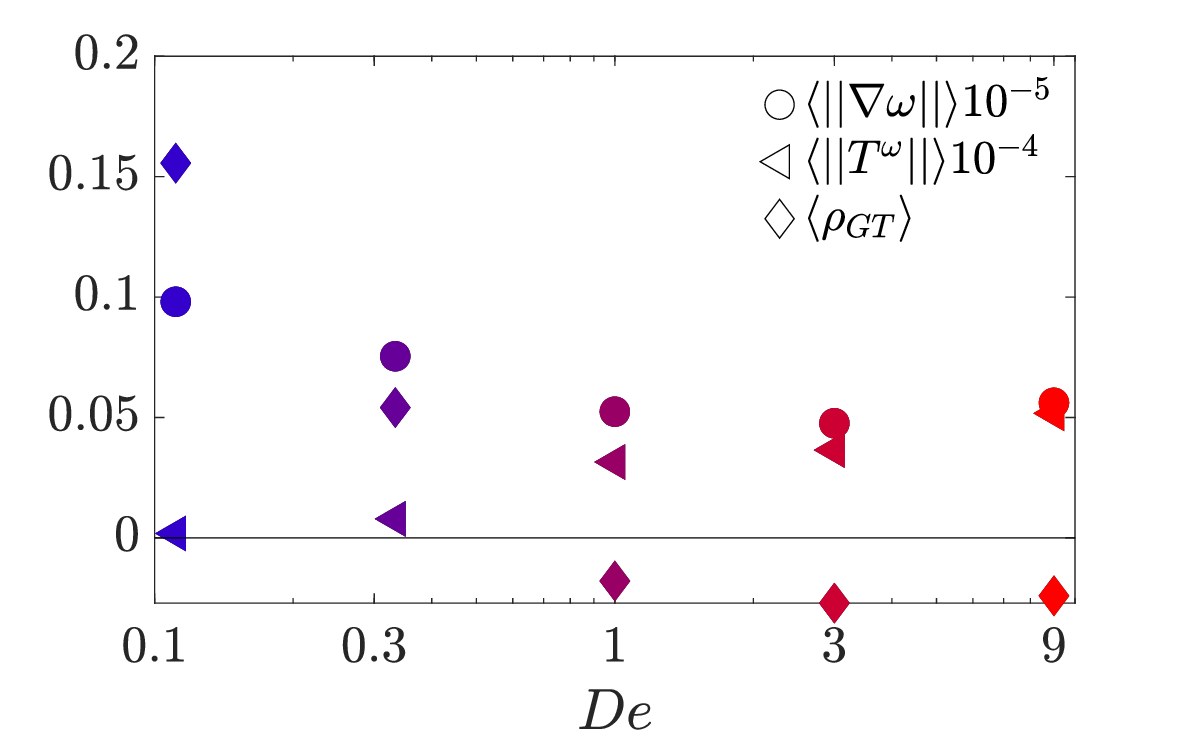}
  \includegraphics[width=0.49\textwidth]{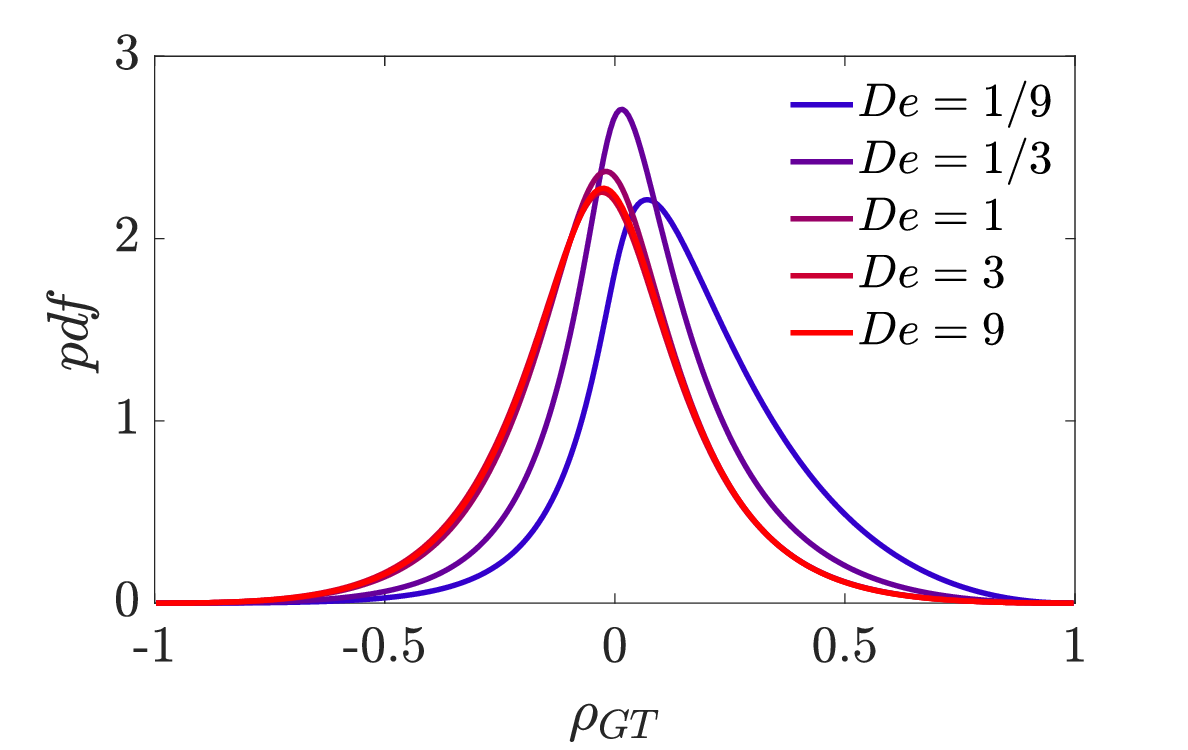}
  \caption{Terms contributing to $\pi^{\delta \omega^2}$ according to equation \ref{eq:pio}. Left: dependence of the average values of $|| T^{\omega} ||$, $ || \bm{\nabla} \bm{\omega} || $ and $ \rho_{GT}$ on $De$. Right: distribution of $\rho_{GT}$ for different values of $De$.}
  \label{fig:piEnst}
\end{figure}
We now connect the influence of $De$ on $\pi^{\delta \omega^2}$, as shown in figure~\ref{fig:VstEpsp}, to changes in the flow alignments. Specifically, we write
\begin{equation}
\pi^{\delta \omega^2} = T^{\omega} : \bm{\nabla} \bm{\omega} = || T^{\omega} ||  \  || \bm{\nabla} \bm{\omega} || \  \rho_{GT},
\label{eq:pio}
\end{equation}
where $-1 \leqslant \rho_{GT} \leqslant 1$ quantifies (loosely) the alignment between the tensors $T_{ij}^{\omega}$ and $\partial \omega_i / \partial x_j$, and we define $||A||^2 = A_{ij} A_{ij}$. The left panel of figure~\ref{fig:piEnst} shows the variation of the mean quantities $\langle || T^{\omega} || \rangle$, $\langle || \bm{\nabla} \bm{\omega} || \rangle$, and $\langle \rho_{GT} \rangle$ with increasing $De$. A clear monotonic decrease in $\langle \rho_{GT} \rangle$ is observed: for $De < 1$, $\langle \rho_{GT} \rangle > 0$, indicating average positive alignment, while for $De \ge 1$, $\langle \rho_{GT} \rangle < 0$, signifying a reversal in the mean alignment direction.
The growth of $\langle \pi^{\delta \omega^2} \rangle$ with $De$ is primarily driven by the increase in $\langle || T^{\omega} || \rangle$. In contrast, $\langle || \bm{\nabla} \bm{\omega} || \rangle$ exhibits a non-monotonic trend, with a minimum around $De \approx 1$. The observed sign reversal in $\langle \pi^{\delta \omega^2} \rangle$ for $De \ge 1$ is thus attributed to the change in alignment between $T_{ij}^{\omega}$ and $\partial \omega_i / \partial x_j$.

Additional insight is provided in the right panel of figure~\ref{fig:piEnst}, which shows the probability distribution of $\rho_{GT}$ for different values of $De$. In all cases, the distribution peaks near $\rho_{GT} = 0$, indicating generally weak alignment. However, for $De < 1$, the mode is slightly positive and positive alignment events ($\rho_{GT} > 0$) are more probable. For $De \ge 1$, the distribution shifts toward negative values, with negative alignment events ($\rho_{GT} < 0$) becoming dominant.

\subsection{Invariants of $A_{ij}$}

\begin{figure}
\centering
\includegraphics[width=0.49\textwidth]{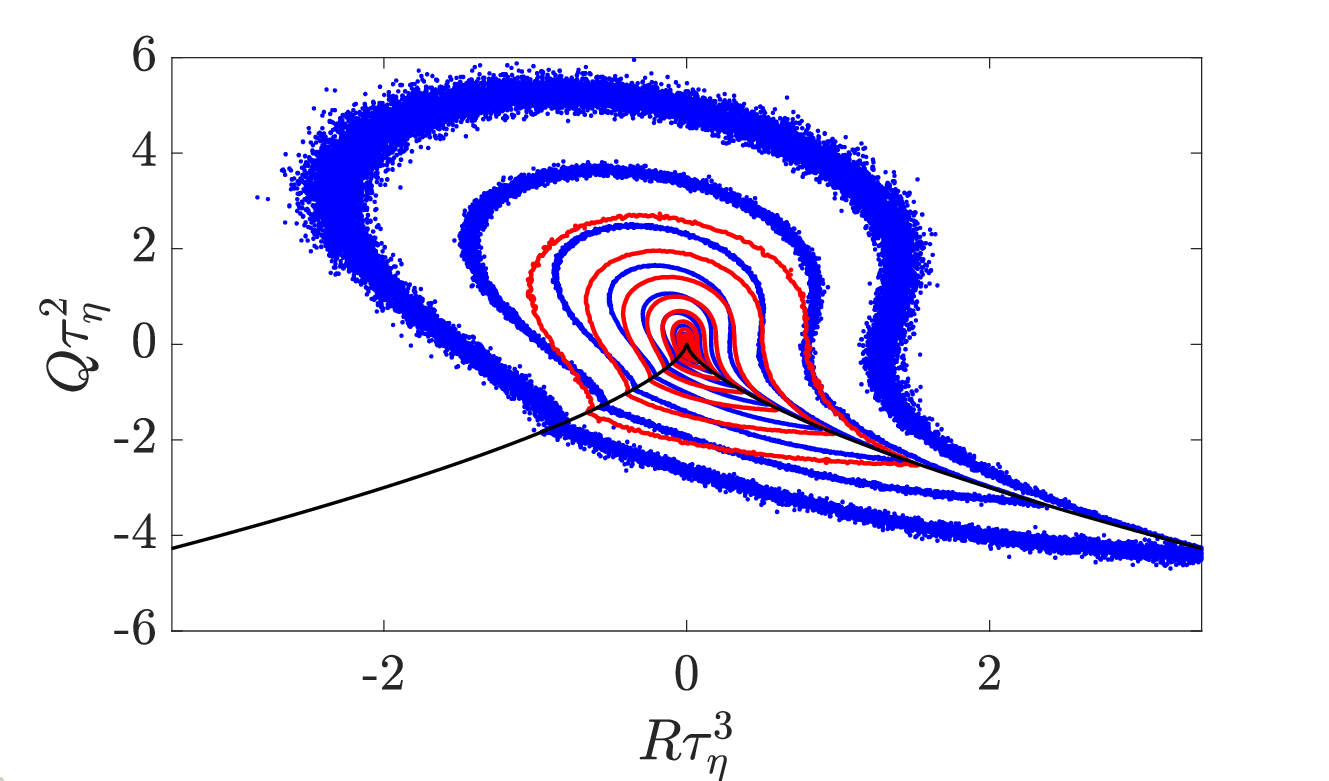}
\includegraphics[width=0.49\textwidth]{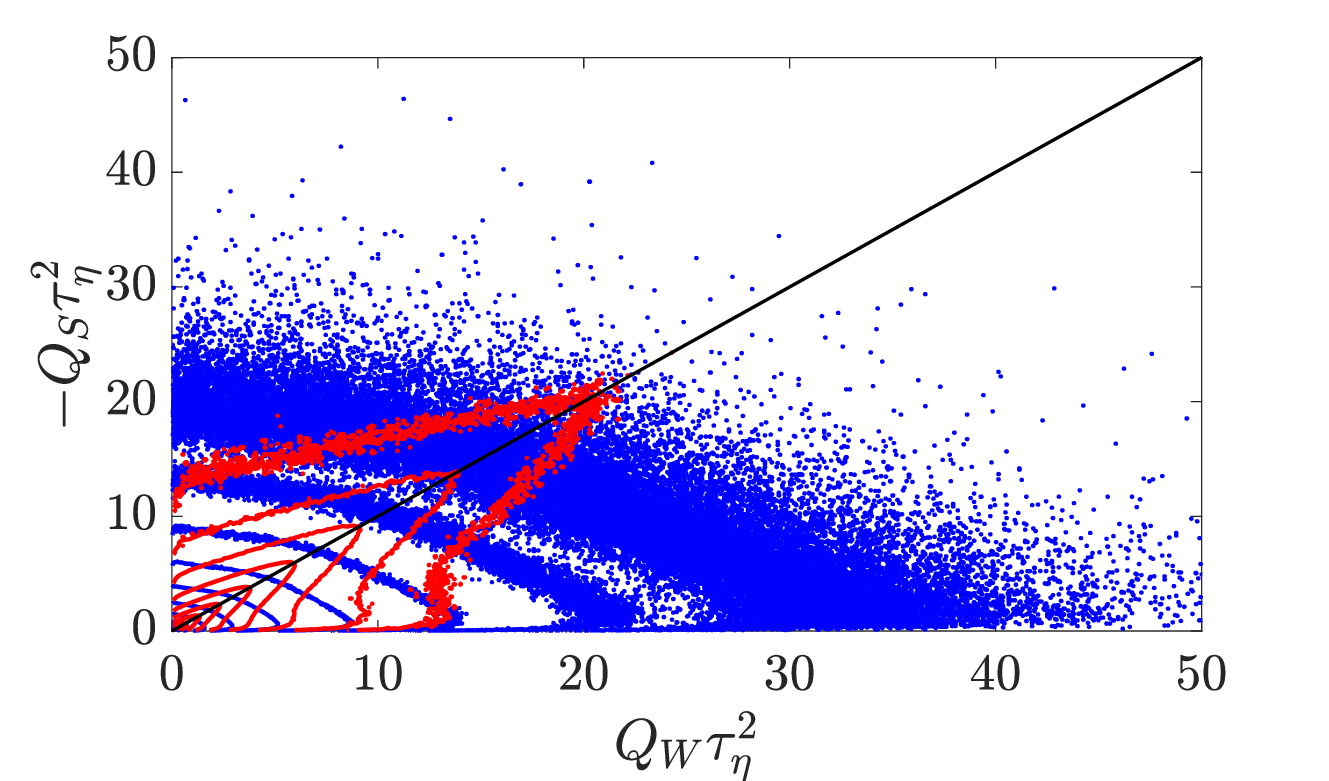}
\includegraphics[width=0.49\textwidth]{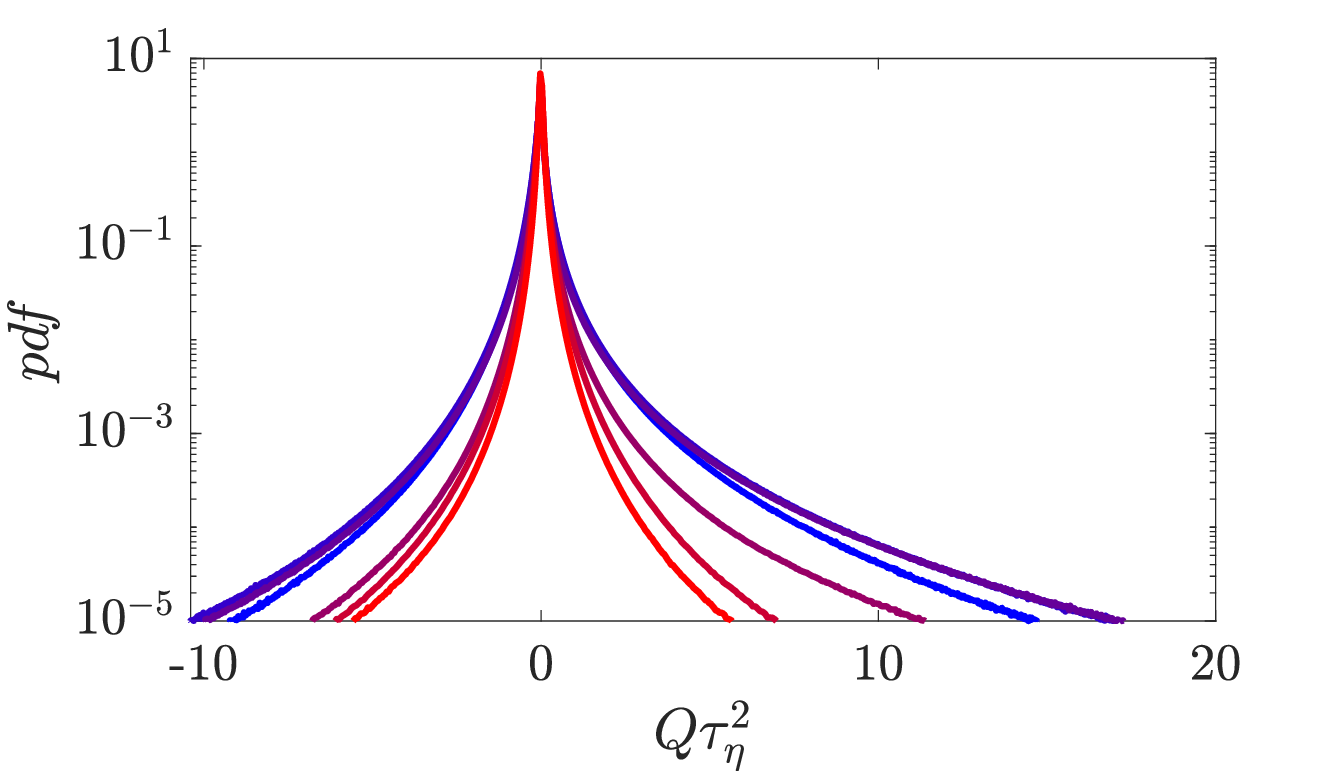}
\includegraphics[width=0.49\textwidth]{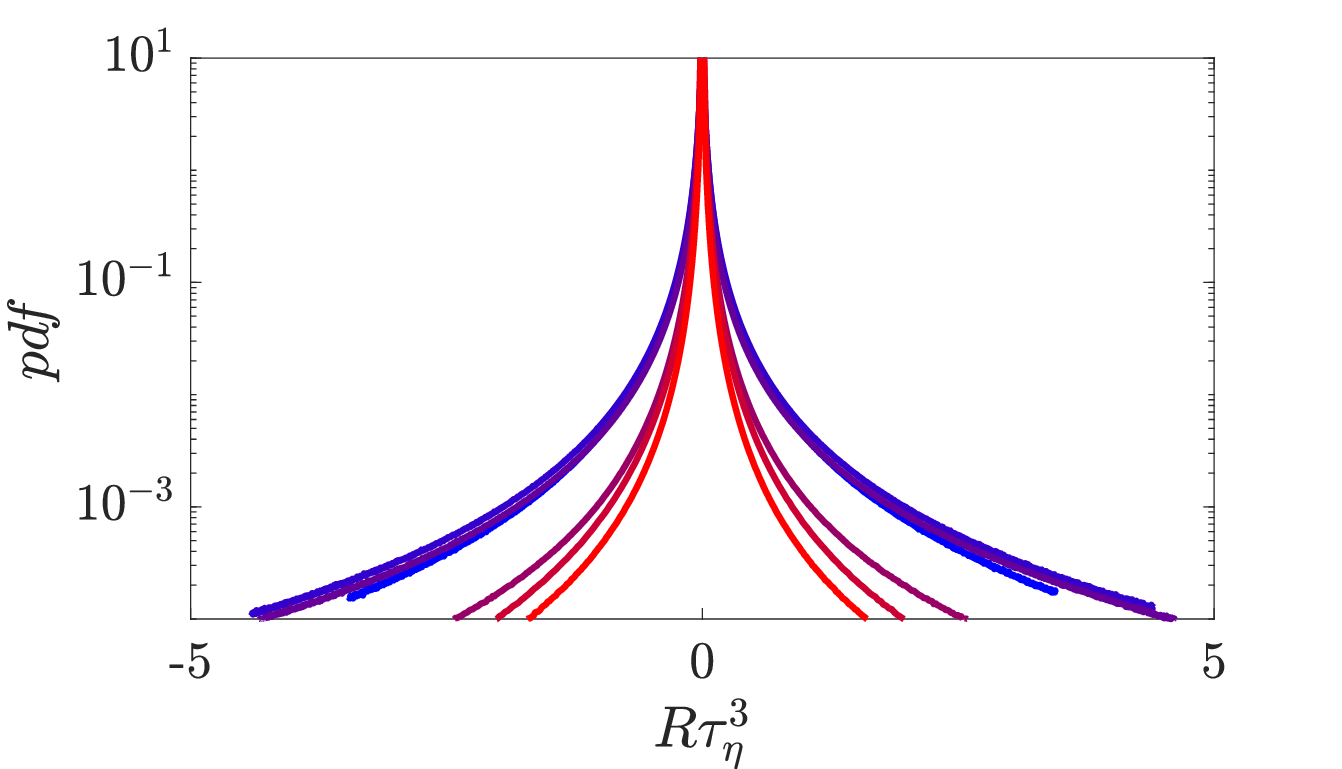}
\includegraphics[trim={0 50 0 0},clip,width=1.0\textwidth]{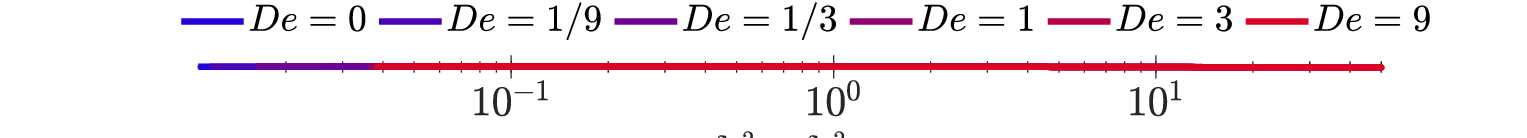}
\caption{Top left: The $Q-R$ map for $De=0$ (blue) and $De=9$ (red). Top right: The $Q_S-Q_W$ map for $De=0$ (blue) and $De=9$ (red). Bottom: distributions of $Q$ (left) and $R$ (right) for different $De$.}
\label{fig:QR_QSQW}
\end{figure}
To further determine the influence of the polymers on the topology of the flow motion, we now focus on two of the invariants of $\gradij$, i.e. $Q$ and $R$ \citep{cantwell-1993}, and plot their distributions in figure \ref{fig:QR_QSQW}. A second-order tensor in three-dimensions $A$ possesses three invariants, which are directly related to its eigenvalues $\lambda_i$ by means of the characteristic equation
\begin{equation}
\lambda^3 - P \lambda^2 + Q \lambda + R = 0.
\label{eq:QR}
\end{equation}
The three invariants are
\begin{equation}
 P = \lambda_1 + \lambda_2 + \lambda_3 = tr(A) = \alpha + \beta + \gamma \  (=0 \  \text{for incompressible flow}),
\end{equation}
\begin{equation}
Q = tr(A^2) = (\lambda_1 \lambda_2 + \lambda_2 \lambda_3 + \lambda_3 \lambda_1 ) =
 - ( \alpha^2 + \beta^2 + \gamma^2 ) + \frac{\omega^2}{4},
\end{equation}
and
\begin{equation}
R = \det(A_{ij}) = -(\lambda_1 \lambda_2 \lambda_3) = -\alpha \beta \gamma - \frac{\omega_i \omega_j S_{ij}}{4},
\end{equation}
where repeated indices are summed over and $tr(A) = A_{ii}$ denotes the trace of the generic tensor $A_{ij}$.
The second invariant $Q$ measures the relative strength of strain and vorticity, with $Q \ll 0$ indicating regions of strong strain, and $Q \gg 0$ indicating regions of intense vorticity. The third invariant is a measure of the relative intensity of the production of vorticity ($R<0$) and strain ($R>0$), see \cite{tsinober-2001}. In figure \ref{fig:QR_QSQW} we also plot the discriminant $\Delta = 27R^2/4+Q^{3} = 0$ curve for equation \eqref{eq:QR} with a black line. Below this curve, $\lambda_i$s are all real and the flow is dominated by strain. Above this curve, however, $\gradij$ has one real and two complex eigenvalues, and enstrophy dominates the flow. When $Q$ is large and positive, strain is locally weak and $R \sim - \omega_i \omega_j S_{ij}$: in this case $R<0$ implies vortex stretching, while $R>0$ implies vortex compression. When instead $Q$ is large and negative then $R \sim -\alpha \beta \gamma$: a negative $R$ implies a region of axial strain ($\alpha>0$; $\beta,\gamma<0$), while a positive $R$ implies a region of biaxial strain ($\alpha,\beta>0$; $\gamma<0$).

In Newtonian HIT, the $Q-R$ joint distribution has a tear-drop shape with a clear point at the right-Vieillefosse tail with $\Delta = 0$, $R>0$ and $Q<0$ \citep{ooi-etal-1999,elsinga-marusic-2010}. 
The largest probability is of events lying in the two quadrants where $QR<0$, meaning that there is strong negative correlation between $Q$ and $R$. In other words, the two most common states are vortex stretching $\omega_i \omega_j S_{ij}>0$ and biaxial strain $\alpha \beta \gamma <0$~\citep{betchov-1956,davidson-2004}. The distributions of $Q$ and $R$ become narrower in PHIT, in agreement with similar observations at small $Re$ by \cite{perlekar-mitra-pandit-2010}. This means that the presence of polymers inhibits the occurrence of vortical and dissipative motions as well as of intense fluid extensions and compressions. This is conveniently visualised in figure \ref{fig:QR_QSQW} by the shrinking distributions of $R$ with increasing $De$. The $Q-R$ joint distribution for $De=9$ shows that there is still a bias for biaxial extensions in PHIT ($\alpha,\beta,-\gamma, -Q,R>0$). However, stretching is largely diminished compared to Newtonian HIT, in agreement with more frequent two-dimensional states. The bottom left panel of figure \ref{fig:QR_QSQW} shows that an increase in $De$ leads to a more symmetric distribution of $Q$ showing a stronger inhibition of events with positive $Q$ (large enstrophy). Moreover, the shrinking of the $Q>0$ tail indicates weaker vorticity while the shrinking of the tails of the $R$ distribution hint at a weaker vortical stretching/compression. This is consistent with the discussion in preceding sections.

The right panel of figure \ref{fig:QR_QSQW} shows the $Q_S-Q_W$ joint distribution, where $Q_S$ and $Q_W$ are the second invariants of the $S_{ij}$ and $W_{ij}$ tensors such that $Q_S + Q_W = Q$:
\begin{equation}
Q_S = -\frac{1}{2} tr(S^2) \ \text{and} \ Q_W = -\frac{1}{2} tr(W^2).
\end{equation}
$Q_S$ and $Q_W$ respectively capture the local rates of strain and rotation. These invariants are related to fluid dissipation $\varepsilon_f^{\delta q^2}$ and fluid enstrophy $\omega^2$ as $\varepsilon_f^{\delta q^2}  = -4 \nu Q_S$ and $\omega^2 = 4Q_W$.
Therefore, the $Q_S-Q_W$ joint distribution also indicates whether the flow is dominated by dissipation or enstrophy. We look at $\mathcal{K} = (-Q_W/Q_S)^{1/2}$ \citep{truesdell-1954}; when $\mathcal{K} = 0$ the flow is extension dominated ($Q_S \gg Q_W)$, when $\mathcal{K} = \infty$ the flow undergoes rigid rotation locally, and is vorticity dominated ($Q_S \ll Q_W$), when $\mathcal{K} = 1$ rotation and stretching are equal, as typical for vortex sheets and shear layers. In Newtonian HIT, events with $Q_W>-Q_S$ are more frequent, meaning that the flow is mainly dominated by rigid rotations. In PHIT, instead, the shape of the distribution changes and events with $Q_W = - Q_S $ ($\mathcal{K} = 1$) are favoured. This is in line with the above observation that polymers favour two-dimensional strain states. 
 Note that the influence of polymers on the distribution of $Q_S$ ($-\varepsilon_f^{\delta q^2}$) and $Q_W$ ($\omega^2$) is rather different. This is visualised in the bottom left panel of figure \ref{fig:QR_QSQW}, which considers $Q$, with $Q_S$ and $Q_W$ contributing to the negative and positive values of $Q$, respectively. As $De$ increases, the positive tail of the $Q$ distribution (associated with rigid-body rotation, $Q_W$) narrows more significantly than the negative tail (associated with strain-dominated regions, $Q_S$). This indicates that polymers preferentially suppress intense rotational motions over strong extensional events.
 
\begin{figure}
\centering
\includegraphics[width=0.49\textwidth]{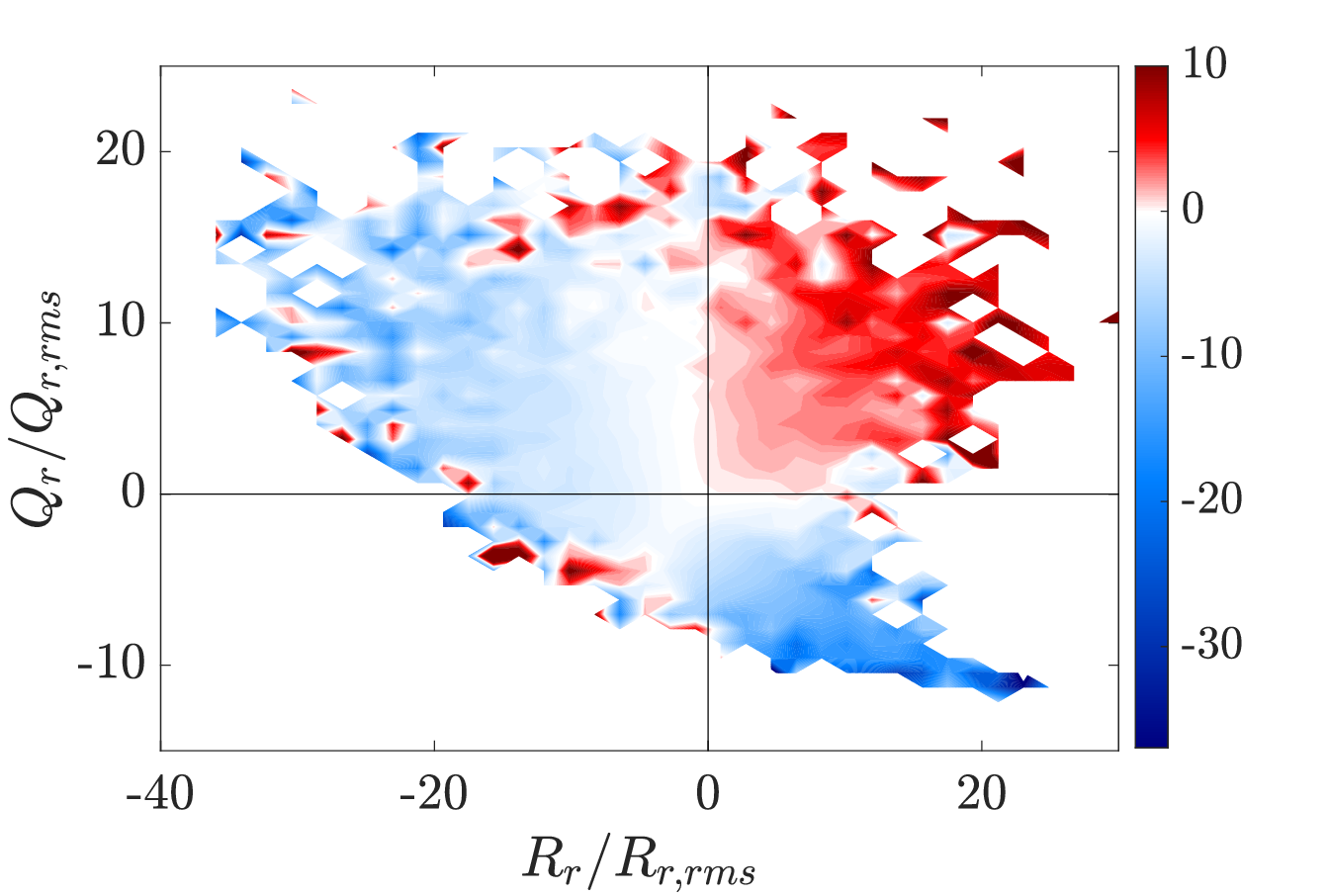}
\includegraphics[width=0.49\textwidth]{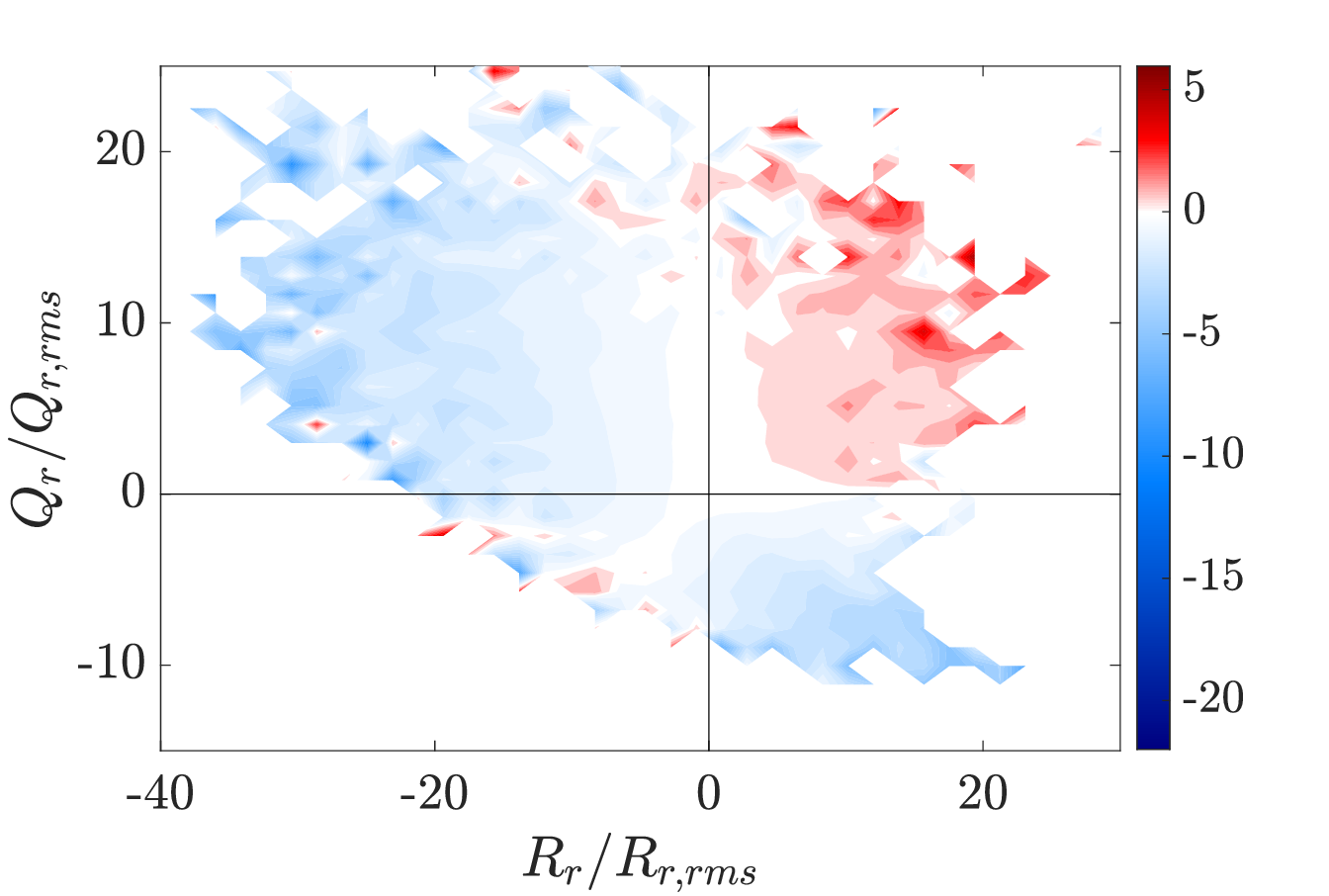}
\caption{Joint conditional average $\langle \Phi_f^{\delta q^2}(r) | Q_r, R_r \rangle$, for $De=0$ (left) and $De=1$ (right). The scale is set to $r=0.1$, which is within the elastic range of scales. The conditional flux is normalised with $\langle \varepsilon_f^{\delta q^2} \rangle r$. Positive values of $\Phi_f^{\delta q^2}$ denote inverse energy cascade, while negative values denote forward energy cascade.}
\label{fig:Flux_Q_R}
\end{figure}
We conclude this section by briefly relating the polymer-induced modifications of local flow topology to their impact on the energy cascade, as discussed in \S\ref{sec:energy}. Here, we focus on the behaviour of the nonlinear flux term $\Phi_f^{\delta q^2}$ within the elastic range of scales. Figure \ref{fig:Flux_Q_R} displays the joint conditional average $\langle \Phi_f^{\delta q^2}(r) | Q_r, R_r \rangle$, where
\begin{equation*}
  Q_r = \frac{1}{\mathcal{V}(r)} \int_{\mathcal{V}(r)} Q \text{d}\Omega \qquad \text{and} \qquad R_r = \frac{1}{\mathcal{V}(r)} \int_{\mathcal{V}(r)} R \text{d} \Omega.
\end{equation*}
The scale $r = 0.1$ is chosen to be well within the elastic range for sufficiently large $De$. Results are shown for the Newtonian case ($De = 0$, left panel) and the viscoelastic case ($De = 1$, right panel).

In both cases, $\Phi_f^{\delta q^2}(r)$ exhibits the characteristic teardrop-shaped structure in the $Q_r$–$R_r$ plane. A pronounced inverse cascade occurs in the quadrant where $Q_r > 0$ and $R_r > 0$, corresponding to vortex compression events \citep{yao-etal-2024b}. Conversely, forward cascade activity dominates the quadrants associated with vortex stretching ($Q_r > 0$, $R_r < 0$) and biaxial strain ($Q_r < 0$, $R_r > 0$), in line with the known role of strain in promoting forward energy transfer \citep{carbone-bragg-2020, jhonson-2021}.
While the overall structure remains similar between Newtonian and polymer-laden turbulence, at $De = 1$ the region associated with vortex compression and positive $\Phi_f^{\delta q^2}(r)$ becomes more sharply localised in the normalised $Q_r/Q_{r,\mathrm{rms}}$–$R_r/R_{r,\mathrm{rms}}$ plane. In contrast, the regions linked to vortex stretching and biaxial strain remain more spatially extended. Additionally, whereas the Newtonian case ($De=0$) exhibits the strongest forward cascade intensity in the quadrant with $Q_r < 0$ and $R_r > 0$, this peak is less pronounced in the viscoelastic case. In this case, forward cascade intensity becomes more balanced between vortex stretching and biaxial strain events. These findings suggest that, within the elastic range, polymers attenuate the inverse cascade by suppressing vortex compression and weaken the forward cascade mainly through the attenuation of biaxial strain events.

\section{Conclusion}
\label{sec:conclusions}

In this work, we have investigated the cascades of energy, enstrophy and helicity in homogeneous and isotropic polymeric turbulence. The study is based on direct numerical simulations of a dilute polymeric solution at $Re_\lambda \approx 460$, with $1/9 \le De \le 9$. We have extended the formulation introduced by \cite{baj-etal-2022} and derived the exact scale-by-scale budget equations for $\aver{\delta q^2}$, $\aver{\delta \omega^2}$ and $\aver{\delta h}$, for polymeric turbulent flows. The equations are general and remain valid for a generic inhomogeneous and anisotropic turbulent flow as well. They capture the mechanisms of production, transfer and dissipation in the combined space of scales and positions. Compared to purely Newtonian flows, the polymer/fluid interaction introduces additional sink/source process and an alternative scale-space transfer for the three quantities. The newly derived equations have been then tailored to homogeneous and isotropic turbulence for the present problem.

For $De \ge 1$ polymers effectively modify the cascades of energy and helicity, the two inviscid invariants of the $3D$ Navier--Stokes equations. Both $\aver{\delta q^2}$ and $\aver{\delta h}$ are injected in the system at the largest scales and transferred to the small scales by two distinct transfer processes driven by fluid inertia and fluid/polymer interactions, where they are dissipated away. 
At large scales, the nonlinear cascade dominates, while at small scales the polymer-driven transfer takes over. The cross-over scale between the two dominant processes agrees fairly well $r_p^*$ at which $De(r_p^*) \equiv \tau_p/\tau_f(r_p^*)=1$, where $\tau_p$ and $\tau_f(r)$ are the polymer relaxation time and the characteristic turnover time of eddies at scale $r$. When $r<r_p^*$ and $\tau_p>\tau_f(r)$ $(De(r)>1)$ polymers effectively interact with the $r$-fluctuations of the carrier flow. The coexistence of the two transfer mechanisms leads to a multiscaling behaviour of the velocity and helicity structure functions. At small $De$ and large $r$, $\aver{\delta q^2}$ and $\aver{\delta h}$ exhibit the same power law predicted by the Kolmogorov theory, i.e. $\aver{\delta q^2} \sim \aver{\delta h} \sim r^{2/3}$ with $\aver{\delta h}/\aver{\delta q^2} \approx 0.75$, consistently with helicity behaving as a passive scalar in HIT~\citep{borue-orszag-1997}. For $De \ge 1$ and at smaller scales, instead, both $\aver{\delta q^2}$ and $\aver{\delta h}$ deviate from the Kolmogorov predictions, with the latter showing a steeper slope. For $De = 1$ we measure $\aver{\delta q^2} \sim r^{1.3}$ \citep[in agreement with][]{zhang-etal-2021} and $\aver{\delta h} \sim r^{0.9}$. Accordingly, we observe that compared to the purely Newtonian case, the relative helicity $r \aver{\delta h}/2 \aver{\delta q^2}$ increases with $De$ at all scales, indicating that polymers favour events that break mirror symmetry at small scales.

A closer look at the energy fluxes reveals that polymers deplete the nonlinear cascade by weakening both the direct and inverse extreme events of both  cascades, leading to a less skewed flux distribution as $De$ increases. For $r \gtrapprox r_p^*$, the nonlinear flux distribution collapses reasonably well for all $De$ once the quantities are normalised with their standard deviation. 
Moving to the polymeric flux, we observe that the amount of energy carried by the polymer-driven cascade increases with $De$ at all scales.
Similar to the nonlinear flux, on average the polymeric flux transfers energy from larger to smaller scales, but local events with intense backscatter exist which transfer energy from small to large scales. However, the probability of very intense inverse transfer decreases as $De$ increases.

Unlike energy and helicity, enstrophy is not an inviscid invariant of the $3D$ Navier--Stokes equations. The budget equation for $\aver{\delta \omega^2}$, indeed, features a source term $\langle V_{st} \rangle$, which is related to the enstrophy produced by vortex stretching-like processes~\citep{davidson-2004}. The conventional notion of cascade used for energy and helicity, thus, does not apply in this case, as most of the enstrophy is directly generated at the small scales. However, vortex stretching is active in the inertial range which results in an increasing rate of transfer of $\aver{\delta \omega^2}$ at small $r$, being maximum at the end of the inertial range. For $De \ge 1$, the transfer via the polymeric route is comparable with the fluid nonlinear contribution and, unlike for $\aver{\delta q^2}$ and $\aver{\delta h}$, there is no transition with the two mechanisms coexisting in the entire inertial range. Overall, the net effect of polymers is to largely weaken vortex stretching, with fluid-polymer interactions becoming the primary enstrophy source at high $De$. The underlying local flow topology further reveals that vortex stretching modulation is a result of polymers promoting events with a two-dimensional state of straining, like shear and planar extensional flows. Accordingly, we observe that polymers favour events where rotation and stretching are equally strong (being typical of vortex sheets and shear layers), rather than those dominated by rigid rotation.

Having characterised the influence of polymer additives on the simultaneous transfer of energy, enstrophy and helicity in polymeric homogeneous isotropic turbulence, the present study will serve as a stepping stone for similar investigations in more complex settings that may serve to elucidate the underlying mechanism of polymeric drag reduction. With a look towards applications, the next step is to introduce inhomogeneity and/or anisotropy in the flow, and use the presented formulation to investigate the influence of the polymers on the transfers of $\aver{\delta q^2}$, $\aver{\delta \omega^2}$ and $\aver{\delta h}$ in the combined space of scales and positions. In order of complexity, we mention for example shear polymeric turbulence \citep{robert-etal-2010, warwaruk-ghaemi-2024}, polymer-laden turbulent channels \citep{min-etal-2003, izbassarov_rosti_brandt_tammisola_2021a, rota2024unified}, and jets \citep{guimaraes-etal-2020, soligo_rosti_2023a}.

\section*{Acknowledgments}
The authors acknowledge the computer time provided by the Scientific Computing and Data Analysis section of Research Support Division at OIST, and by HPCI, under the Research Project grants \textit{hp210269}, \textit{hp220099}, \textit{hp230018}, \textit{hp250021} and \textit{hp250035}.

\section*{Funding} 
The research was supported by the Okinawa Institute of Science and Technology Graduate University (OIST) with subsidy funding to M.E.R. from the Cabinet Office, Government of Japan. M.E.R. acknowledges funding from the Japan Society for the Promotion of Science (JSPS), grant 24K17210 and 24K00810.

\section*{Declaration of Interests} 
The authors report no conflict of interest.

\appendix
\section{Energy spectra}
\label{sec:spec}

\begin{figure}
  \centering
  \includegraphics[width=0.7\textwidth]{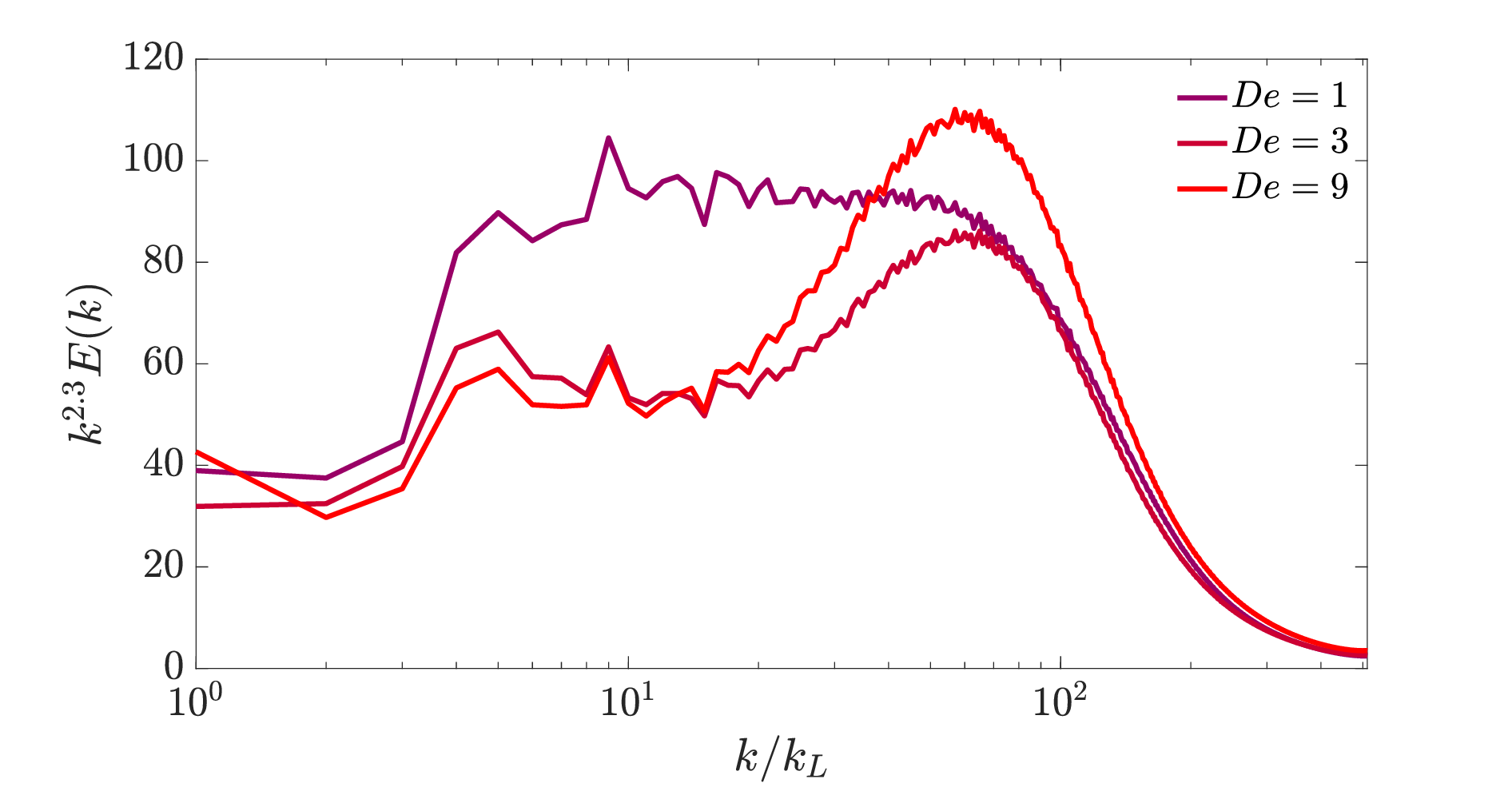}
  \caption{Compensated energy spectra $k^{2.3}E(k)$ for $1 \le De \le 9$}
  \label{fig:spec}
\end{figure}

Figure~\ref{fig:spec} shows the scale-compensated energy spectra, $k^{2.3}E(k)$, for $1 \le De \le 9$. In all cases, the compensated spectra display a well-defined plateau over an intermediate range of scales where polymer dynamics dominate, confirming the emergence of the $E(k) \sim k^{-2.3}$ scaling. The extent of this plateau is maximal at $De = 1$ and gradually decreases with increasing $De$, consistent with the progressive decoupling between the fluid and polymeric phases.

It is worth stressing that this elastic self-similar behaviour is more clearly discernible in Fourier space ($E(k)$) rather than in physical space ($S_2(r)$), and the transition between regimes becomes evident earlier in the spectral domain. This can be attributed to the fact that second-order structure function is, loosely speaking, an integrated representation of the energy spectrum. As a result, the $S_2$ scaling exponent observed at a given scale $r$ reflects the cumulative contribution from all wavenumbers larger than $k \gtrsim 2\pi / r$. \citet{rosti-perlekar-mitra-2023} showed that the range over which the spectrum follows the $k^{-2.3}$ scaling becomes narrower for $De > 1$, while the classical $k^{-5/3}$ inertial range correspondingly expands. Based on this reasoning, the second-order structure function exhibits a combination of the $1.3$ and classical $2/3$ scalings at intermediate $De>1$. The classical classical $2/3$ scaling is fully recovered only at sufficiently large $De$, where the inertia-dirven cascade dominates across all intermediate scales.

\bibliographystyle{../jfm}

\end{document}